\documentclass[aps,prd,amssymb,nofootinbib,floatfix,a4paper,eqsecnum,twocolumn]{revtex4}  

\usepackage{mathrsfs}
\usepackage{latexsym,amsmath,amsfonts,amssymb}
\usepackage{bm}
\usepackage{graphicx}

\allowdisplaybreaks

\newcommand{\beq}{\begin{equation}}
\newcommand{\eeq}{\end{equation}}
\newcommand{\bea}{\begin{eqnarray}}
\newcommand{\eea}{\end{eqnarray}}

\newcommand{\eq}[1]{\eqref{#1}}
\newcommand{\g}{{\gamma}}
\newcommand{\e}{\widehat{\mathcal E}_{\rm eff}}

\newcommand{\pinf}{p_{\infty}}
 
\newcommand{\ie}{{\it i.e.}}

\newif\ifusesec
\usesectrue  

\begin{document}

\title{Sixth post-Newtonian local-in-time dynamics of binary systems}

\author{Donato Bini$^{1,2}$, Thibault Damour$^3$, Andrea Geralico$^1$}
  \affiliation{
$^1$Istituto per le Applicazioni del Calcolo ``M. Picone,'' CNR, I-00185 Rome, Italy\\
$^2$INFN, Sezione di Roma Tre, I-00146 Rome, Italy\\
$^3$Institut des Hautes \'Etudes Scientifiques, 91440 Bures-sur-Yvette, France
}

\date{\today}

\begin{abstract}
Using a recently introduced method   [Phys.\ Rev.\ Lett.\  {\bf 123}, 231104 (2019)], which splits the conservative dynamics
of gravitationally interacting binary systems into a non-local-in-time part and a local-in-time one, 
we compute the local part of the dynamics at the sixth post-Newtonian (6PN) accuracy. 
Our strategy combines several theoretical formalisms: post-Newtonian, post-Minkowskian, multipolar-post-Minkowskian, 
effective-field-theory, gravitational self-force, effective one-body, and Delaunay averaging.   
The full functional structure of the local 6PN Hamiltonian (which involves 151 numerical coefficients)
is derived, but contains  four undetermined numerical coefficients.  
Our 6PN-accurate results are complete at orders $G^3$ and $G^4$, and the
derived $O(G^3)$ scattering angle agrees, within our 6PN accuracy, with the computation of 
[Phys.\ Rev.\ Lett.\  {\bf 122}, no. 20, 201603 (2019)]. All our results are expressed in several different
gauge-invariant ways. We highlight, and make a crucial use of, several aspects of the hidden simplicity of the 
mass-ratio dependence of the two-body dynamics.
\end{abstract}

\maketitle

\section{Introduction}

A new method for analytically computing the conservative dynamics of gravitationally interacting binary systems
has been recently introduced \cite{Bini2019}. This method draws its efficiency from combining in a specific way results
coming from several different theoretical formalisms:
 post-Newtonian (PN), post-Minkowskian (PM), multipolar-post-Minkowskian (MPM), effective-field-theory (EFT), gravitational self-force (SF),
effective one-body (EOB), and Delaunay averaging. We have recently applied this method to the 
derivation of the fifth post-Newtonian (5PN), and fifth-and-a-half post-Newtonian (5.5PN) dynamics \cite{Bini2020}.
Here, we extend the application of this method to the sixth post-Newtonian (6PN) level.

Let us recall the main idea and the various complementary steps of our strategy. As the main purpose of the
present paper is to present our new, 6PN-level results, we will be as brief as possible in guiding the reader
through the results presented below. For more details and references, see Refs. \cite{Bini2019,Bini2020}.

The main idea of our strategy is to decompose, from the start, the 
the total reduced\footnote{The reduced two-body action is defined as the two-worldline action obtained by integrating out
the mediating field from the original particle-plus-field action. It was introduced in electromagnetism by Schwarzschild, Tetrode and Fokker 
(see Ref. \cite{Wheeler:1949hn} for references and further developments). Its generalization to the gravitational two-body interaction was introduced 
in the PN context in Ref.  \cite{Infeld:1957zz}, and in the PM context in Ref.\cite{Damour:1995kt}.}
two-body conservative action ($S_{\rm tot}$)  in two separate pieces: a nonlocal-in-time part ($S_{\rm nonloc, f}$) 
and a local-in-time part ($S_{\rm loc, f}$). This decomposition is done at some given PN accuracy,  say $n$PN, and yields (when $n\geq4$)
an action of the form
\bea
\label{Sdecomp}
&&S_{\rm tot}^{\leq n \rm PN}[x_1(s_1), x_2(s_2)]=S_{\rm loc, f}^{\leq n \rm PN}[x_1(s_1), x_2(s_2)] \nonumber\\
&&\qquad\qquad +S_{\rm nonloc, f}^{\leq n \rm PN}[x_1(s_1), x_2(s_2)]\,. 
\eea
Here each action piece is a time-symmetric functional of the worldlines of the two bodies, say $x_1(s_1)$ and $x_2(s_2)$.
The meaning of the additional subscript f (which stands for ``flexibility factor'') will be discussed below.

The fact that the PN-approximated dynamics of a gravitationally interacting system must include,  starting at the 4PN level,
 a nonlocal-in-time part was discovered in Ref. \cite{Blanchet:1987wq} by using the  
 PN-matched~\cite{Blanchet:1987wq,Blanchet:1989ki,Damour:1990ji,Blanchet:1998in,Poujade:2001ie}  
multipolar-post-Minkowskian (MPM) formalism~\cite{Blanchet:1985sp}. The description of the 4PN-level nonlocal-in-time
(henceforth abbreviated as ``nonlocal'') dynamics by an action was initiated in Ref.~\cite{Foffa:2011np} 
(later refined in Ref.~\cite{Galley:2015kus}) within the EFT approach to the dynamics \cite{Goldberger:2004jt}
of binary systems and their coupling to radiation \cite{Goldberger:2009qd,Ross:2012fc}. However, the nonlocal action considered
in Refs. \cite{Foffa:2011np,Galley:2015kus} is a Schwinger-Keldysh-type, in-in, action, with doubled fields,  that is not 
 appropriate to the Tetrode-Fokker-type approach we are using. The corresponding appropriate time-symmetric 4PN-level
nonlocal action was first written down in Ref. \cite{Damour:2014jta}. See 
Refs. \cite{Bernard:2015njp,Marchand:2017pir,Foffa:2019rdf,Foffa:2019yfl} for later discussions of this 4PN nonlocal action.

The extension of the nonlocal  action to the 5PN level was obtained in Refs. \cite{Damour2010,Damour:2015isa},
with extension to the 5.5PN level in the latter reference. The derivation of these nonlocal actions in Refs. \cite{Damour:2014jta,Damour:2015isa}
was obtained by combining information from the MPM formalism, with special properties of the 1PN-accurate interaction of
a gravitationally system with an external tidal field \cite{Damour:1990pi,Damour:1991yw}.
Here, we need the extension of the nonlocal part of the action to the 6PN level. As emphasized in Refs. \cite{Foffa:2019eeb,Blanchet:2019rjs},
the EFT approach \cite{Goldberger:2004jt,Goldberger:2009qd,Ross:2012fc,Levi:2018nxp} is useful in this respect  and gives a guide for writing
the nonlocal part of the action beyond the leading order. We are, however, confused by the meaning of some of the equations presented
in Refs. \cite{Foffa:2019eeb,Blanchet:2019rjs} because they seem to refer to non conservative systems that should be treated by
a doubled-field  Schwinger-Keldysh-type, while we are interested in the Tetrode-Fokker-type time-symmetric action
for conservative systems. There is also a lack of explicit proof (beyond the 5PN level, which was explicitly treated in Ref. \cite{Damour:2015isa},
see also the Appendix A of Ref. \cite{Blanchet:2019rjs}) that the multipole moments to be used in the tail-transported nonlocal action
are the same as the ``canonical'' (or ``algorithmic'') moments, $M_L$, $S_L$, parametrizing the fully nonlinear multipolar structure
of gravitationally radiating systems in the MPM formalism~\cite{Blanchet:1985sp}. In addition, a consistent 6PN-level evaluation
of the nonlocal action requires (as will be made clear below) that the multipole moments, $M_L$, $S_L$, 
parametrizing the exterior MPM gravitational field be expressed as functionals of the source variables. The MPM formalism
succeeded in doing this task, and its appropriate results will be used below. Recent work \cite{Blanchet:2019rjs} provides
some partial checks of this circle of ideas at the level of the logarithmic terms associated with
nonlocal correlations\footnote{The fact that nonlocal interactions generate logarithmic terms was pointed out in
Refs. \cite{Damour:2009sm,Blanchet:2010zd}.}, in the restricted case of circular motions. 
Our work here will provide further checks concerning elliptic motions.

\section{Nonlocal action at the 6PN order}

The starting point for our method is to have in hand an explicit expression for the nonlocal part of the action, $S_{\rm nonloc, f}$.
At the 6PN accuracy, the nonlocal  action can be linearly decomposed into its $4+5+6$PN piece, and its 5.5PN piece
\beq \label{Snonloc00}
S_{\rm nonloc, f}^{ \leq 6 \rm PN}= S_{\rm nonloc, f}^{ 4+5+6 \rm PN}+  S_{\rm nonloc}^{ 5.5 \rm PN}\,.
\eeq
The 5.5PN piece (which is independent of the flexibility factor $f$) has already been treated in Ref. \cite{Bini2020} 
and will not be further discussed here.
In view of the work recalled above, the $4+5+6$PN  piece reads
\beq 
\label{Snonloc_0}
S_{\rm nonloc, f}^{ 4+5+6\rm PN}[x_1(s_1), x_2(s_2)]= -\int dt \, H_{\rm nonloc, f}^{4+5+6\rm PN}(t)\,, 
\eeq
with
\begin{eqnarray} 
\label{Snonloc}
H_{\rm nonloc, f}^{4+5+6 \rm PN}(t) &=& \frac{G^2 {\cal M}}{c^3} {\rm Pf}_{2 r_{12}^f(t)/c} \int  \frac{ dt'}{|t-t'|} {\cal F}_{ \rm 2PN}^{\rm split}(t,t')\,.\nonumber\\
\end{eqnarray} 
Here, ${\cal M}$ denotes the total ADM conserved mass-energy of the binary system, 
\beq
r_{12}^f(t)= f(t) r_{12}^h(t)\,,
\eeq
is a flexed version of the radial distance between the two bodies ($ r_{12}^h(t)$ denoting the harmonic-coordinate distance
and $f(t)$ being a function of the instantaneous state of the system),
while ${\cal F}_{ \rm 2PN}^{\rm split}(t,t')$ is
the time-split version of the fractionally 2PN-accurate gravitational-wave energy flux (absorbed and) emitted  by the (conservative) system. It can be decomposed  as
\begin{eqnarray}
{\cal F}_{ \rm 2PN}^{\rm split}(t,t')&=&\frac{G}{c^5} \left[F_{ I_2}^{\rm split}(t,t')+\eta^2F_{I_3, J_2}^{\rm split}(t,t')\right. \nonumber\\
&& \left.+\eta^4F_{ I_4, J_3}^{\rm split}(t,t')\right]\,,
\end{eqnarray}
with
\begin{eqnarray}
\label{flux2PNdef}
F_{  I_2}^{\rm split}(t,t')&=& \frac15 I_{ab}^{\rm (3)}(t) I_{ab}^{\rm (3)}(t') \,,\nonumber\\
F_{ I_3, J_2}^{\rm split}(t,t')&=& \frac1{189 } I_{abc}^{\rm (4)}(t) I_{abc}^{\rm (4)}(t') +\frac{16}{45 } J_{ab}^{\rm (3)}(t) J_{ab}^{\rm (3)}(t')\,,\nonumber\\
F_{I_4, J_3}^{\rm split}(t,t')&=& \frac{1}{9072}I_{abcd}^{\rm (5)}(t) I_{abcd}^{\rm (5)}(t') \nonumber\\
&&+\frac{1}{84}J_{abc}^{\rm (4)}(t) J_{abc}^{\rm (4)}(t')\,,
 \end{eqnarray}
where $\eta\equiv1/c$ and the superscript in parenthesis  denotes  repeated time-derivatives. The multipole moments $I_L$, $J_L$ denote
here the values of the canonical moments $M_L$, $S_L$ parametrizing (in a minimal, gauge-fixed way) the exterior field
(and therefore the relevant coupling between the system and a long-wavelength external radiation field) when they are
reexpressed as explicit functionals of the instantaneous state of the binary system. We employ here the notation used\footnote{In
more recent developments~\cite{Blanchet:2013haa} the notation $I_L$, $J_L$ refers to slightly different source-related moments, with
a difference starting at order $\frac1{c^5}$ which is, anyway, not relevant to the present work.} in the early works
on the PN-matched MPM formalism \cite{Blanchet:1989ki,Damour:1990ji,Damour:1994pk} in which the source-related values
of the algorithmic multipole moments, $M_L=I_L[{\rm source}]$, $S_L=J_L[{\rm source}]$, were obtained with 1PN fractional
accuracy. The latter accuracy suffices for the contribution involving $F_{ I_3, J_2}^{\rm split}(t,t')$ 
(and a fortiori $F_{I_4, J_3}^{\rm split}(t,t')$). However, for the first contribution involving $F_{  I_2}^{\rm split}(t,t')$
we need the 2PN-accurate value of the quadrupole moment expressed in terms of the material source~\cite{Blanchet:1995fr,Blanchet:1995fg}.
We need also to use the explicit form of the 2PN-accurate dynamics of a binary system in harmonic coordinates \cite{DD1981a,D1982},
and its relation \cite{Damour:1990jh} to the 2PN-accurate Hamiltonian in Arnowitt-Deser-Misner coordinates \cite{Schaefer:1986rd}.

The nonlocal Hamiltonian can be further decomposed into
\beq \label{Hnonlocf}
H_{\rm  nonloc, f}^{4+5+6 \rm PN}(t) =  H_{\rm  nonloc, h}^{4+5+6 \rm PN} + \Delta^{\rm f-h} H(t)\,,
\eeq
where, replacing ${\cal M} = \frac{H}{c^2}$ where $H$ is the Hamiltonian\footnote{At the present level, we can use the 2PN-accurate
Hamiltonian.}, and introducing an intermediate length scale $s$,
\begin{eqnarray} 
\label{Hnonloch}
 H_{\rm nonloc, h}^{4+5+6 \rm PN}(t)&=&
-\frac{G^2 H_{}}{c^{5}}{\rm Pf}_{2s/c}\int \frac{d\tau}{|\tau|}{\mathcal F}^{\rm split}_{\rm 2PN}(t,t+\tau)\nonumber\\
&+&2\frac{G^2 H_{}}{c^{5}}{\mathcal F}^{\rm split}_{\rm 2PN}(t,t)  \ln \left( \frac{r_{12}^h(t)}{s}\right)
\,,
\end{eqnarray} 
and
\beq \label{DHfh}
\Delta^{\rm f-h} H(t)= + 2\frac{G^2 H_{}}{c^{5}}{\mathcal F}^{\rm split}_{\rm 2PN}(t,t)  \ln \left( f(t)\right)\,.
\eeq
The corresponding local Hamiltonians are defined so that
\beq
H_{\rm tot}=  H_{\rm loc, h}+ H_{\rm nonloc, h}=  H_{\rm loc, f}+ H_{\rm nonloc, f}\,.
\eeq
In view of Eq. \eq{Hnonlocf}, we have
\beq
H_{\rm loc, h}= H_{\rm loc, f} + \Delta^{\rm f-h} H(t)\,,
\eeq
where it should be noted that $\Delta^{\rm f-h} H(t)$ is (like $f(t)$) a local function of the dynamical variables.

Depending on the various sections of this paper, we shall work either with the ``h-route" nonlocal Hamiltonian 
$H_{\rm nonloc, h}^{4+5+6 \rm PN}$, or the flexed ``f-route" local Hamiltonian  $H_{\rm loc, f}^{4+5+6 \rm PN}$.
As discussed in \cite{Bini2020}, the use of a suitable flexibility factor $f(t)$ within our strategy allows one to cleanly 
separate the determination of the local Hamiltonian $H_{\rm loc, f}$ from the nonlocal physics. 
The present paper will focus on the explicit
computation of the f-route local Hamiltonian $H_{\rm loc, f}$ (under the sole assumption that 
$f(t) = O(\nu)$). We leave to a separate work a full study of the
complementary nonlocal Hamiltonian $H_{\rm nonloc, f}$, and the determination of the flexibility factor $f(t)$.

\section{Computing the Delaunay average of the nonlocal-in-time h-route Hamiltonian} \label{sec:Hnonloch}

The first stage of our strategy consists of computing the Delaunay average of the nonlocal h-route Hamiltonian
$H_{\rm nonloc, h}$, Eq. \eqref{Hnonloch}. This computation is conveniently separated into several successive steps:
(1) computing the 2PN-accurate multipole moments entering ${\cal F}_{ \rm 2PN}^{\rm split}(t,t')$;
 (2) using a generic 2PN quasi-Keplerian parametrization of the motion;
(3) computing the quasi-Keplerian parameters in harmonic coordinates; (4)  computing the quasi-Keplerian parameters in EOB coordinates;
(5) evaluating the multipole moments along the orbit; and finally, (6) computing the Delaunay-average of the h-route nonlocal Hamiltonian 
 in harmonic coordinates.

In this section, we shall use as (rescaled) energy and angular momentum variables
\beq \label{defbarEj}
\bar E \equiv \frac{H-Mc^2}{\mu} \,,\qquad  \; j \equiv \frac{J}{G M \mu}\,.
\eeq
Beware that we shall also use other rescaled energy variables in other sections.

\subsection{The 2PN-accurate multipole moments in harmonic coordinates}

In this subsection, $x^i$ and $v^i \equiv \frac{ d x^i}{dt}$ denote the harmonic-coordinate {\it relative} center-of-mass
 position and  velocity of a two-body system.
One also uses the shorthand notation $L_i \equiv\epsilon_{ijk}x^jv^k$.
Using the standard notation for the symmetric and tracefree part of a tensor $T$, $T_{\langle ijk\ldots \rangle}$, and for the tensor product of two or more vectors $x_ix_jx_k\ldots =x_{ijk\ldots}$, the following results hold
\begin{eqnarray}
x_{\langle ijk \rangle}&=&x_{ijk}-\frac35 x^2 x_{(i}\delta_{jk)}\,,\nonumber\\
x_{\langle ij}v_{k\rangle}&=& x_{(ij}v_{k)}-\frac15 x^2 \delta_{(ij}v_{k)}-\frac25 ({\mathbf v}\cdot {\mathbf x})\delta_{(ij}x_{k)}\,,\nonumber\\
x_{\langle i}v_{jk\rangle}&=& v_{(ij}x_{k)}-\frac15 v^2 \delta_{(ij}x_{k)}-\frac25 ({\mathbf v}\cdot {\mathbf x})\delta_{(ij}v_{k)}\,,\nonumber\\
x_{\langle ijkl \rangle}&=&x_{ijkl}-\frac67 x^2 \delta_{(ij}x_{kl)}+\frac{3}{35}x^4 \delta_{(ij} \delta_{kl)}\,,\nonumber\\
L_{\langle i}x_{jk\rangle}&=& L_{(i}x_{jk)}-\frac15 x^2 \delta_{(ij}L_{k)}\,,\nonumber\\
L_{\langle i}x_{jkl\rangle}&=&L_{( i}x_{jkl)}-\frac37 x^2 \delta_{(ij}L_{k}x_{l)}\,,
\end{eqnarray}
where $x^2={\mathbf x}\cdot {\mathbf x}= x^i x^i$, ${\mathbf v}\cdot {\mathbf x} = v^i x^i$, etc., and where parentheses denote
 symmetrization (with weight one).

The mass quadrupole moment, $I_{ij}$  at the 2PN accuracy~\cite{Blanchet:1995fr,Blanchet:1995fg}, the mass octupole moment $I_{ijk}$, 
and mass hexadecapole moment, $I_{ijkl}$ at the 1PN accuracy \cite{Blanchet:1989ki} , the spin quadrupole moment, $J_{ij}$, and the spin octupole moment, $J_{ijk}$, at the 1PN  accuracy \cite{Damour:1990ji,Damour:1994pk},  have the following expressions
\cite{Arun:2007rg}

\begin{eqnarray}
I_{ij}&=&\mu [C_1 x_{\langle ij \rangle}+C_2 v_{\langle ij \rangle}+C_3 x_{\langle i}v_{j \rangle}]\,,\nonumber\\
I_{ijk}&=&\mu [B_1 x_{\langle ijk \rangle}+B_2x_{\langle ij}v_{k\rangle}+B_3x_{\langle i}v_{jk\rangle}]\,,\nonumber\\
I_{ijkl}&=&\mu  (1-3\nu)x_{\langle ijkl \rangle}\,,\nonumber\\
J_{ij}&=& \mu[D_1 L_{\langle i}x_{j\rangle}+D_2L_{\langle i}v_{j\rangle}]\,,\nonumber\\
J_{ijk}&=&\mu (1-3\nu)L_{\langle i}x_{jk\rangle}\,,
\end{eqnarray}
where the various parameters $C_1,C_2,\ldots$ etc. are listed in Table \ref{table_flux_relations}.


\begin{table*}
\caption{\label{table_flux_relations} Parameters entering the multipolar moments used in the 2PN flux.
}
\begin{ruledtabular}
\begin{tabular}{ll}
\hline
$C_1$ &$ 1+ \eta^2 \left[\frac{29}{42}  (1-3\nu) v^2 -\frac17 (5-8\nu) \frac{G M}{r}\right]$\\
&$+
 \eta^4  \left[\frac{G M}{r} v^2 \left(\frac{2021}{756}-\frac{5947}{756}\nu-\frac{4833}{756}\nu^2\right)
+\frac{G^2 M^2}{r^2} \left(\frac{355}{252}-\frac{953}{126}\nu+\frac{337}{252}\nu^2\right)\right. $\\
&$ \left.
+v^2 \left(\frac{253}{504}-\frac{1835}{504}\nu+\frac{3545}{504}\nu^2\right)
+\frac{G M}{r} \dot r^2 \left(-\frac{131}{756}+\frac{907}{756}\nu-\frac{1273}{756}\nu^2\right)\right]$\\
$C_2$ &  $
 \eta^2 r^2\left\{\frac{11}{21} (1-3\nu)  
+ \eta^2  \left[\frac{G M}{r} \left(\frac{106}{27}-\frac{335}{189}\nu-\frac{985}{189}\nu^2\right)+v^2\left(\frac{41}{126}-\frac{337}{126}\nu+\frac{733}{126}\nu^2\right)
+\dot r^2 \left(\frac{5}{63}-\frac{25}{63} \nu+\frac{25}{63}\nu^2\right)\right]\right\}$\\
$C_3$ &$ 2 \eta^2 r\dot r  
\left\{ -\frac{2}{7}+\frac{6}{7}\nu+ \eta^2 
\left[v^2 \left(-\frac{13}{63}+\frac{101}{63}\nu-\frac{209}{63}\nu^2\right)
+\frac{G M}{r} \left(-\frac{155}{108}+\frac{4057}{756}\nu+\frac{209}{108}\nu^2\right)\right]\right\}$\\
$B_1$ & $\sqrt{1-4\nu}\left\{-1+ \eta^2  \left[\frac{GM}{r} \left(\frac56 -\frac{13}{6}\nu\right)+v^2 \left(-\frac56 +\frac{19}{6}\nu\right)\right]\right\}$\\
$B_2$ &$ \sqrt{1-4\nu}(1-2\nu)\,  \eta^2  r \dot r $\\
$B_3$ &$- \sqrt{1-4\nu}(1-2\nu)\,  \eta^2  r^2 $\\
$D_1$ &$ \sqrt{1-4\nu}\left\{-1+ \eta^2  \left[\frac{GM}{r} \left(-\frac{27}{14}-\frac{15}{7}\nu\right)+v^2\left(-\frac{13}{28}+\frac{17}{7}\nu\right)\right]\right\}$\\
$D_2$ &$ \sqrt{1-4\nu}  r \dot r \left(-\frac{5}{28}-\frac{5}{14}\nu\right) \eta^2 $\\
\end{tabular}
\end{ruledtabular}
\end{table*}

\subsection{Generic 2PN quasi-Keplerian parametrization of elliptic motion (valid in all coordinates)}
 
The 2PN quasi-Keplerian parametrization \cite{Damour:1988mr,SW1993,Memmesheimer:2004cv} of elliptic motion,
in polar coordinates, $(r, \phi)$,  is the following
\begin{eqnarray} \label{QK}
r&=& a_r (1-e_r \cos u)\,,\nonumber\\
\ell &=& n (t-t_P)=u-e_t \sin u + f_t \sin V+g_t(V-u)\,,\nonumber\\
\bar \phi &=&\frac{\phi-\phi_P}{K}=V+f_\phi \sin 2V+g_\phi \sin 3V\,,
\end{eqnarray}
where
\beq
V(u)=2 \,{\rm arctan}\left[\sqrt{\frac{1+e_\phi}{1-e_\phi}}\tan \frac{u}{2}  \right]\,.
\eeq

Here $a_r$ is the semi-major axis of the orbit, $e_t,e_r,e_\phi$ are three  kinds of eccentricities, $K$ is the periastron advance and $n = \frac{2\pi}{T}$ is the 
circular frequency of the radial motion. This representation is valid (at 2PN) in any (usual) coordinate systems:  harmonic, ADM, or EOB.
The gauge-invariant quantities $K$ and $n$ are numerically the same in all coordinates, while the quasi-Keplerian elements 
$a_r, e_t,e_r,e_\phi$ depend on the coordinate system. We will distinguish them by decorating them with
 an extra label; for example $e_t^h$ for the harmonic coordinate expression, $e_t^e$ for the EOB coordinate expression, etc.
To ease the notation, we will omit the extra label specification  when it is clear from the context what are the coordinates used.
Most of the time we will (as in our previous works) use rescaled versions of many physical quantities. Notably, we use a 
dimensionless radial distance $r= r^{\rm phys}/(G M \eta^2)$ and a dimensionless radial period $T= T^{\rm phys}/(G M \eta^3)$.

We recall that
\begin{eqnarray}
V'(u) &=&  \frac{\sqrt{1-e_\phi^2}}{1-e_\phi \cos u}\,, \nonumber\\
\sin V&=&\frac{\sqrt{1-e_\phi^2}\, \sin u}{1-e_\phi \cos u} \,,\nonumber\\
\cos V&=&\frac{\cos u -e_\phi}{1-e_\phi \cos u}\,.
\end{eqnarray}
These relations imply, for example, the following explicit expression for $\ell(u)$
\begin{eqnarray}
\ell &=&u-e_t \sin u + f_t \frac{\sqrt{1-e_\phi^2}\, \sin u}{1-e_\phi \cos u}  +g_t(V-u)\nonumber\\
&=& (1-g_t)u-e_t \sin u + f_t \sqrt{1-e_\phi^2}\frac{\sin u}{1-e_\phi \cos u}\nonumber\\
&+ &  2g_t \, {\rm arctan}\left[\sqrt{\frac{1+e_\phi}{1-e_\phi}}\tan \frac{u}{2}  \right]\,, 
\end{eqnarray}
or, equivalently, replacing $u$ in terms of $V$, the explicit expression for $\ell(V)$
\begin{eqnarray}
\ell &=& 2(1-g_t)\,  {\rm arctan}\left[\sqrt{\frac{1-e_\phi}{1+e_\phi}} \tan \frac{V}{2}
\right]\nonumber\\
&-& e_t \sqrt{1-e_\phi^2}\frac{\sin V}{1+e_\phi \cos V}\nonumber\\
&+&  f_t \sin V+g_t V \,.
\end{eqnarray}

\subsection{2PN expressions of the orbital parameters in harmonic coordinates}

To get gauge-invariant expressions for the Keplerian elements one needs to relate them to the
conserved 2PN-accurate energy and angular momentum \cite{DD1981}.
We made use of explicit (3PN-accurate)  results in the literature \cite{Memmesheimer:2004cv,Arun:2007rg}.
[Note that, at the 3PN level, one needs to transform away some harmonic-gauge-related logarithms.]

We list in  Table \ref{h_orbital_elements2PN} the 2PN-accurate expressions of the {\it harmonic-coordinate} orbital parameters, 
as functions of the conserved energy and angular momentum of the system, as defined in Eq. \eq{defbarEj}. We use the shorthand notation
\beq
e_{\rm N}^2(\bar E,j) \equiv 1+2\bar Ej^2\,.
\eeq

\begin{table*}
\caption{\label{h_orbital_elements2PN}  2PN expressions of the harmonic-coordinates orbital parameters, as functions of the conserved energy and angular momentum of the system, Eq. \eq{defbarEj}.}
\begin{ruledtabular}
\begin{tabular}{l|l}
\hline
$n$  &$ (-2\bar E)^{3/2}\left[1+\eta^2\frac{-15+\nu}{8}(-2\bar E) +\eta^4\frac{(-2\bar E)^2}{128}\left(555+30\nu+11\nu^2 +\frac{192}{\sqrt{-2\bar E}j}(-5+2\nu)  \right) \right]$\\
$a_r$ & $\frac{1}{(-2\bar E)}\left\{1+\eta^2\frac{-7+\nu}{4}(-2\bar E )+\eta^4\frac{(-2\bar E)^2}{16}\left[ 1+\nu^2+\frac{16}{(-2\bar E)j^2}(7\nu-4) \right] \right\}$\\
$e_t^2$&$e_{\rm N}^2 +\frac{(- 2\bar E)}{4}\eta^2 [-8+8\nu-(-2\bar E)j^2(-17+7\nu)]
+\frac{(-2\bar E)^2}{8}\left[4(3+18\nu+5\nu^2)-(-2\bar E)j^2(112-47\nu+16\nu^2)
\right. $\\
&$ \left. -\frac{16}{(-2\bar E)j^2}(-4+7\nu)-24 \sqrt{-2\bar E}j(-5+2\nu)+\frac{24}{\sqrt{-2\bar E} j}(-5+2\nu)\right]\eta^4  $\\
$e_r^2$&$ e_{\rm N}^2+ 2\bar Ej^2 \left[\frac54\left(3-\nu \right) (-2\bar E ) +\frac{-6+\nu}{j^2} \right] \eta^2$\\
&$  +\frac{(-2\bar E)^2}{8}\left[2( 30+74\nu+\nu^2)-(80-45\nu+4\nu^2)(-2\bar E) j^2-\frac{32}{(-2\bar E)j^2}(-4+7 \nu)\right]\eta^4  $\\
$e_\phi^2$ &$ e_{\rm N}^2+ 2\bar Ej^2 \left[\left(\frac{15}{4}-\frac{1}{4}\nu \right) (-2\bar E)   -\frac{6}{j^2} \right]\eta^2$\\
&$  
-\frac{(-2\bar E)^2}{16 (-2\bar E)j^2}\left[-416+91\nu+15\nu^2-2(-2\bar  E)j^2(-20+17\nu+9\nu^2) +(-2\bar E)^2j^4(160-31\nu+3\nu^2)\right]\eta^4   $\\
$f_t$&$ -\frac{(-2\bar E)^{3/2} e_{\rm N}}{8j }\nu\left(-15+\nu \right)\eta^4$\\
$g_t$&$ -\frac32 \frac{(-2\bar E)^{3/2}}{j} \left(-5+2\nu \right)\eta^4$\\
$f_\phi$&$\eta^4 \frac{e_{\rm N}^2}{8j^4} \left(1+19\nu -3\nu^2 \right)$\\
$g_\phi$&$-\eta^4 \frac{1}{32}\frac{e_{\rm N}^3}{j^4}    \nu\left(-1+3\nu \right)$\\
$K$&$ 1+\frac{3}{j^2}\eta^2+ \frac{\eta^4}{4}\left[\frac{3 (-2\bar E) }{ j^2}(-5+2\nu)+\frac{15  }{ j^4}(7-2\nu)\right]$\\
\end{tabular}
\end{ruledtabular}
\end{table*}

It is also useful to have the inverse expressions, \ie,  $\bar E$ and $j$ expressed in terms of  $a_r$ and $e_t$: \begin{widetext}
\begin{eqnarray}
\bar E&=&-\frac{1}{2a_r} -\frac{(-7+\nu)}{ 8 a_r^2} \eta^2 +\left[-\nu^2+7\nu- 25 -\frac{(-32+56\nu)}{(1-e_t^2)}\right] \frac{\eta^4}{16 a_r^3} \,,\nonumber\\
j&=& \sqrt{a_r(1-e_t^2)} +\left[-(-3+\nu) (1-e_t^2)^{1/2}-\frac{(1-\nu)}{(1-e_t^2)^{1/2}}\right]\frac{\eta^2}{\sqrt{a_r}}\nonumber\\
&+&
\left[\frac12 (-5-3\nu) (1-e_t^2)^{1/2}+\frac{15}{2}-3\nu+\frac12 \frac{(6\nu+\nu^2+3)}{(1-e_t^2)^{1/2}}+
\frac{1}{2} \frac{(-15+6\nu)}{(1-e_t^2)} -\frac12 \frac{(-7+12\nu+\nu^2)}{(1-e_t^2)^{3/2}}\right]\frac{\eta^4}{a_r^{3/2}}\,,
\end{eqnarray}
from which one gets (we defined the usual periastron advance parameter $k \equiv K-1$)
\begin{eqnarray} 
n&=&\frac{1}{a_r^{3/2}}+ \frac{\left(-\frac92 +\frac12\nu\right)\eta^2}{a_r^{5/2}}+ \left[\frac{147}{8}-\frac{25}{8}\nu+\frac{3}{8}\nu^2+\frac{3(-5+2\nu)}{2\sqrt{1-e_t^2}}+\frac{3(-4+7\nu)}{2(1-e_t^2)}\right]\frac{\eta^4}{a_r^{7/2}}\,, \nonumber\\
k &=&\frac{3}{(1-e_t^2)}\frac{\eta^2}{ a_r}+\left[ \frac{ -\frac{87}{4}+\frac{15}{2}\nu }{ (1-e_t^2)}+\frac{\frac{129}{4}-\frac{27}{2}\nu}{ (1-e_t^2)^2} \right]\frac{\eta^4}{a_r^2}\,,\nonumber\\
e_r &=&e_t\left[1 +\left(4-\frac32\nu\right)\frac{\eta^2}{a_r}+ \left(-\frac{29}{8}\nu+\frac{3}{8}\nu^2+\left(\frac{15}{2}-3\nu\right)\frac{1}{(1-e_t^2)^{1/2}}+\frac{(4-7\nu)}{(1-e_t^2)}\right)\frac{\eta^4}{a_r^2}\right]\,,\nonumber\\
e_\phi  &=& e_t\left[1+(4-\nu)\frac{\eta^2}{a_r} +\left(-\frac{(96\nu-240)}{32 (1-e_t^2)^{1/2}}+\frac{3\nu (\nu-27)}{32}
-\frac{  (5\nu+9) (3\nu-32)}{32(1-e_t^2)}\right)\frac{\eta^4}{a_r^2}\right]\,,\nonumber\\
f_t &=& -\frac{e_t\nu (-15+\nu)}{8a_r^2\sqrt{1-e_t^2}} \eta^4\,, \nonumber\\
g_t &=& -\frac{3(-5+2\nu)}{2a_r^2\sqrt{1-e_t^2}}\eta^4\,,\nonumber\\
f_\phi &=&   -\frac{ e_t^2 (-1-19\nu+3\nu^2)}{8 a_r^2(1-e_t^2)^2}\eta^4\,,\nonumber\\
g_\phi &=&  -\frac{e_t^3\nu (-1+3\nu)}{32a_r^2(1-e_t^2)^2}\eta^4\,.
\end{eqnarray}
\end{widetext}

\subsection{2PN quasi-Keplerian orbital parameters  in EOB coordinates}

The 2PN-accurate quasi-Keplerian representation \eq{QK} is also valid in EOB coordinates. As we shall need to transform the
harmonic-coordinate  2PN expressions of the orbital parameters into their EOB counterparts, it is very useful to express
both as functions of the conserved energy and angular momentum of the system. 
The relations $a_r(\bar E, j)$ and $e_r(\bar E, j)$ in EOB coordinates are easily obtained by 
 evaluating  the reduced energy $\bar E=\widehat H \equiv \frac{H - Mc^2}{\mu}$, and angular momentum $j\equiv \frac{J}{G M \mu}$
at the periastron ($r=a_r(1-e_r)$, $u=0$, $p_r=0$) and the apoastron ($r=a_r(1+e_r)$, $u=\pi$, $p_r=0$).
The resulting expressions are listed in Table \ref{table_orb_param_EOB} below.
\begin{table*}
\caption{\label{table_orb_param_EOB} 2PN expressions of the  EOB-coordinates orbital parameters, as functions of the conserved energy and angular momentum of the system, Eq. \eq{defbarEj}.}
\begin{ruledtabular}
\begin{tabular}{l|l}
\hline
$n$&$ (-2\bar E)^{3/2}\left[1+\eta^2\frac{-15+\nu}{8}(-2\bar E) +\eta^4\frac{(-2\bar E)^2}{128}\left(555+30\nu+11\nu^2 +\frac{192}{\sqrt{-2\bar E}j}(-5+2\nu)  \right) \right]$\\
$a_r$  &$ \frac{1}{(-2\bar E)}+\frac{\nu-3}{4}\eta^2-\frac{\eta^4}{8j^2}[8(4-\nu)+\bar E j^2 (1+\nu^2)]$\\
$e_t$& $e_{\rm N}+\frac{\eta^2 \bar E}{2e_{\rm N}}[4+j^2\bar E (17+\nu)]+\frac{\eta^4\bar E}{e_{\rm N}^3}\left[e_{\rm N}^4 (2\nu-5)\frac{6\bar E}{j\sqrt{-2\bar E}}-\frac{\bar E^3 j^4}{8}(\nu^2-102\nu-607)
+\frac{\bar E j^2}{2}(19\nu+90)+5\bar E(\nu-3)+\frac{2}{j^2}(\nu-4)  \right]$\\
$e_r$& $e_{\rm N}+\frac{\eta^2 \bar E}{2e_{\rm N}}[-8+j^2\bar E (\nu-7)]-\frac{\eta^4\bar E}{8 e_{\rm N}^3 j^2}\left[
(\nu^2+42\nu-79)\bar E^3 j^6-(5\nu-52)\bar E^2 j^2
-80(\nu-5)\bar E j^2 +32(4-\nu)\right]$\\
$e_\phi$& $e_{\rm N}+\frac{\eta^2 \bar E}{2e_{\rm N}}[-12+j^2\bar E (\nu-15)]+\frac{\eta^4\bar E}{8 e_{\rm N}^3 j^2}\left[
(\nu^2+90\nu-415)j^6\bar E^3
-4(107\nu-30)j^4 \bar E^2-40(13\nu-15)j^2 \bar E-16(9\nu-13)\right]$\\
$f_t$& $0$ \nonumber\\
$g_t$&$ \frac{6(5-2\nu)\bar E^2}{j\sqrt{-2\bar E}}\eta^4$\\
$f_\phi$&$-\frac{e_{\rm N}^2(6\nu-1)}{8j^4}\eta^4$\\
$g_\phi$&$0$ \\
$K$&$ 1+\frac{3}{j^2}\eta^2+ \frac{\eta^4}{4}\left[\frac{3 (-2\bar E) }{ j^2}(-5+2\nu)+\frac{15  }{ j^4}(7-2\nu)\right]$
\end{tabular}
\end{ruledtabular}
\end{table*}
From these relations one finds in particular
\begin{eqnarray}
e_t &=&e_r\left[1-  3\frac{\eta^2}{a_r}-\frac{\eta^4}{2a_r^2} \left(\frac{2(9e_r^2-5-\nu)}{ (1-e_r^2)}\right.\right. \nonumber\\
&&\left.\left. -\frac{3(2\nu-5)}{\sqrt{1-e_r^2}}\right)\right]\,, \nonumber\\
e_\phi &=&e_r\left[1 +\frac{\eta^2}{a_r}+\eta^4\frac{(6-7\nu-e_r^2)}{a_r^2(1-e_r^2)}  \right]  \,. 
\end{eqnarray}

\subsection{Evaluating the multipole moments along the orbit}

Let us turn to the definition \eqref{flux2PNdef} of the 2PN  split-flux.
The various multipole moments are functions of $r(t)$ and $\phi(t)$ and their derivatives up to the fifth order.
In order to compute the  Delaunay average of the nonlocal Hamiltonian \eqref{Snonloc} it is convenient to work with the 
``mean anomaly'', \ie, the angular variable $\ell$,
with respect to which all scalar functions are periodic with period $2 \pi$. [Note that $ d \ell/dt = n=$cst.]
Therefore, we first compute the multipole moments as functions of $u$ and $u'$, and then replace $u=u(\ell,e_t, \nu)$, $u'=u'(\ell',e_t, \nu)$. Finally, we will take the partie finie in $\ell'$ and the average over $\ell$.

We need  to invert the 2PN-accurate generalized Kepler equation
\beq
\label{rel_l_2PN}
\ell =u-e_t \sin u + f_t \sin V+g_t(V-u)\,,
\eeq
where $V=V(u; e_t,a_r)$.
At 1PN (\ie, when neglecting $f_t =O(\eta^4)= g_t $) this inversion  is well known, because Eq. \eq{rel_l_2PN}
then reduces to the usual Kepler equation. Namely, 
\beq
\label{u_1PN}   
u = \ell+\sum_{n=1}^N c_n^{\rm 1PN}(e_t) \sin(n \ell) + O(\eta^4)\,,
\eeq
with the notation 
\beq
c_n^{\rm 1PN}(e_t)\equiv\frac{2}{n}{\rm BesselJ}(n,n e_t)\,.
\eeq
Evidently, the exact inversion of Kepler's equation necessitates to take the upper limit $N=\infty$, but all our
computations are done with a finite upper limit $N$, chosen large enough to end up with the required accuracy on
the eccentricity expansion of the redshift $z_1$.

The inversion of Eq. \eqref{rel_l_2PN} at 2PN is obtained  first by expressing $f_t$, $g_t$ and $V$ as functions of $e_t$ and $a_r$, and then by looking for an $O(\eta^4)$-modified relation of the type 
\beq
\label{rel_u_test}
u = \ell+\sum_{n=1}^N \left(c_n^{\rm 1PN}(e_t)+\frac{\eta^4}{a_r^2} \phi_n(e_t;\nu)\right) \sin(n \ell) \,.
\eeq
Substituting Eq. \eqref{rel_u_test} into Eq. \eqref{rel_l_2PN}, and expanding in series of $e_t$, one straightforwardly obtains  
the expressions listed in Table \ref{table_relations phi_n}, where terms  only up to $e_t^{10}$ (included) are shown.
\begin{table*}
\caption{\label{table_relations phi_n} The expressions of the various $\phi_j(e_t;\nu)$ entering Eq. \eqref{rel_u_test}.}
\begin{ruledtabular}
\begin{tabular}{l|l}
\hline
$\phi_1(e_t;\nu)$ &$\left(-\frac{15}{2}+\frac{9}{8}\nu+\frac18 \nu^2\right) e_t+\left(-\frac{15}{8}+\frac{93}{64}\nu-\frac{3}{64}\nu^2\right) e_t^3+\left(-\frac{255}{128}-\frac{1}{512}\nu^2+\frac{423}{512}\nu\right) e_t^5+\left(\frac{17995}{24576}\nu-\frac{127}{73728}\nu^2-\frac{5425}{3072}\right) e_t^7$\\
&$+\left(\frac{1290347}{1966080}\nu-\frac{1741}{1966080}\nu^2-\frac{158029}{98304}\right) e_t^9 $\\
$\phi_2(e_t;\nu)$ &$ \left(\frac{15}{16}\nu-\frac{75}{8}+\frac{3}{16}\nu^2\right) e_t^2+\left(-\frac{11}{96}\nu^2+\frac{55}{32}\nu\right) e_t^4+\left(\frac{139}{256}\nu-\frac{115}{64}+\frac{3}{256}\nu^2\right) e_t^6+\left(-\frac{9}{2560}\nu^2+\frac{4913}{7680}\nu-\frac{1127}{768}\right) e_t^8
$\\
&$ +\left(-\frac{349}{276480}\nu^2-\frac{12595}{9216}+\frac{3475}{6144}\nu\right) e_t^{10}$\\
$\phi_3(e_t;\nu)$ &$ \left(\frac{17}{64}\nu^2+\frac{49}{64}\nu-\frac{95}{8}\right) e_t^3+\left(\frac{2273}{1024}\nu+\frac{745}{256}-\frac{231}{1024}\nu^2\right) e_t^5+\left(\frac{2613}{40960}\nu-\frac{2253}{1024}+\frac{2229}{40960}\nu^2\right) e_t^7
+\left(\frac{209349}{327680}\nu-\frac{19533}{16384}-\frac{3539}{327680}\nu^2\right) e_t^9 $\\
$\phi_4(e_t;\nu)$ &$  \left(\frac{71}{192}\nu^2-\frac{975}{64}+\frac{35}{64}\nu\right) e_t^4+\left(-\frac{129}{320}\nu^2+\frac{955}{128}+\frac{49}{16}\nu\right) e_t^6+\left(\frac{387}{2560}\nu^2-\frac{2827}{768}-\frac{6107}{7680}\nu\right)e_t^8
+\left(-\frac{23495}{32256}+\frac{741}{896}\nu-\frac{8639}{241920}\nu^2\right) e_t^{10}$\\
$\phi_5(e_t;\nu)$ &$  \left(\frac{1167}{5120}\nu+\frac{523}{1024}\nu^2-\frac{5049}{256}\right) e_t^5+\left(\frac{542539}{122880}\nu-\frac{50195}{73728}\nu^2+\frac{44521}{3072}\right) e_t^7
+\left(-\frac{2439751}{344064}-\frac{16009117}{6881280}\nu+\frac{473695}{1376256}\nu^2\right) e_t^9$\\
$\phi_6(e_t;\nu)$ &$  \left(-\frac{1645}{64}+\frac{899}{1280}\nu^2-\frac{65}{256}\nu\right) e_t^6+\left(\frac{117137}{17920}\nu+\frac{6451}{256}-\frac{19851}{17920}\nu^2\right) e_t^8
+\left(-\frac{20475}{4096}\nu-\frac{397035}{28672}+\frac{100713}{143360}\nu^2\right) e_t^{10}$\\
$\phi_7(e_t;\nu)$ &$ \left(-\frac{844747}{860160}\nu-\frac{723943}{1504}+\frac{355081}{368640}\nu^2\right) e_t^7+\left(\frac{56824273}{1376256}-\frac{6902693}{3932160}\nu^2+\frac{270188581}{27525120}\nu\right) e_t^9 $\\
$\phi_8(e_t;\nu)$ &$ \left(-\frac{55697}{26880}\nu+\frac{47259}{35840}\nu^2-\frac{271981}{6144}\right) e_t^8+\left(-\frac{3959051}{1451520}\nu^2+\frac{273379}{18432}\nu+\frac{50476225}{774144}\right) e_t^{10}$\\
$\phi_9(e_t;\nu)$ &$  \left(\frac{16541017}{9175040}\nu^2-\frac{101441437}{27525120}\nu-\frac{80363041}{1376256}\right) e_t^9$\\
$\phi_{10}(e_t;\nu)$ &$ \left(-\frac{59782109}{774144}-\frac{7804319}{1290240}\nu+\frac{5719087}{2322432}\nu^2\right) e_t^{10}$\\ 
\end{tabular}
\end{ruledtabular}
\end{table*}
The 2PN coefficients $\phi_n(e_t;\nu)$ depend at most quadratically on $\nu$, and are even (respectively odd) polynomials in $e_t$, when
$n$ is even (respectively odd).

As a preparation for using the Delaunay-averaging technique, it is useful to express the motion
in terms of the two independent angles entering the action-angle description of equatorial motion: the angle $\ell$ measuring
the periodicity in the radial motion, and the angle $g$ measuring the mean periastron precession. These two angles
are canonically conjugated to two corresponding action variables, traditionally denoted as $L$ and $G$. [Modulo some rescalings,
the link between the Delaunay action variables $L,G$ and the usual action variables is $L=I_3=I_r+I_\phi$ and $G=I_\phi$.]
The radial motion is entirely expressed in terms of the sole angle $\ell$, while one must separate in the azimuthal motion
(given, on shell, by $\phi(\ell )=K\ell +W(\ell)$, where $K\equiv 1+k$) the contributions coming from $\ell$ and from $g$ (which is
equal to $k \ell$ on shell): 
\beq
\phi(\ell, g )=\ell+g +W(\ell)\,.
\eeq
Here, $W(\ell)$ is a periodic function of $\ell$, say 
\begin{eqnarray}
W(\ell)&=& \sum_{k=1}^N \left[P^{(0)}_k(e_t, \nu)+\frac{\eta^2}{a_r} P^{(2)}_k(e_t, \nu)\right. \nonumber\\
&&\left. +
\frac{\eta^4}{a_r^2} P^{(4)}_k(e_t, \nu)
\right]\sin(k\ell) \nonumber\\
&\equiv&\sum_{k=1}^N c_k(e_t,\nu,a_r,\eta)\sin(k\ell)
\,.
\end{eqnarray}
The structure of the angular motion is then of the form
\beq
\phi(\ell, g)= \ell+g+\sum_{k=1}^N c_k(e_t,\nu,a_r,\eta)\sin(k\ell)\,,
\eeq
where we recall that one must consider the angle $g$ (mean periastron argument) 
as an independent angular variable. On-shell we have (remembering the notation $K \equiv1+k$)
\bea \label{lgonshell}
\ell^{\rm on shell} &=& \ell_0 + n \, t \,,\nonumber\\
 g^{\rm on shell} &=& k\,\ell = g_0 + k\, n\, t \,.
\eea
One then computes 
\beq
e^{i\phi(\ell,g)}=e^{i\ell}e^{ig}e^{iW(\ell)}\,,\qquad 
e^{-i\phi(\ell,g)}=e^{-i\ell}e^{-ig}e^{-iW(\ell)}\,,
\eeq
with  
\beq
e^{ iW(\ell)}=1+e_t \sum_{k=-N}^N d^{(1)}_k e^{ik\ell}+e_t^2 \sum_{k=-N}^N d^{(2)}_k e^{ik\ell}+\ldots
\eeq
expanded in series of $e_t$.
The Cartesian coordinates of the relative (equatorial) motion are then expressed as the
following doubly-periodic functions of $\ell$ and $g$:
\begin{eqnarray}
x(\ell,g) &=&\frac12 r_h(\ell)(e^{i\phi(\ell,g)}+e^{-i\phi(\ell,g)})\,,\nonumber\\
y(\ell,g) &=&\frac1{2i} r_h(\ell)(e^{i\phi(\ell,g)}-e^{-i\phi(\ell,g)})\,.
\end{eqnarray}
Finally, one computes all the on-shell time derivatives entering the
definition of the multipole moments by using, in view of Eqs. \eq{lgonshell}, $d \ell/dt=n$ and 
$dg/dt=k\, n=O(\eta^2)+O(\eta^4)$.

\subsection{Computing the Delaunay-average of the nonlocal Hamiltonian in harmonic  coordinates}

We are now ready to sketch the computation of the Delaunay-average of the h-route nonlocal Hamiltonian, \ie, 
the average of the action-angle Hamiltonian $H_{\rm nonloc, h}^{4+5+6 \rm PN}(L,G,\ell,g)$ over the two angles $\ell,g$:
\beq
\langle H_{\rm nonloc, h}^{4+5+6 \rm PN} \rangle\equiv \oint  \frac{d\ell}{2 \pi} \frac{dg}{2\pi}\, H_{\rm nonloc, h}^{4+5+6 \rm PN}(L,G,\ell,g) \,.
\eeq
We recall that
\begin{eqnarray} 
\label{delta_H_nonloc}
&&H_{\rm  nonloc, h}^{4+5+6 \rm PN}(t)=
-\frac{G^2 H_{\rm 2PN}}{c^{5}}{\rm Pf}_{2s/c}\int \frac{d\tau}{|\tau|}{\mathcal F}^{\rm split}_{\rm 2PN}(t,t+\tau)\nonumber\\
&&\qquad\qquad+2\frac{G^2 H_{\rm 2PN}}{c^{5}}{\mathcal F}^{\rm split}_{\rm 2PN}(t,t)  \ln \left( \frac{r_{12}^h(t)}{s}\right)
\,,
\end{eqnarray} 
with 
\beq
\frac{H_{\rm 2PN}}{Mc^2}= 1+\nu\bar E\eta^2
=1-\frac{\nu}{2a_r^h}\eta^2+\frac{\nu}{8(a_r^h)^2}(7-\nu)\eta^4+O(\eta^6)\,.
\eeq
One must, in principle, express the nonlocal integrand entering Eq. \eq{delta_H_nonloc} in terms
of two quadruplets of Delaunay variables, $(L,G,\ell,g)$ and $(L',G',\ell',g')$, where the first quadruplet refers to the state of the
system at time $t$ while the second refers to the state at the shifted time $t'=t+\tau$. [The Delaunay variables $(L,G,\ell,g)$
are action-angle variables for the main part of the Hamiltonian, to which $H_{\rm  nonloc, h}^{4+5+6 \rm PN}$ is added
as a first-order perturbation. In practice, it suffices to use the 2PN-accurate Hamiltonian.] This yields an integrand which can
be expressed as a multi-Fourier series of the general form
\bea \label{Fsplitnonloc}
&&{\mathcal F}^{\rm split}(L,G,\ell,g; L',G',\ell',g')= \nonumber\\
&&\sum_{p,m,p',m'} C_{p,m,p',m'} e^{i(p \ell+m g + p'\ell'+m' g')}\,,
\eea
where the relative integers $p,m,p',m'$ are summed from $-\infty$ to  $+\infty$, and where the
coefficients $ C_{p,m,p',m'}$ are functions of $(L,G,L',G')$.

As shown in Ref. \cite{Damour:2014jta},
one can first use a nonlocal shift of the phase-space variables to replace the second quadruplet by its on-shell value in terms
of $(L,G,\ell,g)$ and of the time shift $\tau$. In other words, we can insert in Eq. \eq{Fsplitnonloc} 
\beq
(L',G',\ell',g') \mapsto (L,G,\ell + n(L,G) \tau,g + k(L,G) n(L,G) \tau)\,,
\eeq
where we used the simple equations of motion of the Delaunay variables $(L,G,\ell,g)$.

After this replacement the crucial nonlocal $\tau$ integral in Eq. \eq{delta_H_nonloc} can be explicitly evaluated by using the
basic formula (where $\gamma$ denotes Euler's constant)
\begin{eqnarray}
&&{\rm Pf}_{2s/c}\int_{-\infty}^{+\infty} \frac{d\tau}{|\tau|} e^{i(p'+m'k)n\tau} =\nonumber\\
&&\qquad\qquad -2  \ln \left( |(p'+m'k)n|\frac{2s e^\gamma}{c} \right)\,.
\end{eqnarray}
Using the latter formula for evaluating the $\tau$ integral in Eq. \eq{delta_H_nonloc} yields a result which is a function of 
$(L,G,\ell,g)$. Adding the local term $2\frac{G^2 H_{\rm 2PN}}{c^{5}}{\mathcal F}^{\rm split}_{\rm 2PN}(t,t)  \ln \left( \frac{r_{12}^h(t)}{s}\right)$, we can finally evaluate the double average over the two angular variables $\ell$ and $g$.

For conceptual clarity, we have assumed here that we were using the (2PN-accurate) Delaunay action variables
$L,G$ as arguments in the double-Fourier expansion of the nonlocal integrand \eq{Fsplitnonloc}.
However, in practice, it suffices to use the Keplerian elements $a_r^h$ and $e_t^h$ entering the 
2PN-accurate quasi-Keplerian representation \eq{QK} of the elliptic motion. At the end of the day,
the method presented above leads to an explicit expression for the Delaunay-averaged h-route nonlocal
Hamiltonian of the form
\beq \label{Hh}
\langle H_{\rm nonloc, h}^{h}\rangle \equiv \langle H_{\rm  nonloc, h}^{4+5 +6\rm PN} \rangle= F^h(a_r^h,e_t^h)
 \,,
\eeq
with
\begin{eqnarray} \label{Fh}
F^h(a_r^h,e_t^h)&=&\frac{\nu^2}{(a_r^h)^5}\left[{\mathcal A}^{\rm 4PN}(e_t^h)+{\mathcal B}^{\rm 4PN}(e_t^h)\ln a_r^h\right]\nonumber\\
&+&
\frac{\nu^2}{(a_r^h)^6}\left[{\mathcal A}^{\rm 5PN}(e_t^h)+{\mathcal B}^{\rm 5PN}(e_t^h)\ln a_r^h\right]
\nonumber\\
&+&
\frac{\nu^2}{(a_r^h)^7}\left[{\mathcal A}^{\rm 6PN}(e_t^h)+{\mathcal B}^{\rm 6PN}(e_t^h)\ln a_r^h\right]\,.\nonumber\\
\end{eqnarray} 
The coefficients entering this decomposition are independent of the intermediate scale $s$, and are obtained as  
expansions in powers of $e_t^h$ that we have computed up to the order $O((e_t^h)^{10})$ included. 
The values of the 4PN and 5PN coefficients have been given in Ref. \cite{Bini2020}.
We list the 6PN  coefficients ${\mathcal A}^{\rm 6PN}(e_t^h)$, $ {\mathcal B}^{\rm 6PN}(e_t^h)$,
 in Table \ref{table_F_h}.  [We use here  $G=1=c$, and we recall that $a_r^h$ has been adimensionalized by $GM$.]

\begin{table*}
\caption{\label{table_F_h} 6PN coefficients of the averaged nonlocal Hamiltonian (with scale $2 r_{12}^h/c$) in harmonic coordinates}
\begin{ruledtabular}
\begin{tabular}{ll }
Coefficient & Expression \\
\hline
${\mathcal A}^{\rm 6PN}(e_t^h)$ & $-8\nu^2+\frac{1238}{7}+\frac{6578}{105}\nu
+\left(\frac{32}{5}\nu^2+\frac{1173532}{2835}+\frac{38266}{63}\nu\right)\gamma$\\
& $+\left(\frac{3125212}{2835}-\frac{35362}{945}\nu+\frac{622648}{945}\nu^2\right)\ln(2)
+\left(-\frac{2673}{7}\nu^2-243+\frac{29889}{28}\nu\right)\ln(3)$\\
& $+\left[\frac{252377}{315}+\frac{34409}{90}\nu-\frac{75104}{135}\nu^2
+\left(\frac{577921}{105}\nu-\frac{260872}{945}+\frac{2815}{3}\nu^2\right)\gamma
+\left(-\frac{8467217}{315}\nu^2+\frac{39588209}{945}\nu-\frac{8908616}{945}\right)\ln(2)\right.$\\
& $\left.
+\left(-\frac{837621}{140}\nu+\frac{2071089}{560}+\frac{3052323}{560}\nu^2\right)\ln(3)
+\left(\frac{9765625}{9072}-\frac{9765625}{1512}\nu+\frac{9765625}{1008}\nu^2\right)\ln(5)\right](e_t^h)^2$\\
& $+\left[\frac{13092029}{1680}\nu-\frac{1114139}{648}-\frac{52844503}{7560}\nu^2
+\left(\frac{35023}{4}\nu^2+\frac{1693451}{140}\nu-\frac{1047607}{315}\right)\gamma\right.$\\
& $
+\left(-\frac{235370381}{1260}\nu+\frac{1392676751}{3780}\nu^2+\frac{55700171}{945}\right)\ln(2)
+\left(-\frac{929042703}{8960}\nu+\frac{18495459}{160}\nu^2+\frac{989253}{1120}\right)\ln(3)$\\
& $\left.
+\left(\frac{904296875}{5376}\nu-\frac{25390625}{864}-\frac{166015625}{756}\nu^2\right)\ln(5)\right](e_t^h)^4$\\
& $+\left[\frac{598387141}{10080}\nu-\frac{1164531499}{30240}\nu^2-\frac{411052325}{13608}
+\left(\frac{4585927}{1890}+\frac{154063}{4}\nu^2-\frac{33879}{20}\nu\right)\gamma\right.$\\
& $
+\left(-\frac{1792419163}{5670}+\frac{34283848589}{34020}\nu-\frac{101367784649}{34020}\nu^2\right)\ln(2)
+\left(\frac{6204343797}{4480}\nu-\frac{7349467203}{35840}-\frac{80018658513}{35840}\nu^2\right)\ln(3)$\\
& $
+\left(\frac{99453653125}{746496}-\frac{198307890625}{435456}\nu+\frac{2578115234375}{1741824}\nu^2\right)\ln(5)$\\
& $\left.
+\left(-\frac{96889010407}{124416}\nu+\frac{96889010407}{82944}\nu^2+\frac{96889010407}{746496}\right)\ln(7)\right](e_t^h)^6$\\
& $+\left[\frac{40216186627}{161280}\nu-\frac{107831692469}{725760}-\frac{739210729}{5376}\nu^2
+\left(\frac{3745525}{32}\nu^2+\frac{477783529}{10080}-\frac{81947429}{960}\nu\right)\gamma\right.$\\
& $
+\left(\frac{2521461747193}{816480}+\frac{496432848889}{12960}\nu^2-\frac{9704902576499}{544320}\nu\right)\ln(2)
+\left(\frac{501042990087}{573440}+\frac{1148629836951}{81920}\nu^2-\frac{9521506906299}{2293760}\nu\right)\ln(3)$\\
& $
+\left(-\frac{19630345703125}{3981312}\nu^2-\frac{677663178125}{3096576}-\frac{81800483984375}{111476736}\nu\right)\ln(5)$\\
& $\left.
+\left(-\frac{15648846319351}{11943936}-\frac{68694308378563}{3981312}\nu^2+\frac{145243406006069}{15925248}\nu\right)\ln(7)\right](e_t^h)^8$\\
& $+\left[-\frac{97074060217}{201600}+\frac{405149913757}{537600}\nu-\frac{609781084333}{1612800}\nu^2
+\left(\frac{182458353}{640}\nu^2-\frac{617936773}{1920}\nu+\frac{52105469}{280}\right)\gamma\right.$\\
& $
+\left(-\frac{9975543985969843}{27216000}\nu^2+\frac{300764760469259}{1814400}\nu-\frac{40023535388969}{1701000}\right)\ln(2)$\\
& $
+\left(-\frac{1205456030071641}{22937600}\nu+\frac{910396452443931}{114688000}+\frac{240491992467807}{7168000}\nu^2\right)\ln(3)$\\
& $
+\left(\frac{972942783453125}{222953472}\nu-\frac{114839474809375}{111476736}+\frac{2227079123046875}{222953472}\nu^2\right)\ln(5)$\\
& $\left.
+\left(\frac{8183105251126477}{1592524800}-\frac{67003553703806461}{1990656000}\nu+\frac{33107071745082307}{318504960}\nu^2\right)\ln(7)\right](e_t^h)^{10}$\\
\hline
${\mathcal B}^{\rm 6PN}(e_t^h)$ & $-\frac{16}{5}\nu^2-\frac{19133}{63}\nu-\frac{586766}{2835}
+\left(\frac{130436}{945}-\frac{2815}{6}\nu^2-\frac{577921}{210}\nu\right)(e_t^h)^2
+\left(-\frac{35023}{8}\nu^2-\frac{1693451}{280}\nu+\frac{1047607}{630}\right)(e_t^h)^4$\\
& $+\left(-\frac{154063}{8}\nu^2+\frac{33879}{40}\nu-\frac{4585927}{3780}\right)(e_t^h)^6
+\left(\frac{81947429}{1920}\nu-\frac{477783529}{20160}-\frac{3745525}{64}\nu^2\right)(e_t^h)^8$\\
& $+\left(\frac{617936773}{3840}\nu-\frac{182458353}{1280}\nu^2-\frac{52105469}{560}\right)(e_t^h)^{10}$\\
\end{tabular}
\end{ruledtabular}
\end{table*}

\section{Deriving the h-route nonlocal EOB Hamiltonian} \label{Heffnonloc}

An important ingredient of our method is to translate the h-route nonlocal averaged Hamiltonian computed in the previous
section into a canonically equivalent  EOB Hamiltonian.
This is done by parametrizing the corresponding h-route nonlocal EOB Hamiltonian by means of the usual
EOB potentials, in some fixed EOB gauge. At this stage of our computation, it is most convenient to
use the $p_r$ gauge (introduced in Ref. \cite{Damour:2000we}).

Explicitly, we look for a rescaled squared effective EOB Hamiltonian of the general form (where $u=1/r=GM/r^{\rm phys}$, $p_r=p_r^{\rm phys}/\mu$, $p_\phi=p_\phi^{\rm phys}/(G M \mu)\equiv j$)
\begin{eqnarray} \label{Hf2}
\widehat H_{\rm eff}^2&=&A(u;\nu)\Big(1+p_\phi^2 u^2+A(u;\nu) \bar D(u;\nu) p_r^2 \nonumber\\
&& +\widehat Q(u,p_r;\nu) \Big)\,,
\end{eqnarray}
with potentials $A(u;\nu)$, $\bar D(u;\nu)$ and 
\begin{eqnarray}
\widehat Q(u,p_r;\nu)&=&p_r^4 q_4(u;\nu)+p_r^6 q_6(u;\nu)\nonumber\\
&+&p_r^8 q_8(u;\nu)+p_r^{10}q_{10}(u;\nu)+\ldots \,.
\end{eqnarray}
All the potentials  $A(u;\nu)$, $\bar D(u;\nu)$, $\widehat Q(u,p_r;\nu)$ reduce to their Schwarzschild values when $\nu \to 0$:
 $A(u;0)=1-2u$, $\bar D(u;0)=1$, $\widehat Q(u,p_r;0)=0$, and can be expanded in powers of $\nu$ away
 from the test-mass limit:
\begin{eqnarray}
A(u;\nu)&=&1-2u +\nu a^{\nu^1}(u)+\nu^2 a^{\nu^2}(u)+\nu^3 a^{\nu^3}(u)+\ldots\nonumber\\ 
\bar D(u;\nu)&=&1+\nu \bar d^{\nu^1}(u)+\nu^2 \bar d^{\nu^2}(u)+\nu^3 \bar d^{\nu^2}(u)+\ldots\nonumber\\
q_4(u;\nu)&=& \nu q_{4}^{\nu^1}(u)+\nu^2 q_{4}^{\nu^2}(u)+\nu^3 q_{4}^{\nu^3}(u)+\ldots\nonumber\\
q_6(u;\nu)&=& \nu q_{6}^{\nu^1}(u)+\nu^2 q_{6}^{\nu^2}(u)+\nu^3 q_{6}^{\nu^3}(u)+\ldots\nonumber\\
q_8(u;\nu)&=& \nu q_{8}^{\nu^1}(u)+\nu^2 q_{8}^{\nu^2}(u)+\nu^3 q_{8}^{\nu^3}(u)+\ldots\,.
\end{eqnarray}
Each EOB potential can be decomposed in a local part and a nonlocal one:
\bea
&& A=A^{\rm loc,h}+A^{\rm nonloc,h}= A^{\rm loc,f}+A^{\rm nonloc,f}, \nonumber\\
&& \bar D=\bar D^{\rm loc,h}+\bar D^{\rm nonloc,h}=\bar D^{\rm loc,f}+\bar D^{\rm nonloc,f}\,, \nonumber\\
&& \widehat Q=\widehat Q^{\rm loc,h}+\widehat Q^{\rm nonloc,h}=\widehat Q^{\rm loc,f}+ \widehat Q^{\rm nonloc,f}\,.
\eea
The nonlocal parts start at 4PN.They can be treated as first-order perturbations of the local parts, 
which start at 2PN (and also at 2PM).
We, indeed, recall that the EOB formalism has the remarkable feature to describe both the 1PN-accurate dynamics
and the 1PM one,  by a Schwarzschild effective metric. This means that all the local contributions to 
$A-(1-2u)$, $\bar D-1$ and $Q$ start  at order $u^2=O(G^2)$ or more (and contain a factor $\nu$). 
[The main $A$ potential actually
starts to deviate from $1-2u$ by a term $2 \nu u^3$.]
For clarity, we have indicated that the precise values of both the local and nonlocal EOB potentials will
depend on the choice of the flexibility factor $f(t)$ used in defining the Pf scale $r_{12}^f=f(t) r_{12}^h$ entering the
nonlocal action \eq{Snonloc}. The h-route is defined by choosing the default value $f=1$, while the (tuned) f-route
is defined by choosing a value $f=1+ \nu O(\eta^2)$ determined in the way explained at 5PN in Ref. \cite{Bini2020}.
As a consequence the difference between the h and f values of any quantity starts at 5PN and at the second-self-force (2SF)
order, \ie, the  order  $O(\nu^2)$ in $\widehat H_{\rm eff}^2$. This corresponds to the order  $O(\nu^3)$ in the
usual Hamiltonian. Indeed, the universal EOB energy map says that the usual center-of-mass Hamiltonian
of the system, $H=Mc^2 + \ldots$ is given by
\beq \label{eobmap}
H=Mc^2\sqrt{1+2\nu(\widehat H_{\rm eff}-1)}\,,
\eeq
where one should note the factor $\nu$ in front of $\widehat H_{\rm eff}$.

To simplify the notation, we shall denote the nonlocal part of the {\it squared} effective EOB Hamiltonian as
\beq
\delta^{\rm h,f} \widehat H_{\rm eff}^2 \equiv  \left[\widehat H_{\rm eff}^2 \right]_{\rm nonloc,h,f}^{4+5+6 \rm PN}\,.
\eeq
It is related to the corresponding (h-route or f-route) nonlocal Hamiltonian via
\beq
\label{deltaHeob}
 H_{\rm nonloc, h, f}^{4+5+6 \rm PN}=\frac{\mu M}{2H \widehat H_{\rm eff}} \delta^{\rm h,f} \widehat H_{\rm eff}^2\,.
\eeq
The prefactor on the right-hand side of  Eq. \eqref{deltaHeob} is given (at the 2PN accuracy) by 
\begin{eqnarray}
\frac{\mu M}{2H \widehat H_{\rm eff}}&=&M \Big( \frac12\nu+\frac{\nu(\nu+1)}{4a_r^e}\eta^2\nonumber\\
&+&\frac{\nu(-\nu+3\nu^2-1)}{16(a_r^e)^2}\eta^4+O(\eta^6) \Big)\,,
\end{eqnarray}
where $a_r^e$ denotes the EOB-coordinate semi-major axis.

With this notation, the squared effective EOB Hamiltonian reads
\beq
\widehat H_{\rm eff}^2=\widehat H_{\rm eff,loc, h,f}^2  +\delta^{\rm h,f}  \widehat H_{\rm eff}^2\,,
\eeq
where
\begin{eqnarray}
\label{Heffquadloc}
\widehat H_{\rm eff,loc, h,f}^2&=&A^{\rm loc, h,f}[1+A^{\rm loc, h,f}\bar D^{\rm loc,h,f}p_r^2\nonumber\\
&& +p_\phi^2 u^2+\widehat Q^{\rm loc, h,f}]\,,
\end{eqnarray}
and
\begin{eqnarray}
\delta^{\rm h,f} \widehat H_{\rm eff}^2&=&[1+2(1-2u)p_r^2+p_\phi^2 u^2]\delta^{\rm h,f} A\nonumber\\
&&+ (1-2u)^2 p_r^2 \delta^{\rm h,f} \bar D\nonumber\\
&&+(1-2u)\delta^{\rm h,f} \widehat Q\,.
\end{eqnarray}
At the 4+5+6PN accuracy, the expressions for the nonlocal EOB potentials read
\begin{eqnarray} \label{ADQnonloc}
\delta A&=&a_5^{\rm nonloc}u^5+a_6^{\rm nonloc}u^6+a_7^{\rm nonloc}u^7\,,\nonumber\\
\delta \bar D &=& \bar d_4^{\rm nonloc}u^4+\bar d_5^{\rm nonloc}u^5+\bar d_6^{\rm nonloc}u^6\,,\nonumber\\
\delta \widehat Q&=&p_r^4 (q_{43}^{\rm nonloc}u^3+q_{44}^{\rm nonloc}u^4+q_{45}^{\rm nonloc}u^5)\nonumber\\
&+& p_r^6 (q_{62}^{\rm nonloc}u^2+q_{63}^{\rm nonloc}u^3+q_{64}^{\rm nonloc}u^4)\nonumber\\
&+& p_r^8 (q_{81}^{\rm nonloc}u+q_{82}^{\rm nonloc}u^2+q_{83}^{\rm nonloc}u^3)\nonumber\\
&+& p_r^{10} (q_{10,0}^{\rm nonloc}+ q_{10,1}^{\rm nonloc}u+q_{10,2}^{\rm nonloc}u^2)\,,
\end{eqnarray}
etc., where each coefficient will be decomposed in ``constant'',
and ``logarithmically running'' parts according to the scheme: $a_5^{\rm nonloc}=a_5^{\rm nl,c}+a_5^{\rm nl,log}\ln(u)$, etc.
To ease the notation, we have suppressed on each nonlocal quantity the extra label h or f specifying whether this is computed
by the h-route or the f-route. A term $\propto q_{2p,q}^{\rm nonloc} u^q$ belongs to the $n$-PN approximation
with $n=p+q-1$. Note that, contrary to the local EOB potentials that must start at order $u^2$ at least, the nonlocal ones,
being obtained by matching a nonlocal action by means of a nearzone eccentricity (or $p_r$) expansion, include,
at high orders in $p_r$ powers of $u$ that are smaller than 2.

Having clarified the meaning of the nonlocal parts of the EOB potentials, we can now determine the values of
the h-route nonlocal EOB potentials, $a_n^{\rm nonloc, h}$, $\bar d_n^{\rm nonloc, h}$, $q_{2p,q}^{\rm nonloc, h}$
that are gauge-equivalent to the h-route nonlocal Hamiltonian computed in the previous section.
These values are determined by writing the equality between the corresponding Delaunay-averaged perturbed Hamiltonians,
namely
\beq \label{eob=h}
\langle H^{\rm eob}_{\rm nonloc, h}\rangle = \langle H^{h}_{\rm nonloc, h}\rangle \,,
\eeq
where the left-hand side is the Delaunay average of the EOB-parametrized Hamiltonian
\beq
\langle H^{\rm eob}_{\rm nonloc, h}\rangle  \equiv  \frac{\mu M}{2H \widehat H_{\rm eff}} \oint \frac{d\ell}{2 \pi} \frac{dg}{2\pi} \, \delta^{\rm h} \widehat H_{\rm eff}^2 \,,
\eeq
and where the right-hand side is the function $ F^h(a_r^h,e_t^h)$ computed in the previous section, see Eq. \eq{Hh}.
The equality Eq. \eq{eob=h} expresses the requirement that the EOB nonlocal dynamics is canonically equivalent to the 
original nonlocal dynamics, described by Eq. \eq{Snonloc} (see Ref.~\cite{Damour:2015isa}).
The computations needed to evaluate $\langle H^{\rm eob}_{\rm nonloc, h}\rangle$ are similar to the computations
described above (with the simplifying feature that one only works with an Hamiltonian given as a function of the
instantaneous state of the system). One uses the EOB version of the 2PN-accurate quasi-Keplerian representation of elliptic motions,
as decribed in the previous section.

Finally, the identification Eq. \eq{eob=h} uniquely determines, from the knowledge of the function $ F^h(a_r^h,e_t^h)$, Eq. \eq{Fh},
 all the coefficients parametrizing the nonlocal EOB potentials Eq. \eq{ADQnonloc}. We give the resulting values in
Table \ref{eob_loc_f_coeffs}, up to the eight power of $p_r$. Indeed, we will not need in the following the coefficient
$q_{10}(u;\nu)$ of $p_r^{10}$.

\begin{table*}
\caption{\label{eob_loc_f_coeffs} h-route nonlocal EOB coefficients.}
\begin{ruledtabular}
\begin{tabular}{l|l}
\hline
$a_7^{\rm nl,c}$ & $
\left(\frac{206740}{567}\ln(2)+\frac{12664}{105}-\frac{4617}{14}\ln(3)-\frac{5044}{405}\gamma\right)\nu$\\
&$
+\left(-\frac{1139672}{945}\ln(2)+\frac{10132}{105}+\frac{10449}{7}\ln(3)+\frac{101272}{315}\gamma\right)\nu^2$\\
&$
+\left(-\frac{112}{5}+32\gamma+\frac{1214624}{945}\ln(2)-\frac{4860}{7}\ln(3)\right)\nu^3
\,,$\\
$a_7^{\rm nl,log}$&$
-\frac{2522}{405}\nu+\frac{50636}{315}\nu^2+16\nu^3
\,,$\\
$d_6^{\rm nl,c}$ &$
\left(-\frac{6381680}{189}\ln(2)+\frac{2043541}{2835}+\frac{1765881}{140}\ln(3)-\frac{64096}{45}\gamma+\frac{9765625}{2268}\ln(5)\right)\nu$\\
&$
+\left(\frac{28429312}{189}\ln(2)-\frac{3576231}{70}\ln(3)+\frac{167906}{105}+\frac{302752}{105}\gamma-\frac{9765625}{378}\ln(5)\right)\nu^2$\\
&$
+\left(-\frac{9908480}{63}\ln(2)-\frac{744704}{945}+\frac{9765625}{252}\ln(5)+\frac{2944}{3}\gamma+\frac{1275021}{28}\ln(3)\right)\nu^3
\,,$\\
$d_6^{\rm nl,log}$&$
-\frac{32048}{45}\nu+\frac{151376}{105}\nu^2+\frac{1472}{3}\nu^3
$\\
$q_{45}^{\rm nl,c}$&$
\left(\frac{70925884}{63}\ln(2)+\frac{13212013}{5670}-\frac{3873663}{16}\ln(3)-\frac{8787109375}{27216}\ln(5)-\frac{617716}{315}\gamma\right)\nu$\\
&$
+\left(\frac{92560887}{280}\ln(3)-\frac{12619052648}{2835}\ln(2)-\frac{1437979}{63}+\frac{632344}{315}\gamma+\frac{7755859375}{4536}\ln(5)\right)\nu^2$\\
&$
+\left(-\frac{177316}{35}+\frac{11263031264}{2835}\ln(2)+\frac{16544}{9}\gamma-\frac{4091796875}{2268}\ln(5)+\frac{2908467}{20}\ln(3)\right)\nu^3
$\\
$q_{45}^{\rm nl,log}$&$
-\frac{308858}{315}\nu+\frac{316172}{315}\nu^2+\frac{8272}{9}\nu^3
$\\
$q_{64}^{\rm nl,c}$&$
\left(-\frac{211076833264}{14175}\ln(2)-\frac{137711989}{28350}-\frac{9678652821}{5600}\ln(3)+\frac{447248}{1575}\gamma+\frac{153776136875}{23328}\ln(5)\right.$\\
&$\left.
+\frac{96889010407}{116640}\ln(7)\right)\nu$\\
&$
+\left(\frac{44592947739}{2800}\ln(3)+\frac{2411178384736}{42525}\ln(2)-\frac{126070663}{4725}-\frac{26848}{175}\gamma-\frac{796015515625}{27216}\ln(5)\right.$\\
&$\left.
-\frac{96889010407}{19440}\ln(7)\right)\nu^2$\\
&$
+\left(-\frac{40513708}{4725}-\frac{109566260523}{5600}\ln(3)+\frac{1424826953125}{54432}\ln(5)+\frac{96889010407}{12960}\ln(7)+\frac{2368}{5}\gamma\right.$\\
&$\left.
-\frac{431564554688}{8505}\ln(2)\right)\nu^3
$\\
$q_{64}^{\rm nl,log}$&$
\frac{223624}{1575}\nu-\frac{13424}{175}\nu^2+\frac{1184}{5}\nu^3
$\\
$q_{83}^{\rm nl,c}$&$
\left(\frac{5196312336176}{35721}\ln(2)+\frac{17515638027261}{313600}\ln(3)-\frac{63886617280625}{1016064}\ln(5)-\frac{29247366220639}{933120}\ln(7)\right.$\\
&$\left.
-\frac{709195549}{132300}\right)\nu$\\
&$
+\left(-\frac{177055674739808}{297675}\ln(2)-\frac{43719724468071}{156800}\ln(3)+\frac{366449151015625}{1524096}\ln(5)+\frac{26506549233199}{155520}\ln(7)\right.$\\
&$\left.
-\frac{1746293}{70}\right)\nu^2$\\
&$
+\left(\frac{57604236136064}{99225}\ln(2)+\frac{10467583300341}{39200}\ln(3)-\frac{73366198046875}{381024}\ln(5)-\frac{7709596970957}{38880}\ln(7)\right.$\\
&$\left.
-\frac{154862}{21}\right)\nu^3
$\\
$q_{83}^{\rm nl,log}$&$0$\\
\end{tabular}
\end{ruledtabular}
\end{table*}

\section{Computing the 1SF time-averaged redshift to eighth order in eccentricity and deriving its EOB counterpart} \label{SF}

The second pillar of our method is to combine the information extracted from the analytical knowledge of the nonlocal part of the dynamics
with a knowledge obtained from self-force calculations, which gives information about  the {\it total}, local plus nonlocal, near-zone dynamics,
at the first order in mass ratio $q=\frac{m_1}{m_2}$ beyond the test-mass limit.
Indeed, Refs.~\cite{LeTiec:2011ab,Barausse:2011dq,Tiec:2015cxa} have found a relation between the $m_1$-dependence
of the Hamiltonian of a two-body system, and the (regularized) redshift \cite{Detweiler:2008ft,Barack:2011ed} $z_1 = d s_1/dt$ 
of particle 1 in the gravitational field created by the two particles. We have developed efficient tools in previous work 
\cite{Bini:2013zaa,Bini:2013rfa} 
for tapping information by such self-force computations. The current limitation of this technique (for non-spinning bodies) is not
the PN accuracy (which can be pushed to extremely high levels \cite{Bini:2015bla,Kavanagh:2015lva}) 
but rather the order of expansion in the
eccentricity of the considered elliptic motion of a small mass $m_1$ around a large mass $m_2$.
Here, we have extended our previous results \cite{Bini:2015bfb,Bini:2016qtx,Bini2019,Bini2020} by 
computing the first-order-self-force (1SF) correction to the time-averaged redshift 
$\langle z_1 \rangle= \langle ds_1/dt \rangle$ \cite{Barack:2011ed} of  body 1 to the eighth order in eccentricity and through the 9.5PN 
accuracy. Obtaining the eight order in eccentricity is, by itself, a major technical endeavour, and is crucial to allow us to 
inform the terms $\sim q_8(u;\nu) p_r^8$  in the total EOB effective Hamiltonian, and thereby to reach the 6PN approximation. 

The gauge-invariant  1SF observable we are using is defined as follows. One initially considers the
averaged redshift $\langle z_1 \rangle$ as a function of the two adimensionalized frequencies of an elliptic motion:
$ \widehat \Omega_r =G m_2 \Omega_r$ and $\widehat \Omega_\phi = G m_2 \Omega_\phi$ and of the mass ratio $q=m_1/m_2$. 
The 1SF expansion of the latter function yields: 
\beq
\langle z_1 \rangle(\widehat \Omega_r, \widehat \Omega_\phi, q)= \langle z_1 \rangle(\widehat \Omega_r, \widehat \Omega_\phi, 0)+ q \, \delta  z_1(\widehat \Omega_r, \widehat \Omega_\phi) + O(q^2)\,,
\eeq
where the $q=0$ term is the test-mass (Schwarzschild) result. The 1SF redshift is the function $\delta  z_1(\widehat \Omega_r, \widehat \Omega_\phi)$,
which can be alternatively expressed as a function of the unperturbed (Schwarzschild-backgound) semi-latus rectum $p^{\rm phys} \equiv G m_2 \,p$ and
eccentricity $e$. Denoting $u\equiv \frac1{p}$, the function $\delta z_1(u,e)$ is obtained as an expansion in powers of $e$, say
\beq
\delta z_1(u,e)= \delta z_1^{e^0}(u) + e^2 \delta z_1^{e^2}(u) + \ldots +  e^8 \delta z_1^{e^8}(u) + O(e^{10})\,,
\eeq
where each coefficient $\delta z_1^{e^{2n}}(u)$ is computed as a PN expansion (\ie, an expansion in powers of $u$) up to some order.

At the 4PN approximation, the functions $\delta z_1^{e^{2n}}(u)$ have been determined  up to the order $O(e^{20})$ in Refs. \cite{Hopper:2015icj,Bini:2016qtx}. Higher-PN order computations of the functions $\delta z_1^{e^{2}}(u)$ and $\delta z_1^{e^{4}}(u)$ were done in Refs. \cite{Bini:2015bfb,Bini:2016qtx} through the 9.5PN order (i.e., up to $u^{9.5}$), while the term $\delta z_1^{e^{6}}(u)$ was computed to the same accuracy in our recent 5PN-level works \cite{Bini2019,Bini2020}.
For the present 6PN-level work, we needed to extend this determination to the function $\delta z_1^{e^8}(u)$.
Our result for this function (up to the 9.5PN order) reads:
\begin{eqnarray} \label{dz1e8}
\delta z_1^{e^8}(u)&=&C_3 u^3 +C_4 u^4+(C_5^{\rm c}+C_5^{\ln{}}\ln u)u^5\nonumber\\
&+&(C_6^{\rm c}+C_6^{\ln{}}\ln u)u^6+C_{13/2}u^{13/2}\nonumber\\
&+&(C_7^{\rm c}+C_7^{\ln{}}\ln u)u^7+C_{15/2}u^{15/2}\nonumber\\
&+&(C_8^{\rm c}+C_8^{\ln{}}\ln u+C_8^{\ln^2{}}\ln^2 u)u^8+C_{17/2}u^{17/2}\nonumber\\
&+&(C_9^{\rm c}+C_9^{\ln{}}\ln u+C_9^{\ln^2{}}\ln^2 u)u^9\nonumber\\
&+& (C_{19/2}^{\rm c}+C_{19/2}^{\ln{}}\ln u)u^{19/2}\nonumber\\
&+&O_{\ln(u)}(u^{10})\,,
\end{eqnarray}
where the coefficients $C_i$ are listed in Table \ref{z1_e_8coeffs}.


\begin{table*}
\caption{\label{z1_e_8coeffs} List of the  various coefficients entering the self-force based  expression of $\delta z_1^{e^8}(u)$. }
\begin{ruledtabular}
\begin{tabular}{l|l}
\hline
$C_3$&$\frac{15}{64}$ \\
$C_4$&$\frac{3001}{384}-\frac{287}{4096}\pi^2$ \\
$C_5^{\rm c}$&$\frac{4597}{96}-\frac{162109375}{2304}\ln(5)-\frac{11332791}{1280}\ln(3)+\frac{55}{6}\gamma+\frac{15967961}{90}\ln(2)-\frac{474715}{196608}\pi^2$\\
$C_5^{\ln{}}$&$\frac{55}{12}$ \\
$C_6^{\rm c}$&$-\frac{9863051}{40320}+\frac{96889010407}{442368}\ln(7)-\frac{64481546637}{114688}\ln(3)-\frac{5977}{240}\gamma
-\frac{16605499789}{5040}\ln(2)
+\frac{1466047}{196608}\pi^2+\frac{4761539921875}{3096576}\ln(5)$\\
$C_6^{\ln{}}$&$-\frac{5977}{480}$\\
$C_{13/2}$&$+\frac{7527343}{145152}\pi $ \\
$C_7^{\rm c}$&$-\frac{18761241007}{870912}-\frac{12511111253459}{1492992}\ln(7)-\frac{75643996671875}{5225472} \ln(5)-\frac{53971661}{45360} \gamma $\\
&$
+\frac{13950859695883}{408240} \ln(2)+\frac{32462513613}{2240} \ln(3)-\frac{368710657}{33554432} \pi^4+\frac{205074667027}{113246208} \pi^2$\\
$C_7^{\ln{}}$&$-\frac{53971661}{90720}$ \\
$C_{15/2}$&$-\frac{107115666451}{162570240} \pi $ \\
$C_8^{\rm c}$&$-\frac{6488211537}{11200} \gamma \ln(3)-\frac{45307496529}{11200} \ln(2) \ln(3)
+\frac{111806640625}{12096} \ln(5) \gamma+\frac{111806640625}{12096} \ln(5) \ln(2)
-\frac{10769592586}{525} \gamma \ln(2)$\\
&$
+\frac{1919773074129997}{10059033600}+\frac{111806640625}{24192} \ln(5)^2+\frac{1922666600157935849}{14014218240} \ln(7)$\\
&$
-\frac{6488211537}{22400} \ln(3)^2+\frac{22363}{45} \gamma^2-\frac{555027930119}{14175} \ln(2)^2-\frac{5263490413}{453600} \gamma
-\frac{396348077586606421}{1571724000} \ln(2)-\frac{4261220414023638519}{35323904000} \ln(3)$\\
&$ 
-\frac{2926}{3} \zeta(3)+\frac{2572903425668796875}{103004504064} \ln(5)+\frac{472342810483}{805306368} \pi^4-\frac{83426620549601}{4529848320} \pi^2$\\
$C_8^{\ln{}}$&$-\frac{5263490413}{907200} +\frac{22363}{45} \gamma +\frac{111806640625}{24192}   \ln(5)-\frac{5384796293}{525} \ln(2)  -\frac{6488211537}{22400}\ln(3)$\\
$C_8^{\ln^2{}}$ &$+\frac{22363}{180}$\\
$C_{17/2}$&$+\frac{76704522232619}{15088550400} \pi$ \\
$C_9^{\rm c}$&$\frac{9273891051462598777}{3814050240000}-\frac{7041196288536323}{687194767360} \pi^4-\frac{534085235454726901}{2536715059200} \pi^2-\frac{6382001}{350} \gamma^2
+\frac{148431462289177}{198450} \ln(2)^2+\frac{76287}{5} \zeta(3)$\\
&$
+\frac{445208365512387}{10035200} \ln(3)^2-\frac{3307792499609375}{32514048} \ln(5)^2-\frac{8816899947037}{663552} \ln(7)^2-\frac{297870709952219425357}{273277255680} \ln(7)
$\\&$
+\frac{242219572992492481181}{143026884000} \ln(2)+\frac{344698525788968065625}{360515764224} \ln(5)-\frac{863597247149654361801}{1607237632000} \ln(3)
-\frac{8816899947037}{331776} \gamma \ln(7)$\\
&$
-\frac{8816899947037}{331776} \ln(2) \ln(7)+\frac{40278774263897}{99225} \gamma \ln(2)+\frac{445208365512387}{5017600} \gamma \ln(3)
+\frac{1075881868211907}{5017600} \ln(2) \ln(3)$\\
&$
-\frac{3307792499609375}{16257024} \ln(5) \gamma+\frac{42671896046383}{174636000} \gamma
-\frac{3307792499609375}{16257024} \ln(5) \ln(2)$\\ 
$C_9^{\ln{}}$&$+\frac{42653978392783}{349272000} -\frac{8816899947037}{663552}  \ln(7)-\frac{6382001}{350} \gamma  +\frac{40278774263897}{198450} \ln(2)  +\frac{445208365512387}{10035200}   \ln(3)-\frac{3307792499609375}{32514048}  \ln(5)$\\
$C_9^{\ln^2{}}$&$-\frac{6382001}{1400}$\\
$C_{19/2}^{\rm c}$&$\frac{3162423854803}{95256000} \pi \gamma-\frac{3345263881047}{784000} \pi \ln(3)+\frac{1900712890625}{72576} \pi \ln(5)
-\frac{5131372911332653}{95256000} \pi \ln(2)-\frac{44111568271901365400513}{154661262852096000} \pi$\\
&$-\frac{29555363129}{2721600} \pi^3$\\
$C_{19/2}^{\ln{}}$&$+\frac{3162423854803}{190512000} \pi $\\
\end{tabular}
\end{ruledtabular}
\end{table*}

The gauge-invariant information contained in the 1SF-accurate (first order in mass ratio)
function $\delta z_1^{e^8}$ can then be converted (by extending the results of Ref. \cite{Tiec:2015cxa})
 into the corresponding $O(\nu)$ contribution to the EOB potential 
$q_8(u ; \nu)$ parametrizing the term $q_8(u ; \nu) p_r^8 \in \widehat Q(u, p_r)$.
More precisely, writing as above
\beq
q_8(u;\nu)= \nu q_{8}^{\nu^1}(u)+\nu^2 q_{8}^{\nu^2}(u)+\nu^3 q_{8}^{\nu^3}(u)+\ldots\,,
\eeq
the 1SF  result $\delta z_1^{e^8}(u)$, Eq. \eq{dz1e8}, leads to the determination of the $O(\nu)$ coefficient   $q_{8}^{\nu^1}(u)$
to a reduced (fractional) 5.5PN accuracy. [In view of Eq. \eq{ADQnonloc},  such an accuracy corresponds to an absolute 8.5PN
accuracy of the Hamiltonian, which is more than enough for reaching our aimed 6PN accuracy.] 
We find the following $u^{5.5}$ accurate value for $q_{8}^{\nu^1}(u)$:
\begin{eqnarray} \label{q8nu}
q_8^{\nu^1}(u)&=&B_1 u  +B_2 u^2+B_{5/2}u^{5/2}\nonumber\\
&+& B_3u^3+B_{7/2}u^{7/2}\nonumber\\
&+&(B_4^{\rm c}+B_4^{\ln{}}\ln u)u^4+B_{9/2}u^{9/2}\nonumber\\
&+&(B_5^{\rm c}+B_5^{\ln{}}\ln u)u^5+(B_{11/2}^{\rm c}+B_{11/2}^{\ln{}}\ln u)u^{11/2}\nonumber\\
&+&O_{\ln(u)}(u^{6})\,,
\end{eqnarray}
where the various coefficients are listed in Table \ref{q8_coeffs}.

\begin{table*}
\caption{\label{q8_coeffs} List of the  various coefficients entering the self-force based  expression of $q_8(u)$.}
\begin{ruledtabular}
\begin{tabular}{l|l}
\hline
$B_1$ &$-\frac{27734375}{126}\ln(5)+\frac{6591861}{350}\ln(3)+\frac{21668992}{45}\ln(2)-\frac{35772}{175}$\\ 
$B_2$  &$\frac{13841287201}{17280}\ln(7)-\frac{393786545409}{156800}\ln(3)-\frac{16175693888}{1575}\ln(2)+\frac{875090984375}{169344}\ln(5)+\frac{5790381}{2450}$\\
$B_{5/2}$ &$+\frac{5994461}{12700800}\pi $\\
$B_3$&$-\frac{29247366220639}{933120}\ln(7)-\frac{63886617280625}{1016064}\ln(5)+\frac{5196312336176}{35721}\ln(2)+\frac{17515638027261}{313600}\ln(3)
-\frac{2843819611}{529200}$\\
$B_{7/2}$ &$+\frac{12986592749}{22759833600}\pi$\\
$B_4^{\rm c}$&$\frac{25659132742}{606375}\gamma-\frac{2458476234653610278}{1719073125}\ln(2)$\\
&$
+\frac{835937500}{27}\ln(5) \ln(2)+\frac{835937500}{27}\ln(5)\gamma-\frac{21240840924}{1225}\ln(2) \ln(3)$\\
&$
-\frac{4080804948}{875}\gamma \ln(3)-\frac{3558749575168}{55125}\gamma \ln(2)-\frac{806339890542506373}{1379840000}\ln(3)+\frac{581245383137875}{1824768}\ln(5)$\\
&$
+\frac{946254728855647813}{1642291200}\ln(7)+\frac{417968750}{27}\ln(5)^2-\frac{2040402474}{875}\ln(3)^2-\frac{60734915396608}{496125}\ln(2)^2$\\
&$
-\frac{303760055}{11010048}\pi^2-\frac{68634427305713}{185220000}$\\
$B_4^{\ln{}}$&$+\frac{12829566371}{606375} -\frac{1779374787584}{55125}\ln(2)  -\frac{2040402474}{875} \ln(3)+\frac{417968750}{27}\ln(5)$\\
$B_{9/2}$ &$+\frac{341391869291507}{18435465216000} \pi $\\
$B_5^{\rm c}$&$-\frac{4856666007821}{6306300}\gamma+\frac{6818519203656774203}{556215660}\ln(2)+\frac{156250000000000}{729729}\ln(10)$\\
&$
-\frac{675460646171875}{889056}\ln(5) \ln(2)-\frac{675460646171875}{889056}\ln(5) \gamma+\frac{1155153739426227}{1372000} \ln(2) \ln(3)$\\
&$
+\frac{496094995065267}{1372000}\gamma \ln(3)+\frac{5087539076789248}{3472875}\gamma \ln(2)-\frac{1259557135291}{12960}\ln(2) \ln(7)-\frac{1259557135291}{12960}\gamma \ln(7)$\\
&$
+\frac{400332056150861177697}{1757916160000}\ln(3)+\frac{1508174661184072060625}{1025216704512}\ln(5)-\frac{6373038655368769648873}{1067489280000}\ln(7)$\\
&$
-\frac{1259557135291}{25920}\ln(7)^2-\frac{675460646171875}{1778112}\ln(5)^2+\frac{496094995065267}{2744000}\ln(3)^2+\frac{9330506645499392}{3472875}\ln(2)^2$\\
&$
-\frac{740234446559}{176160768}\pi^2+\frac{82809381657923339131}{13984850880000}$\\
$B_5^{\ln{}}$&$-\frac{4856666007821}{12612600} +\frac{2543769538394624}{3472875}\ln(2) -\frac{1259557135291}{25920}   \ln(7)+\frac{496094995065267}{2744000} \ln(3)-\frac{675460646171875}{1778112} \ln(5)$\\
$B_{11/2}^{\rm c}$&$-\frac{18492868322811}{1097600}\pi \ln(3)+\frac{20550060546875}{222264}\pi \ln(5)-\frac{627504931547331563}{3333960000} \pi \ln(2)$\\
&$
+\frac{1587378124097}{3333960000}\pi \gamma+\frac{37938867020822625604207}{240584186658816000}\pi 
-\frac{14835309571}{95256000}\pi^3$\\
$B_{11/2}^{\ln{}}$&$+\frac{1587378124097}{6667920000} \pi  $\\
\end{tabular}
\end{ruledtabular}
\end{table*}

\section{Determining the local part of the EOB potentials at order $\nu^1$} \label{Heffloc1sf}

The next step of our strategy is to derive the {\it local} part of the EOB Hamiltonian by subtracting
the nonlocal part of the EOB potentials (obtained in Sec. \ref{Heffnonloc}) from their complete local-plus-nonlocal parts (obtained 
in Sec. \ref{SF} from self-force computations). As the self-force computation is only accurate to linear order in $\nu$, we thereby determine
the local part of the EOB potentials only at the first order in $\nu$. The nonlocal part we computed was of the h-type (and was
determined exactly in $\nu$). However,
recalling that we will always consider flexibility factors of the type $f=1+O(\nu)$, the h-route and f-route versions of both the nonlocal
and the local Hamiltonians only differ at the second self-force order, \ie, by terms of order $O(\nu^3)$ in the physical Hamiltonians, $H_{\rm nonloc, h,f}$, $H_{\rm loc, h,f}$,
corresponding to terms of order $O(\nu^2)$ in the corresponding squared effective Hamiltonians, $ \widehat H_{\rm eff}^2 $.

The values of the local   EOB potentials at 4+5+6PN, obtained from our results so far, can be written as:
\begin{widetext}
\begin{eqnarray}
\label{local_pot_param}
a_{\rm 4+5+6PN,loc, f}&=&
\left[\left(\frac{2275}{512}\pi^2-\frac{4237}{60}\right)\nu +\left(\frac{41}{32}\pi^2-\frac{221}{6}\right)\nu^2\right] u^5
+ \left[\left(-\frac{1026301}{1575}+\frac{246367}{3072}\pi^2\right)\nu+a_{6, f}^{(\nu)}\right] u^6\nonumber\\
&&+\left[\left(-\frac{2800873}{262144}\pi^4+\frac{608698367}{1769472}\pi^2-\frac{1469618167}{907200}\right)\nu+a_{7,f}^{(\nu)} \right]u^7\,,\nonumber\\
\bar d_{\rm 4+5+6PN,loc, f}&=&
\left[\left(\frac{1679}{9}-\frac{23761}{1536} \pi^2\right)\nu +\left(-260+\frac{123}{16}\pi^2\right) \nu^2\right]u^4
+\left(\frac{331054}{175}\nu-\frac{63707}{512}\nu\pi^2 +\bar d_{5, f}^{(\nu)}\right) u^5\nonumber\\
&&+\left[\left(\frac{229504763}{98304}\pi^2+\frac{135909}{262144}\pi^4-\frac{99741733409}{6350400}\right)\nu+\bar d_{6, f}^{(\nu)}\right]u^6 \,,\nonumber\\
q_{4,\rm 4+5+6PN,loc, f}&=&
\left(20\nu+q_{43}^{(\nu)} \right)u^3+\left[\left(-\frac{93031}{1536}\pi^2+\frac{1580641}{3150}\right)\nu+q_{44, f}^{(\nu)}\right] u^4\nonumber\\
&&+\left[\left(\frac{81030481}{65536}\pi^2-\frac{3492647551}{423360}\right)\nu+q_{45, f}^{(\nu)}\right]u^5\,,\nonumber\\
q_{6, \rm 4+5+6PN,loc, f}&=&\left(-\frac{9}{5}\nu +q_{62}^{(\nu)}\right)u^2 + \left(\frac{123}{10}\nu+q_{63, f}^{(\nu)}\right) u^3\nonumber\\
&&+\left[\left(-\frac{9733841}{327680}\pi^2-\frac{112218283}{294000}\right)\nu+q_{64, f}^{(\nu)}\right]u^4\,,\nonumber\\
q_{8, \rm 5+6PN,loc, f}&=& q_{82}(\nu)u^2 + \left(-\frac{7447}{560}\nu +q_{83}^{(\nu)} \right)u^3 \,, \nonumber\\
q_{10, \rm 6PN,loc, f}&=& q_{10,2}(\nu)u^2\,.
\end{eqnarray}
\end{widetext}

Here, the first coefficients in each line (except in the last two lines) belong to the 4PN level, and are equivalent to results obtained in 
Ref.~\cite{Damour:2015isa}. The explicit values of $q_{43}^{(\nu)}$ and $q_{62}^{(\nu)}$ are:
\bea
\label{local_pot_param2}
 q_{43}^{(\nu)} &&= -83\nu^2+10\nu^3 \,,\nonumber\\
 q_{62}^{(\nu)} &&= -\frac{27}{5}\nu^2+6\nu^3\,.
\eea
As exemplified by these coefficients, we introduced here the general notation $C^{(\nu)}$ to denote all the contributions to any
$\nu$-dependent coefficient $C(\nu)$ that are nonlinear in $\nu$, \ie,
\beq
C(\nu)= C^{\nu^1} \nu + C^{(\nu)}\;;\; {\rm with} \;  C^{(\nu)}=  C^{\nu^2} \nu^2 + \ C^{\nu^3} \nu^3 +\ldots
\eeq
The second coefficients in each line (and the first on the penultimate line) belong to the 5PN level, and were determined in our 
recent work \cite{Bini2019}, modulo two unknown coefficients at order $\nu^2$.
They read
\begin{eqnarray}
\label{local_pot_param3}
a_{6, f}^{(\nu)} &=& a_6^{\nu^2} \nu^2+4 \nu^3\,,\nonumber\\
\bar d_{5,f}^{(\nu)} &=& \bar d_5^{\nu^2}\nu^2+\left(\frac{1069}{3}-\frac{205}{16}\pi^2\right)\nu^3\,,\nonumber\\
q_{44,f}^{(\nu)} &=& \left(-\frac{2075}{3}+\frac{31633}{512}\pi^2\right)\nu^2+\left(640-\frac{615}{32}\pi^2\right)\nu^3\,,\nonumber\\
q_{63,f}^{(\nu)} &=& -\frac{69}{5}\nu^2+116\nu^3-14\nu^4\,,
\end{eqnarray}
where $a_{6, f}^{\nu^2}$ and ${\bar d}_{5, f}^{\nu^2}$ are the only two
 numerical coefficients left undetermined  at 5PN by our method.
 
Finally, the values of the 6PN-level coefficients are determined at the linear-in-$\nu$ level by our
self-force computation and can be written as
\begin{eqnarray}
\label{local_pot_param4}
a_{7, \rm loc, f}(\nu)&=&\left(-\frac{2800873}{262144}\pi^4+\frac{608698367}{1769472}\pi^2\right. \nonumber\\
&&\left.-\frac{1469618167}{907200}\right)\nu+a_{7,f}^{(\nu)} \,,\nonumber\\
\bar d_{6, \rm loc, f}(\nu)&=&\left(\frac{229504763}{98304}\pi^2+\frac{135909}{262144}\pi^4\right. \nonumber\\
&&\left.-\frac{99741733409}{6350400}\right)\nu+\bar d_{6, f}^{(\nu)} \,,\nonumber\\ 
q_{45, \rm loc, f}(\nu)&=&\left(\frac{81030481}{65536}\pi^2-\frac{3492647551}{423360}\right)\nu+q_{45, f}^{(\nu)}\,, \nonumber\\ 
q_{64, \rm loc, f}(\nu)&=&\left(-\frac{9733841}{327680}\pi^2-\frac{112218283}{294000}\right)\nu+q_{64, f}^{(\nu)} \,,\nonumber\\ 
q_{83, \rm loc, f}(\nu)&=& -\frac{7447}{560}\nu +q_{83, f}^{(\nu)}\,.
\end{eqnarray}
At this stage, we have no information about the nonlinear-in-$\nu$ coefficients $a_{7, f}^{(\nu)}$, $\bar d_{6, f}^{(\nu)}$,
$q_{45,f}^{(\nu)}$, $q_{64,f}^{(\nu)}$, and $q_{83,f}^{(\nu)}$. Let us, however, anticipate on the results of the following section,
where we will show how to determine the four $\nu$-nonlinear coefficients $\bar d_{6, f}^{(\nu)}$,
$q_{45,f}^{(\nu)}$, $q_{64,f}^{(\nu)}$, $q_{83,f}^{(\nu)}$, in terms of only two free numerical parameters, namely  
$\bar d_6^{\nu^2}$, and $q_{45,f}^{\nu^2}$. In addition, we will find that $a_{7, f}^{(\nu)}$ is at most cubic in $\nu$.
Our final results will then read:
\begin{eqnarray}
\label{local_pot_param5}
a_{7, f}^{(\nu)} &=& a_7^{\nu^2} \nu^2+a_7^{\nu^3} \nu^3\,,\nonumber\\
\bar d_{6, f}^{(\nu)}&=& \bar d_6^{\nu^2}\nu^2
+\left(\frac{45089}{72}-\frac{44489}{1536}\pi^2-\bar d_5^{\nu^2} -15 a_6^{\nu^2} \right)\nu^3\nonumber\\
&&-48\nu^4, \nonumber\\
q_{45,f}^{(\nu)}&=& q_{45}^{\nu^2}\nu^2+\left(-\frac{474899}{216}+\frac{36677}{1152}\pi^2-\frac{14}{3}\bar d_5^{\nu^2}\right)\nu^3\nonumber\\
&&+\left(-\frac{7375}{6}+\frac{1435}{32}\pi^2\right)\nu^4\,,\nonumber\\
q_{64,f}^{(\nu)} &=& \left(-\frac{21996581}{21000}+\frac{156397}{1280}\pi^2\right)\nu^2\nonumber\\
&&+\left(\frac{6977}{6}-\frac{29665}{256}\pi^2\right)\nu^3\nonumber\\
&&+\left(-\frac{3640}{3}+\frac{287}{8}\pi^2\right)\nu^4\,,\nonumber\\
q_{83,f}^{(\nu)} &=& -\frac{963}{56}\nu^2-\frac{117}{10}\nu^3-147\nu^4+18\nu^5\,.
\end{eqnarray}
In these results, the two coefficients $a_6^{\nu^2}$ and $\bar d_5^{\nu^2}$ come from the 5PN level,
while the new undetermined 6PN-level numerical coefficients are $ a_7^{\nu^2}$, $a_7^{\nu^3}$, $\bar d_6^{\nu^2}$,
and $q_{45}^{\nu^2}$. [The origin of these undetermined coefficients will be discussed below.]

The coefficient $ q_{10,2}(\nu)$ of $p_r^{10} u^2$ cannot be extracted from our $O(e^8)$ self-force results,
but it can be derived from the exact knowledge of the 
2PM ($O(G^2)$) EOB Hamiltonian \cite{Damour:2017zjx}, as will be shown below.
Its value is
\beq
\label{local_pot_param6}
q_{10, 2}(\nu)=-\frac{11}{21}\nu-\frac{11}{7}\nu^2-\frac{20}{7}\nu^3-\frac{5}{3}\nu^4+6\nu^5\,.
\eeq
Note the remarkable fact that the 4+5+6PN-accurate local $O(\nu)$ EOB Hamiltonian is {\it logarithm free}. Not only all the $\ln u$ terms 
present in the nonlocal EOB potentials have
disappeared (as expected because they have been known for a long time to be linked to the time nonlocality), but even the
various numerical logarithms $\ln 2, \ln 3, \ldots$, as well as Euler's constant $\gamma$ have all disappeared. Only rational numbers,
and $\pi^2 \sim \zeta(2)$ enter the $O(\nu)$  local Hamiltonian. In addition, the fractional powers of $u$ have also disappeared
because they only come from the nonlocal 5.5PN action.
For convenience, all these expressions are summarized in Table \ref{loc_eob_coeffs}.

 Note finally that, contrary to the nonlocal EOB potentials shown above,  there are no contributions to the local EOB potentials
 featuring powers of $u$ strictly smaller than 2. This follows from the fact that the PM expansion of the exact potential $Q$ starts
 at order $G^2$  \cite{Damour:2017zjx}. Contributions to $Q$ involving powers $u^n$ with $n<2$
can only enter the nonlocal part of the Hamiltonian, 
where they come from having expanded the nonlocal
Hamiltonian as a formally infinite series of powers of $p_r^2$ \cite{Damour:2015isa}.

\section{Using the mass-ratio dependence of the scattering angle to determine most of the $\nu^{n\geq 2}$ structure
of the 6PN $f$-route local Hamiltonian}

Up to this stage, our method has only determined (besides the full nonlocal part of the Hamiltonian)
the linear-in-$\nu$ part of the local Hamiltonian. The next stage of our method is to use the special $\nu$-dependence
of the scattering angle pointed out in Ref. \cite{Damour2019} to determine
most of the nonlinear dependence on $\nu$ of the local Hamiltonian. [See \cite{Antonelli:2020aeb} for a
generalization of this approach to the dynamics of spinning bodies.]
This is done by going through several steps.

\subsection{Going from the $p_r$-gauge to the energy-gauge}

As a first step, it is convenient to transform the above $p_r$-gauge form of the local EOB
effective Hamiltonian, \eqref{Heffquadloc}, to its (H-type) energy-gauge version, defined by writing
\beq
\label{EG-eff-sq-Ham}
\widehat H^{2 \, \rm EG}_{\rm eff, loc, f}(u, p_r, j;\nu)= H_S^2+(1-2 u) \widehat Q^{\rm EG}_{H\, \rm loc, f}(u, H_S ; \nu)\,,
\eeq
where $H_S$ denotes the (rescaled) Schwarzschild Hamiltonian, \ie, the square root of
\beq \label{HS}
H_S^2(u,p_r,j)=(1-2 u) [1+(1-2 u) p_r^2+j^2 u^2]\,,
\eeq
and where
\begin{eqnarray} \label{QEGH}
&&\widehat Q^{\rm EG}_{H\, \rm  loc, f}(u, H_S ; \nu)=u^2 q^H_{2\rm EG}(H_S;\nu)+u^3 q^H_{3\rm EG}(H_S;\nu)\nonumber\\
&&\qquad+u^4 q^{H \,\rm loc, f}_{4\rm EG}(H_S;\nu)+u^5q^{H \,\rm loc, f}_{5\rm EG}(H_S;\nu)\nonumber\\
&&\qquad+ 
u^6 q^{H \,\rm loc, f}_{ 6\rm EG}(H_S;\nu)+u^7 q^{H \,\rm loc, f}_{7\rm EG}(H_S;\nu)\,. 
\end{eqnarray}
We have added a label ``H'' on $\widehat Q^{\rm EG}_{H\, \rm loc, f}$ and its $u$-expansion coefficients, as a reminder that
we use here the H-version of the energy gauge, by contrast to its E-version \cite{Damour2019}. This means that $\widehat Q^H$
is directly written as a function of the phase-space variable $q,p$, via the argument $H_S(u,p_r,j)$. 
In the E-version of the energy-gauge   $\widehat Q$ is
written as a function of $u$ and the effective energy $\e$:
\begin{eqnarray} 
\label{QEGE}
&&\widehat Q^{\rm EG}_{E\, \rm  loc, f}(u, \e ; \nu)=u^2 q^E_{2\rm EG}(\e;\nu)+u^3 q^E_{3\rm EG}(\e;\nu)\nonumber\\
&&\qquad+u^4 q^{E \,\rm loc, f}_{4\rm EG}(\e;\nu)+u^5q^{E \,\rm loc, f}_{5\rm EG}(\e;\nu)\nonumber\\
&&\qquad+ 
u^6 q^{E \,\rm loc, f}_{ 6\rm EG}(\e;\nu)+u^7 q^{E \,\rm loc, f}_{7\rm EG}(\e;\nu)\,. 
\end{eqnarray}
The difference between the two sequences of expansion coefficients only start at the $u^4 \propto G^4$ level, so that
the first two functions\footnote{When $\g$ is used, as here, to denote the argument of $q^{H \,\rm loc, f}_{n\rm EG}$, it is
understood as a mathematical argument, to be later replaced by $H_S(u,p_r,j)$.} coincide with each other:
 $q^{H }_{2\rm EG}(\gamma;\nu)=q^{E }_{2\rm EG}(\gamma;\nu)$,
  $q^{H }_{3\rm EG}(\gamma;\nu)=q^{E }_{3\rm EG}(\gamma;\nu)$.
We henceforth denote them simply as   $q_{2\rm EG}(\gamma;\nu)$ and $q_{3\rm EG}(\gamma;\nu)$. [See below for
the link between the higher-order coefficients.]
We did not put any extra label ``loc, f'' on the first two coefficients because the effect of the flexibility coefficient $f$ only starts 
at the $G^4$ level. 

The energy-dependent coefficient $q^{H \,\rm loc, f}_{n\rm EG}(\gamma;\nu)$ belongs to the $n$-PM approximation because $u^n= (GM/(r^{\rm phys} c^2))^n$
is proportional to $G^n$. The 2PM coefficient  $q_{2\rm EG}(\gamma;\nu)$ is known exactly. It has been first obtained in 
Ref. \cite{Damour:2017zjx}, and then confirmed in Refs. \cite{Cheung:2018wkq,Bern:2019nnu,Bern:2019crd}.
The 3PM coefficient $q_{3\rm EG}(\gamma;\nu)$ has so far only be derived (as a closed-form function of $\gamma$ and $\nu$)
in Refs. \cite{Bern:2019nnu,Bern:2019crd}. Its 5PN expansion was confirmed in Ref. \cite{Bini2019}, and its 6PN expansion
was recently confirmed in Refs. \cite{Blumlein:2020znm,Cheung:2020gyp,Bini2020}. We will give below the details of our
derivation of the 6PN-accurate value of $q_{3\rm EG}(\gamma;\nu)$.
The higher PM-order coefficients $q^{H\,\rm loc, f}_{n\rm EG}(\gamma;\nu)$ are currently only known in their PN-expanded versions, say
\begin{eqnarray} 
\label{PNqHn}
q^{H\,\rm loc, f}_{n\rm EG}(\gamma;\nu)&=&q_{n\rm EG}^0(\nu)+q_{n\rm EG}^1(\nu)(\gamma^2-1)\nonumber\\
&+&q_{n\rm EG}^2(\nu)(\gamma^2-1)^2+\ldots\,.
\end{eqnarray}
We recall that the properties of the EOB formalism are such that the full potential $Q(u,\gamma;\nu)$ vanishes in the
test-mass limit $\nu \to 0$, so that each PN expansion  coefficient $q_{n\rm EG}^p(\nu)$ must be $\sim \nu + \nu^2 + \ldots$
when $\nu \to 0$.

The PN expansions of all the energy-gauge coefficients  $q^{H, \rm loc, f}_{n\rm EG}(\gamma)$ are determined from the corresponding 
$p_r$-gauge coefficients entering the Hamiltonian
 (notably the 6PN-level ones $a_7^{(\nu)}$, $\bar d_6^{(\nu)}$, $q_{45}^{(\nu)}$, $q_{64}^{(\nu)}$ and  $q_{83}^{(\nu)}$)
 by computing the canonical transformation connecting the two gauges. The structure of this canonical transformation is
 \beq
 g(r, p_r)= (r \, p_r) \left[g_{\rm 2PN} + g_{\rm 3PN}+g_{\rm 4PN}+g_{\rm 5PN} + g_{\rm 6PN}\right]\,,
 \eeq
 where the factor $ r \, p_r$ would describe an identity transformation, and where the leading-order term is at the 2PN (and 2PM) level ,
 and reads
 \beq
 g_{\rm 2PN}=\frac32 \eta^4 \frac{\nu }{r^2}\,.
 \eeq
 The 2PN ($g_{\rm 2PN}$) and 3PN ($g_{\rm 3PN}$) terms  were derived in Ref. \cite{Damour:2017zjx}; the 4PN 
one ($g_{\rm 4PN}$)  was derived in Appendix A of Ref. \cite{Antonelli:2019ytb}; and the 5PN one  ($g_{\rm 5PN}$)
was derived in our recent work \cite{Bini2020}. We have extended the determination of the canonical transformation $g(r,p_r)$
to the 6PN level. This is done by using the method of undetermined coefficients. The looked-for $g_{\rm 6PN}$ is parametrized as
\begin{eqnarray} \label{g6pn}
g_{\rm 6PN} &=& \frac{1}{r^2}\left[\frac{w_1 j^4 }{r^5}p_r^2+\frac{w_2 j^8}{r^8}+w_3p_r^8+\frac{w_4}{r^4}+\frac{w_5}{r}p_r^6\right.\nonumber\\
&+&\frac{w_6j^4}{r^4}p_r^4+\frac{w_7j^2}{r^2}p_r^6+\frac{w_8j^2}{r^4}p_r^2+\frac{w_9j^2}{r^3}p_r^4\nonumber\\
&+&\frac{w_{10}j^6}{r^6}p_r^2+\frac{w_{11}}{r^2}p_r^4+\frac{w_{12}}{r^3}p_r^2+\frac{w_{13}j^2}{r^5}\nonumber\\
&+&\left. \frac{w_{14}j^4}{r^6}+\frac{w_{15}j^6}{r^7}\right]\,,
\end{eqnarray}
with unknown coefficients $w_1,\ldots, w_{15}$. The values of these coefficients are then determined by imposing that 
 the two (effective, squared) Hamiltonians \eqref{Hf2} (with $A=A_{\rm loc, f}$, etc.) and \eqref{EG-eff-sq-Ham} are equivalent 
 (at the 6PN accuracy) through this canonical transformation.
 
The explicit expressions of the 6PN coefficients $w_1\ldots w_{15}$ will be displayed later, in their final form,  in Table \ref{table_gauge_params},
after we determine, using our strategy, all possible unknowns. 

\begin{table*}
\caption{\label{table_gauge_params} Final form of the coefficients $w_1\ldots w_{15}$ parametrizing the 6PN canonical transformation, Eq. \eq{g6pn}.}
\begin{ruledtabular}
\begin{tabular}{l|l}
\hline
$w_1$ &$ -\frac{236879}{80640}\nu-\frac{6753}{1792}\nu^2-\frac{197}{320}\nu^3-\frac{4965}{256}\nu^4-\frac{1417}{128}\nu^5$\\ 
$w_2$ &$ -\frac{33}{512}\nu-\frac{99}{512}\nu^2-\frac{45}{128}\nu^3-\frac{105}{512}\nu^4+\frac{189}{256}\nu^5$\\ 
$w_3$ &$ -\frac{2123}{10752}\nu-\frac{2123}{3584}\nu^2-\frac{965}{896}\nu^3-\frac{965}{1536}\nu^4+\frac{579}{256}\nu^5$\\ 
$w_4$ &$ \left(\frac{1483514111}{9437184}\pi^2-\frac{228466894127}{190512000}+\frac{45303}{1048576}\pi^4\right)\nu+\left(\frac{10486361}{84000}+\frac{1}{12}\bar d_6^{\nu^2} -\frac{132333}{40960}\pi^2+\frac{1}{12}a_6^{\nu^2} -\frac{1}{60} \bar d_5^{\nu^2} -\frac{1}{20}q_{45}^{\nu^2} \right)\nu^2$\\
&$+\left(-\frac{500837}{61440}\pi^2+\frac{213103}{1440}+\frac{3}{20}\bar d_5^{\nu^2} -\frac{5}{4}a_6^{\nu^2}\right)\nu^3+\frac{3}{32}\nu^4+\frac{21}{16}\nu^5$\\ 
$w_5$ &$ -\frac{284849}{80640}\nu-\frac{9951}{1792}\nu^2-\frac{8829}{2240}\nu^3-\frac{5595}{256}\nu^4-\frac{463}{128}\nu^5$\\ 
$w_6$ &$ -\frac{11}{20}\nu-\frac{33}{20}\nu^2-3\nu^3-\frac{7}{4}\nu^4+\frac{63}{10}\nu^5$\\ 
$w_7$ &$ -\frac{869}{1792}\nu-\frac{2607}{1792}\nu^2-\frac{1185}{448}\nu^3-\frac{395}{256}\nu^4+\frac{711}{128}\nu^5$\\ 
$w_8$ &$ \left(-\frac{48669205}{12582912}\pi^2-\frac{781859}{17640}\right)\nu+\left(-\frac{2898667}{20160}+\frac{781985}{49152}\pi^2\right)\nu^2+\left(\frac{119531}{720}-\frac{741625}{49152}\pi^2\right)\nu^3+\left(-\frac{117029}{1152}+\frac{7175}{1536}\pi^2\right)\nu^4+\frac{753}{64}\nu^5$\\ 
$w_9$ &$ -\frac{138127}{26880}\nu-\frac{14067}{1792}\nu^2-\frac{1423}{320}\nu^3-\frac{7199}{256}\nu^4-\frac{1527}{128}\nu^5$\\
$w_{10}$ &$ -\frac{77}{256}\nu-\frac{231}{256}\nu^2-\frac{105}{64}\nu^3-\frac{245}{256}\nu^4+\frac{441}{128}\nu^5$\\ 
$w_{11}$ &$ \left(-\frac{574296619}{125829120}\pi^2-\frac{200657371}{3528000}\right)\nu+\left(-\frac{158915363}{1008000}+\frac{9227423}{491520}\pi^2\right)\nu^2+\left(\frac{491849}{2880}-\frac{1750235}{98304}\pi^2\right)\nu^3+\left(-\frac{329935}{2304}+\frac{16933} {3072}\pi^2\right)\nu^4+\frac{687}{128}\nu^5$\\ 
$w_{12}$ &$ \left(-\frac{4561111909}{5292000}+\frac{15680782981}{110100480}\pi^2\right)\nu+\left(-\frac{2689283}{172032}\pi^2+\frac{134682217}{1764000}+\frac{4}{35}q_{45}^{\nu^2}\right)\nu^2+\left(-\frac{1559441}{4320}-\frac{8}{15}\bar d_5^{\nu^2} +\frac{3831013}{184320}\pi^2\right)\nu^3$\\
&$+\left(\frac{409}{72}-\frac{205}{384}\pi^2\right)\nu^4-\frac{33}{8}\nu^5$\\ 
$w_{13}$ &$ \left(-\frac{2031118237}{7056000}+\frac{5643368761}{110100480}\pi^2\right)\nu+\left(-\frac{416103}{57344}\pi^2+\frac{4901243}{196000}+\frac{3}{70}q_{45}^{\nu^2} \right)\nu^2+\left(-\frac{231013}{1440}-\frac15 \bar d_5^{\nu^2} +\frac{692411}{61440}\pi^2\right)\nu^3$\\
&$+\left(\frac{1343}{48}-\frac{205}{128}\pi^2\right)\nu^4-\frac{9}{2}\nu^5$\\ 
$w_{14}$ &$ \left(-\frac{9733841}{8388608}\pi^2-\frac{2868989}{235200}\right)\nu+\left(-\frac{2745397}{67200}+\frac{156397}{32768}\pi^2\right)\nu^2+\left(\frac{48821}{960}-\frac{148325}{32768}\pi^2\right)\nu^3+\left(-\frac{16513}{768}+\frac{1435}{1024}\pi^2\right)\nu^4+\frac{747}{128}\nu^5$\\ 
$w_{15}$ &$ -\frac{44651}{80640}\nu-\frac{717}{1792}\nu^2+\frac{207}{320}\nu^3-\frac{1425}{256}\nu^4-\frac{433}{128}\nu^5$\\ 
\end{tabular}
\end{ruledtabular}
\end{table*}

\subsection{Computing the f-route local scattering angle}

The next step in the determination of many of the non-linear-in-$\nu$ coefficients in the local EOB Hamiltonian proceeds through 
the computation of the  corresponding scattering angle, $\chi^{\rm loc, f}$. This is most efficiently done in the energy-gauge.

Several procedures (discussed in Refs. \cite{Damour:2017zjx,Damour2019}) can be used to
compute the expansion of $\chi^{\rm loc, f}(\gamma, j)$ in powers of $\frac1{j}\propto G$, at a fixed value of the
EOB effective energy $\gamma\equiv \e$. One uses the fact that, given any (local) Hamiltonian, the corresponding scattering angle 
of hyperboliclike motions is given by the integral 
 ($ u = 1/r$) \cite{Damour:2016gwp}
 \beq \label{chiintegral}
\frac12 (\chi(\gamma,j)+\pi)=-\int_0^{u_{\rm max}} \frac{\partial}{\partial j}p_r(u; \gamma,j) \frac{du}{u^2}\,,
\eeq
where $u_{\rm max}=u_{\rm max}(\gamma,j)=1/r_{\rm min}$ corresponds to the distance of closest approach of the two bodies, and where
 the radial momentum $p_r=p_r(u; \gamma,j)$ is obtained from writing the energy conservation at a given angular momentum. 
 When using the H-version of the energy gauge, Eq. \eq{EG-eff-sq-Ham} directly defines the squared effective Hamiltonian,
$ \widehat H^{2 \, \rm EG}_{\rm eff, loc, f}(u, p_r, j;\nu)$,
 as a function of $p_r, j$ and $u$. To obtain $p_r$ as a function of $\g \equiv \e$ one should then iteratively solve for $p_r$
 (in a PM expanded way, \ie, using the scaling $u\mapsto Gu$ and $j\mapsto G^{-1}j$) the energy conservation law  
 \begin{eqnarray} 
\label{gamma=HEG}
 \gamma^2 &=& \e^2= \widehat H^{2 \, \rm EG}_{\rm eff, loc, f}(u, p_r, j;\nu)\nonumber\\
&=& H_S^2+(1-2 u) \widehat Q^{\rm EG}_{H\, \rm loc, f}(u, H_S ; \nu)\,,
 \end{eqnarray}
 where $H_S(u,p_r,j)$ was defined in Eq.~\eq{HS}, and $\widehat Q^{\rm EG}_{H\,\rm loc, f}(u, H_S ; \nu)$ in Eq. \eq{QEGH}.
 The computation of the function $p_r(\e,j)$ is simpler when using the E-version of the energy gauge, \ie, Eq. \eq{QEGE}.
 Indeed, in that case the EOB mass-shell condition reads
 \beq \label{gamma=EEG}
 - \frac{\e^2}{1-2u} + 1+(1-2 u) p_r^2+j^2 u^2 + \widehat Q^{\rm EG}_{E\,\rm loc, f}(u, \e ; \nu)=0\,,
 \eeq
 which is a {\it linear equation} in $p_r^2(\e,j,u)$ whose exact solution reads (denoting again $\g \equiv \e$)
 \beq \label{prEversion}
 p_r^2(\g,j,u)= \frac{\g^2-(1-2u)\left(1+j^2 u^2 + \widehat Q^{\rm EG}_{E\,\rm loc, f}(u, \g ; \nu)\right)}{(1-2u)^2}\,.
 \eeq
 In both cases, one expands $p_r(\g,j,u)$, as it appears in Eq.\eq{chiintegral}, in powers of $u \mapsto G u$, say
 \beq 
p_r=p_r^{(0)}+Gp_r^{(1)}+G^2p_r^{(2)}+\ldots\,,
\eeq
whose first two terms read
\beq \label{pr0}
p_r^{(0)}=\sqrt{ -1+\gamma^2-j^2u^2}\,,
\eeq
and
\beq
p_r^{(1)}= \frac{ (-1+2\gamma^2-j^2u^2) u }{ \sqrt{ -1+\gamma^2-j^2u^2}} \,.
\eeq
All the integrals that appear in the PM expansion of Eq.\eq{chiintegral} are elementary and are evaluated
(following \cite{Damour:1988mr}) by using Hadamard's partie finie.

The scattering angle is then obtained as a PM expansion of the form
\beq
\label{PM_scat_chi}
\frac1{2} \chi^{\rm loc, f}(\gamma, j;\nu)= \sum_{n\geq1} \frac{\chi_n^{\rm loc, f}(\g;\nu)}{j^n}\,.
\eeq
Here, each $n$-PM-order expansion coefficient $\chi_n^{\rm loc, f}(\g;\nu)$ is determined from the value of the 
corresponding $n$-PM-order energy-gauge coefficient $q^{H\,\rm loc, f}_{n\rm EG}(\gamma;\nu)$, or  
$q^{E\,\rm loc, f}_{n\rm EG}(\gamma;\nu)$, together with the values of the lower PM-order coefficients.

Denoting, for brevity,
 \beq
\Delta \chi_n(\gamma)\equiv\chi_n(\gamma)-\chi_n^{\rm Schw}(\gamma)\,,
\eeq
the scattering-angle coefficients obtained from the  {\it E-version}  $\widehat Q^{\rm EG}_{E\,\rm loc, f}(u, \g ; \nu)$ corrected  
 of the energy gauge (which is simpler to implement in view of the explicit expression \eq{prEversion}) read
\begin{eqnarray} \label{chinvsqn}
\Delta \chi_2(\gamma)&=&-\frac{\pi}{4}q_2(\gamma)\,, \nonumber\\
\Delta \chi_3(\gamma)&=&-\frac{2\gamma^2-1}{\sqrt{\gamma^2-1}}q_2(\gamma)-\sqrt{\gamma^2-1}q_3(\gamma)\,, \nonumber\\
\Delta \chi_4(\gamma)&=&\pi \left[ \frac{3}{16} q_2(\gamma)^2 -\frac{9}{16} (-1+5\gamma^2) q_2(\gamma)\right. \nonumber\\
&&\left.-\frac{3}{8}  (-1+3\gamma^2) q_3(\gamma)-\frac38 q^E_4(\gamma) (\gamma^2-1)  \right]\,, \nonumber\\
\Delta \chi_5(\gamma)&=& \frac{(2\gamma^2-1)}{\sqrt{\gamma^2-1}}q_2(\gamma)^2
+\left[2\sqrt{\gamma^2-1}  q_3(\gamma )\right. \nonumber\\
&&\left.
-\frac{2}{3} \frac{(60\gamma^2-5+64\gamma^6-120\gamma^4)}{(\gamma^2-1)^{3/2}}\right] q_2(\gamma)\nonumber\\
&&
-2 \frac{(8\gamma^4+1-8\gamma^2)}{\sqrt{\gamma^2-1}} q_3(\gamma)\nonumber\\
&&-\frac{4(4\gamma^2-1)\sqrt{\gamma^2-1}}{3}  q^E_4(\gamma)\nonumber\\
&&-\frac{4 (\gamma^2-1)^{3/2}}{3}q^E_5(\gamma)\,,\nonumber\\
\Delta \chi_6(\gamma)&=& -\frac{5}{32}q_2(\gamma)^3
+\left(-\frac{45}{64}+\frac{225}{64}\gamma^2\right) q_2(\gamma)^2\nonumber\\
&&+\left[\left(\frac{45}{16}\gamma^2-\frac{15}{16}\right)q_3(\gamma)\right. \nonumber\\
&&+\left(-\frac{15}{16}+\frac{15}{16}\gamma^2\right)q^E_4(\gamma)\nonumber\\
&&\left.-\frac{17325}{256}\gamma^4+\frac{4725}{128}\gamma^2-\frac{525}{256}\right]q_2(\gamma)\nonumber\\
&&  +\left(\frac{15}{32}\gamma^2-\frac{15}{32}\right)q_3(\gamma)^2\nonumber\\
&&  +\left(-\frac{1575}{64}\gamma^4+\frac{525}{32}\gamma^2-\frac{75}{64}\right)q_3(\gamma)\nonumber\\
&&  +\left(-\frac{525}{64}\gamma^4+\frac{225}{32}\gamma^2-\frac{45}{64}\right) q^E_4(\gamma)\nonumber\\
&&  +\left(-\frac{75}{32}\gamma^4-\frac{15}{32}+\frac{45}{16}\gamma^2\right)q^E_5(\gamma)\nonumber\\
&&  +\left(-\frac{15}{32}-\frac{15}{32}\gamma^4+\frac{15}{16}\gamma^2\right)q^E_6(\gamma)
\,.
\end{eqnarray}
 The first three equations above (for $\chi_2$, $\chi_3$, $\chi_4$) agree with the corresponding ones in Ref. \cite{Damour2019}.
 
While the E-version of the energy-gauge is more simply connected to the scattering angle, the H-version is more simply connected
to the usual $p_r$-gauge EOB Hamiltonian. This is why we use the H-version in practice, as indicated in Eq. \eq{PNqHn}. Let us therefore
complete the above E-type scattering-angle results by the transformation between the E-type coefficients,  $q^{E\,\rm loc, f}_{n\rm EG}(\gamma;\nu)$,
and the H-type ones, $q^{H\,\rm loc, f}_{n\rm EG}(\gamma;\nu)$. We recall that the first two, $q_2$ and $q_3$, are the same.
To write the link between the higher-order ones, it is convenient to provisionally use as common argument
for these functions $x \equiv \g^2$. By writing that Eqs. \eq{gamma=HEG} and \eq{gamma=EEG} define the same mass-shell
constraint one finds (where, for uniformity, we have left the labels E or H on $q^E_2 = q^H_2=q_2$ and 
$q^E_3 = q^H_3=q_3$):

\begin{widetext}
\begin{eqnarray}
q^H_4(x)&=&q^E_4(x) + q_2^E(x) \frac{d q_2^E(x)}{d x}\,,\nonumber\\
q^H_5(x) &=& (q^E_3(x)-2 q^E_2(x))\frac{d q^E_2(x)}{dx} +q^E_5(x)+\frac{d q^E_3(x)}{dx} q^E_2(x)\,,\nonumber\\
q^H_6(x) &=& \frac12\frac{d^2q^E_2(x)}{dx^2}q^E_2(x)^2+q^E_2(x)\left(\frac{d q^E_2(x)}{dx}\right)^2
+(q^E_4(x)-2 q^E_3(x))\frac{d q^E_2(x)}{dx}\nonumber\\
&& +(q^E_3(x)-2 q^E_2(x))\frac{d q^E_3(x)}{dx}+q^E_6(x)+q^E_2(x)\frac{d q^E_4(x)}{dx}\,,\nonumber\\
q^H_7(x) &=& (-2 q^E_2(x)^2+q^E_2(x) q^E_3(x))\frac{d^2q^E_2(x)}{dx^2} +\frac12 \frac{d^2q^E_3(x)}{dx^2} q^E_2(x)^2
+(q^E_3(x)-4 q^E_2(x))\left(\frac{dq^E_2(x)}{dx} \right)^2\nonumber\\
&&
+\left[2\frac{d q^E_3(x)}{dx} q^E_2(x)+q^E_5(x)-2 q^E_4(x)\right] \frac{dq^E_2(x)}{dx}
+(q^E_4(x)-2q^E_3(x))\frac{d q^E_3(x)}{dx} \nonumber\\
&&+(q^E_3(x)-2 q^E_2(x))\frac{dq^E_4(x)}{dx} +q^E_7(x)+q^E_2(x)\frac{d q^E_5(x)}{dx}\,,
\end{eqnarray}
which can also be written in the reverse direction:
\begin{eqnarray}
q^E_4(x)&=&   -\frac{dq^H_2(x)}{dx}q^H_2(x)+q^H_4(x)\,,\nonumber\\
q^E_5(x) &=& \frac{d q^H_2(x) }{dx}[-q^H_3 (x)+2q_2^H (x)]
-\frac{d q^H_3(x)}{dx}q^H_2(x)+q^H_5(x)\,,\nonumber\\
q^E_6(x) &=&-\frac{d q^H_4(x)}{dx}  q_2^H(x)
+\left(\frac{d q^H_2 (x)}{dx}\right)^2 q_2(x)
-\frac{d q^H_2(x)}{dx}(q^H_4(x)-2 q_3^H(x))
+\frac12 \frac{d^2 q^H_2(x)}{dx^2} [q_2^H(x)]^2\nonumber\\
&&-\frac{d q^H_3(x)}{dx} ( q_3^H(x) -2  q_2^H (x))
+q^H_6(x)
\,,\nonumber\\
q^E_7(x) &=& -q^H_2(x)(-q^H_3(x)+2q^H_2(x))\frac{d^2 q^H_2(x)}{dx^2}
+\frac12 \frac{d^2 q^H_3(x)}{dx^2}q^H_2(x)^2
+(q^H_3(x)-4q^H_2(x))\left(\frac{dq^H_2(x)}{dx}\right)^2
\nonumber\\&&
+\left[2\frac{dq^H_3(x)}{dx}q^H_2(x)
-q^H_5(x)+2q^H_4(x)\right]\frac{dq^H_2(x)}{dx}\nonumber\\
&&+(-q^H_4(x)+2q^H_3(x))\frac{dq^H_3(x)}{dx}
+(-q^H_3(x)+2q^H_2(x))\frac{dq^H_4(x)}{dx}
-\frac{dq^H_5(x)}{dx} q^H_2(x)
+q^H_7(x)\,.
\end{eqnarray}
\end{widetext}

\subsection{Using the mass-ratio dependence of the f-route local scattering angle}

Applying the scattering-angle results derived in the previous subsection
 to our PN-expanded parametrization of the H-type energy-gauge coefficients, Eq. \eq{PNqHn},
yields explicit, PN-expanded (6PN-accurate) expressions for the scattering angle. [These can also be
obtained by directly evaluating the integral \eq{chiintegral} in a PN-expanded way].
Let us only give here one specific example:
\begin{eqnarray} \label{chi5locf}
\chi_5^{\rm loc,f} &=& \frac{1}{5 p_\infty^5}-\frac{2}{p_\infty^3}\eta^2+\frac{32-8\nu}{p_\infty}\eta^4\nonumber\\
&&
+\left[320+\left(-\frac{1168}{3}+\frac{41}{8}\pi^2\right)\nu+24\nu^2\right] p_\infty\eta^6\nonumber\\
&&
+\left[640+\left(\frac{5069}{144}\pi^2-\frac{227059}{135}\right)\nu\right. \nonumber\\
&&\left.
+\left(-\frac{287}{24}\pi^2+\frac{7342}{9}\right)\nu^2-40\nu^3\right]p_\infty^3\eta^8\nonumber\\
&&
+\left[\frac{1792}{5}+\left(-\frac{1460479}{525}+\frac{111049}{960}\pi^2\right)\nu\right. \nonumber\\
&&
+\left(\frac{41026}{15}-\frac{40817}{640}\pi^2-\frac{4}{15}\bar d_5^{\nu^2}\right)\nu^2\nonumber\\
&&\left.
+\left(-\frac{11108}{9}+\frac{451}{24}\pi^2\right)\nu^3+56\nu^4 \right] p_\infty^5\eta^{10}\nonumber\\
&&
+\left[\left(\frac{93031}{2304}\pi^2-\frac{498343703}{604800}\right)\nu\right. \nonumber\\
&&
+\left(\frac{2827607}{1152}-\frac{31633}{768}\pi^2\right)\nu^2\nonumber\\
&&
+\left(\frac{205}{16}\pi^2-\frac{253361}{96}\right)\nu^3
+\frac{212879}{384}\nu^4+\frac{63}{64}\nu^5\nonumber\\
&&\left.
-2 q_{3\rm EG}^4-4 q_{4\rm EG}^3-\frac43 q_{5\rm EG}^2\right] p_\infty^7\eta^{12}\,.
\end{eqnarray}
 Here, we used as energy variable  the EOB asymptotic momentum $p_{\infty}$, defined as
\beq
p_{\infty}^2 \equiv \g^2-1\,.
\eeq
This quantity naturally appears in the PM-expanded mass-shell condition, see Eq. \eq{pr0},
and is also a convenient PN-expansion parameter $p_{\infty}^2 \mapsto \eta^2 p_{\infty}^2$.
We recall that the $q_{n\rm EG}^k$'s appearing in Eq. \eq{chi5locf} are  the coefficients of the expansion
in powers of $p_{\infty}^2$ of the H-type $q^{H\,\rm loc, f}_{n\rm EG}(\gamma;\nu)$
coefficients, see Eq. \eq{PNqHn}.

Having in hands the expressions of the $\chi_n^{\rm loc, f}(p_{\infty};\nu)$'s (which we shall indifferently denote as 
$\chi_n^{\rm loc, f}(\g;\nu)$), let us now consider the following
energy-rescaled versions of these coefficients
\beq \label{tildechin}
\widetilde \chi_n^{\rm loc, f}(\g;\nu)\equiv  \left[h(\g;\nu)\right]^{n-1} \chi_n^{\rm loc, f}(\g;\nu)\,,
\eeq
where 
\beq
h(\gamma;\nu) \equiv \sqrt{1 + 2\nu(\gamma-1)} = \frac{H}{M c^2}\,.
\eeq
Ref. \cite{Damour2019} has shown that the {\it total} (local plus nonlocal) scattering angle satisfied the following condition:
\beq
 C_n^{\rm tot} : \widetilde \chi_n^{\rm tot}(\g;\nu) = P^{\gamma}_{d_n}(\nu)\; ; \; {\rm with} \; d_n\equiv \left[\frac{n-1}{2} \right].
 \eeq
 Here, and below, the notation $P^{\gamma}_{k}(\nu)$ denotes a generic polynomial of degree $\leq k$, with ${\gamma}$- 
 (or, equivalently, $p_{\infty}$-) dependent coefficients.
 
In Ref. \cite{Bini2020} we pointed out the simplification brought in the determination of the local Hamiltonian
by choosing a flexibility factor $f(t)$ in the definition of the Pf scale $r_{12}^f= f(t)  r_{12}^h$ such that the condition
$C_n^{\rm tot}$ {\it separately} applies to the nonlocal contribution $\chi_n^{\rm nonloc, f}(\g;\nu)$, and 
to the local one $\chi_n^{\rm loc, f}(\g;\nu)$. [We recall that 
$\chi_n^{\rm tot}(\g;\nu) =\chi_n^{\rm loc, f}(\g;\nu) +\chi_n^{\rm nonloc, f}(\g;\nu)$, because the nonlocal part can
be treated as a first-order perturbation.] We showed there that it was always possible to construct such a flexibility
factor $f=1+O(\frac{\nu}{c^2})$ at the 1PN fractional accuracy. We will show in a separate work that this holds also
at the 2PN fractional accuracy, of relevance to the present study. This choice of such a tuned $f$ allows us to separate
the determination of  the f-route local Hamiltonian, from the discussion of the corresponding nonlocal contribution
to the scattering angle, $\chi_n^{\rm nonloc, f}(\g;\nu)$.

We shall then enforce the condition (with $ d_n\equiv \left[\frac{n-1}{2} \right]$)
\beq \label{Cn1}
 C_n^{\rm loc, f} : \widetilde \chi_n^{\rm loc, f}(\g;\nu) = P^{\gamma}_{d_n}(\nu)\; ; 
 \eeq
 \ie,
 \beq \label{Cn2}
 C_n^{\rm loc, f} : \widetilde \chi_n^{\rm loc, f}(\g;\nu) = c_{n 0}(\g) + c_{n 1}(\g) \nu +\ldots+  c_{n d_n}(\g) \nu^{d_n}\; . 
 \eeq
 This condition yields strong constraints on the $\nu$-dependence of the various coefficients in the local Hamiltonians (in any gauge),
 and allows one to determine most of the coefficients entering the (usual) local Hamiltonian $H^{\rm loc, f}(r,p_r,j)$.

Applying the condition $ C_n^{\rm loc, f}$  for $n=2,\ldots , 7$, we could determine the nonlinear $\nu$-dependence of
the coefficients entering the 6PN-accurate $p_r$-gauge effective Hamiltonian $\widehat H^2_{\rm eff \, loc, f}$, except for
the following {\it four} numerical coefficients
\beq
a_7^{\nu^2}, a_7^{\nu^3}, \bar d_6^{\nu^2}, \,{\rm and}\;  q_{45}^{\nu^2}\,.
\eeq
We recall that, at the 5PN level, we could determine the nonlinear $\nu$-dependence of the EOB potentials
except for {\it two} numerical coefficients: $a_6^{\nu^2}$, and $\bar d_5^{\nu^2}$.
We list in Table \ref{loc_eob_coeffs} the knowledge of the coefficients parametrizing the f-route 
 local  EOB potentials. We note that among the 52  coefficients entering the 5+6PN local EOB
 potentials our method allowed to determine 46. 
To complete the previous information we also list in Tables \ref{HEG_coeffs}, \ref{EEG_coeffs}
the parameters entering the H-type and E-type energy-gauge (squared) effective  Hamiltonian for $n\ge 3$ and $n\ge 4$, respectively.
[We recall that $q^E_3 = q^H_3$.]

The situation is even more impressive if one considers
 the usual Hamiltonian, expressed in terms of the effective one by Eq. \eq{eobmap}, as a function of $u, p_r$, and $p^2\equiv p_r^2+j^2u^2$,
\beq \label{Heobfin}
H^{\rm loc, f, 6PN}=\sum_{0\leq k \leq 7, 0\leq l\leq 5, k+l\leq7}  C^{(2 l)}_{2k}(\nu) p^{2k} p_r^{2l}u^{7-k-l}\,.
\eeq
Indeed, the $\nu$-dependence of this 6PN-level Hamiltonian (see Table \ref{table_Hreal}) contains $151$ 
(or $147$, if we consider that four coefficients start at $O(\nu^2)$)  numerical coefficients, 
 and our method determines $151-4=147$ (or $147-4=143$)  of them.
The  $\nu$-dependent coefficients  $C^{(2 l)}_{2k}(\nu)= \sum_n C^{(2 l)}_{2k, n} \nu^n$ are listed in Table \ref{table_Hreal}.


\begin{table*}
\caption{\label{loc_eob_coeffs} List of the $f$-route EOB potentials in $p_r$-gauge.}
\begin{ruledtabular}
\begin{tabular}{l|l}
\hline
$a_5^{\rm loc,f}$&$
\left(-\frac{4237}{60}+\frac{2275}{512}\pi^2\right)\nu+\left(\frac{41}{32}\pi^2-\frac{221}{6}\right)\nu^2
$\\
$a_6^{\rm loc,f}$&$
\left(-\frac{1026301}{1575}+\frac{246367}{3072}\pi^2\right)\nu+a_6^{\nu^2}\nu^2+4\nu^3
$\\
$a_7^{\rm loc,f}$&$
\left(-\frac{2800873}{262144}\pi^4+\frac{608698367}{1769472}\pi^2-\frac{1469618167}{907200}\right)\nu+a_7^{\nu^2}\nu^2+a_7^{\nu^3}\nu^3
$\\
$\bar d_4^{\rm loc,f}$&$
\left(\frac{1679}{9}-\frac{23761}{1536}\pi^2\right)\nu+\left(\frac{123}{16}\pi^2-260\right)\nu^2
$\\
$\bar d_5^{\rm loc,f}$&$
\left(\frac{331054}{175}-\frac{63707}{512}\pi^2\right)\nu+\bar d_5^{\nu^2}\nu^2+\left(-\frac{205}{16}\pi^2+\frac{1069}{3}\right)\nu^3
$\\
$\bar d_6^{\rm loc,f}$&$
\left(\frac{229504763}{98304}\pi^2+\frac{135909}{262144}\pi^4-\frac{99741733409}{6350400}\right)\nu+\bar d_6^{\nu^2}\nu^2+\left(\frac{45089}{72}-\frac{44489}{1536}\pi^2-\bar d_5^{\nu^2}-15a_6^{\nu^2}\right)\nu^3-48\nu^4
$\\
$q_{43}^{\rm loc,f}$&$
20\nu-83\nu^2+10\nu^3
$\\
$q_{44}^{\rm loc,f}$&$
\left(\frac{1580641}{3150}-\frac{93031}{1536}\pi^2\right)\nu+\left(-\frac{2075}{3}+\frac{31633}{512}\pi^2\right)\nu^2+\left(640-\frac{615}{32}\pi^2\right)\nu^3
$\\
$q_{45}^{\rm loc,f}$&$
\left(\frac{81030481}{65536}\pi^2-\frac{3492647551}{423360}\right)\nu+q_{45}^{\nu^2}\nu^2+\left(-\frac{14}{3}\bar d_5^{\nu^2}+\frac{36677}{1152}\pi^2-\frac{474899}{216}\right)\nu^3+\left(\frac{1435}{32}\pi^2-\frac{7375}{6}\right)\nu^4
$\\
$q_{62}^{\rm loc,f}$&$
-\frac{9}{5}\nu-\frac{27}{5}\nu^2+6\nu^3
$\\
$q_{63}^{\rm loc,f}$&$
\frac{123}{10}\nu-\frac{69}{5}\nu^2+116\nu^3-14\nu^4
$\\
$q_{64}^{\rm loc,f}$&$
\left(-\frac{9733841}{327680}\pi^2-\frac{112218283}{294000}\right)\nu+\left(\frac{156397}{1280}\pi^2-\frac{21996581}{21000}\right)\nu^2+\left(\frac{6977}{6}-\frac{29665}{256}\pi^2\right)\nu^3+\left(\frac{287}{8}\pi^2-\frac{3640}{3}\right)\nu^4
$\\
$q_{82}^{\rm loc,f}$&$
\frac{6}{7}\nu+\frac{18}{7}\nu^2+\frac{24}{7}\nu^3-6\nu^4
$\\
$q_{83}^{\rm loc,f}$&$
-\frac{7447}{560}\nu-\frac{963}{56}\nu^2-\frac{117}{10}\nu^3-147\nu^4+18\nu^5
$\\
$q_{10,2}^{\rm loc,f}$&$
-\frac{11}{21}\nu-\frac{11}{7}\nu^2-\frac{20}{7}\nu^3-\frac{5}{3}\nu^4+6\nu^5
$\\
\end{tabular}
\end{ruledtabular}
\end{table*}


\begin{table*}
\caption{\label{HEG_coeffs} List of the $\pinf^2$-expansion coefficients  of  the  $u$-coefficients in the H-type energy-gauge (squared) effective  Hamiltonian for $n\ge 3$, see Eqs. \eqref{QEGH} and \eq{PNqHn}.}
\begin{ruledtabular}
\begin{tabular}{l|l}
\hline
$ q_{3\rm EG}^0$ &$5\nu$\\
$ q_{3\rm EG}^1$ &$27\nu-\frac{23}{4}\nu^2 $\\
$ q_{3\rm EG}^2$ &$  \frac{1021}{80}\nu-\frac{445}{16}\nu^2+\frac{49}{8}\nu^3 $\\
$ q_{3\rm EG}^3$ &$-\frac{213}{2240}\nu-\frac{2409}{320}\nu^2+\frac{437}{16}\nu^3  -\frac{407}{64}\nu^4$\\ 
$ q_{3\rm EG}^4$ &$ -\frac{50557}{80640}\nu+\frac{1149}{1792}\nu^2+\frac{489}{320}\nu^3-\frac{6735}{256}\nu^4+\frac{835}{128}\nu^5 $\\ 
$ q_{4\rm EG}^0$ &$ \nu \left(\frac{175}{3}-\frac{41}{32}\pi^2\right)-\frac72 \nu^2$\\
$ q_{4\rm EG}^1$ &$ \left(\frac{5632}{45}-\frac{33601}{6144}\pi^2\right)\nu+\left(-\frac{405}{4}+\frac{123}{64}\pi^2\right)\nu^2+\frac{13}{2}\nu^3$\\ 
$q_{4\rm EG}^2 $ &$ \left(-\frac{93031}{12288}\pi^2+\frac{699761}{7200}\right)\nu+\left(\frac{31633}{4096}\pi^2-\frac{77443}{480}\right)\nu^2+\left(-\frac{615}{256}\pi^2+130\right)\nu^3-\frac{293}{32}\nu^4$\\ 
$q_{4\rm EG}^3 $ &$ \left(-\frac{40415563}{1411200}-\frac{9733841}{4194304}\pi^2\right)\nu
+\left(-\frac{6742919}{67200}+\frac{156397}{16384}\pi^2\right)\nu^2
+\left(\frac{80153}{480}-\frac{148325}{16384}\pi^2\right)\nu^3
+\left(-\frac{7223}{48}+\frac{1435}{512}\pi^2\right)\nu^4
+\frac{185}{16}\nu^5$\\ 
$q_{5\rm EG}^0 $ &$ \left(-\frac{29917}{6144}\pi^2+\frac{44357}{360}\right)\nu+\left(-\frac{2387}{24}+\frac{205}{64}\pi^2\right)\nu^2+\frac{9}{4}\nu^3$\\ 
$q_{5\rm EG}^1 $ &$\left(\frac{15540691}{25200}-\frac{2590847}{61440}\pi^2\right)\nu
+\left(\frac{1}{5}\bar d_5^{\nu^2}-\frac{15581}{80}+\frac{347673}{20480}\pi^2\right)\nu^2
+\left(-\frac{1763}{256}\pi^2+\frac{5131}{24}\right)\nu^3
-\frac{93}{16}\nu^4 $\\ 
$q_{5\rm EG}^2 $ &$ \left(-\frac{807638471}{1764000}+\frac{30033990443}{440401920}\pi^2\right)\nu
+\left(\frac{3}{35}q_{45}^{\nu^2}+\frac{6466655}{114688}\pi^2-\frac{299020817}{588000}\right)\nu^2
+\left(-\frac{7356287}{245760}\pi^2+\frac{289783}{1440}-\frac{2}{5}\bar d_5^{\nu^2}\right)\nu^3$\\
&$+\left(-\frac{41833}{128}+\frac{5535}{512}\pi^2\right)\nu^4
+\frac{657}{64}\nu^5$\\ 
$ q_{6\rm EG}^0$ &$ \left(\frac{541363}{10240}\pi^2-\frac{69733}{350}\right)\nu+\left(\frac{11717}{60}+a_6^{\nu^2}+\frac{1}{5}\bar d_5^{\nu^2}+\frac{17857}{5120}\pi^2\right)\nu^2+\left(\frac{326}{3}-\frac{287}{64}\pi^2\right)\nu^3-\frac{11}{8}\nu^4$\\ 
$ q_{6\rm EG}^1$ &$ \left(\frac{45303}{524288}\pi^4-\frac{195178823647}{47628000}+\frac{100876235443}{165150720}\pi^2\right)\nu
+\left(\frac{1}{6}\bar d_6^{\nu^2}+\frac{11}{70}q_{45}^{\nu^2}+\frac{7}{6}a_6^{\nu^2}+\frac{596127373}{588000}+\frac{593223}{28672}\pi^2+\frac{1}{6}\bar d_5^{\nu^2}\right)\nu^2$\\
&$
+\left(-\frac{431999}{30720}\pi^2-\frac{78703}{90}-\frac{9}{10}\bar d_5^{\nu^2}-\frac{5}{2}a_6^{\nu^2}\right)\nu^3
+\left(-\frac{28463}{96}+\frac{205}{16}\pi^2\right)\nu^4
+\frac{73}{16}\nu^5$\\ 
$q_{7\rm EG}^0$& $\left(-\frac{5556443}{524288}\pi^4-\frac{37569023551}{4762800}+\frac{1163042866561}{990904320}\pi^2\right)\nu 
+\left(\frac{83453959}{29400}-\frac{675041}{8960}\pi^2+a_7^{\nu^2}+\frac{25}{6}a_6^{\nu^2}+\frac{1}{14}q_{45}^{\nu^2}+\frac{1}{6}\bar d_6^{\nu^2}-\frac{1}{30}\bar d_5^{\nu^2}\right)\nu^2$\\
&$
+\left(a_7^{\nu^3}+\frac{373067}{12288}\pi^2-\frac{1}{2}\bar d_5^{\nu^2}-\frac{5}{2}a_6^{\nu^2}-\frac{222749}{144}\right)\nu^3
+\left(-\frac{3021}{32}+\frac{615}{128}\pi^2\right)\nu^4
+\frac{13}{16}\nu^5$\\
\end{tabular}
\end{ruledtabular}
\end{table*}


\begin{table*}
\caption{\label{EEG_coeffs} List of the $\pinf^2$-expansion coefficients (similarly to Eq.~\eq{PNqHn}) of the E-type energy gauge (squared) effective  Hamiltonian, Eq. \eqref{QEGE}, for $n\ge 4$.}
\begin{ruledtabular}
\begin{tabular}{l|l}
\hline
$ q_{4E,\rm EG}^0$ &$   \left(\frac{175}{3}-\frac{41}{32}\pi^2\right)\nu-\frac{7}{2}\nu^2 $\\
$ q_{4E,\rm EG}^1$ &$   \left(\frac{5632}{45}-\frac{33601}{6144}\pi^2\right)\nu+\left(-\frac{441}{4}+\frac{123}{64}\pi^2\right)\nu^2+\frac{13}{2}\nu^3 $\\ 
$q_{4E,\rm EG}^2 $ &$  \left(\frac{699761}{7200}-\frac{93031}{12288}\pi^2\right)\nu+\left(-\frac{90403}{480}+\frac{31633}{4096}\pi^2\right)\nu^2+\left(-\frac{615}{256}\pi^2+\frac{601}{4}\right)\nu^3-\frac{293}{32}\nu^4 $\\ 
$q_{4E,\rm EG}^3 $ &$   \left(-\frac{9733841}{4194304}\pi^2-\frac{40415563}{1411200}\right)\nu+\left(\frac{156397}{16384}\pi^2-\frac{7498919}{67200}\right)\nu^2+\left(\frac{102833}{480}-\frac{148325}{16384}\pi^2\right)\nu^3+\left(\frac{1435}{512}\pi^2-\frac{8789}{48}\right)\nu^4+\frac{185}{16}\nu^5 $\\ 
$q_{5E,\rm EG}^0 $ &$  \left(-\frac{29917}{6144}\pi^2+\frac{44357}{360}\right)\nu+\left(\frac{205}{64}\pi^2-\frac{2747}{24}\right)\nu^2+\frac{9}{4}\nu^3 $\\ 
$q_{5E,\rm EG}^1 $ &$   \left(\frac{15540691}{25200}-\frac{2590847}{61440}\pi^2\right)\nu+\left(-\frac{29501}{80}+\frac{347673}{20480}\pi^2+\frac{1}{5}\bar d_5^{\nu^2}\right)\nu^2+\left(-\frac{1763}{256}\pi^2+\frac{6499}{24}\right)\nu^3-\frac{93}{16}\nu^4 $\\ 
$q_{5E,\rm EG}^2 $ &$   \left(-\frac{807638471}{1764000}+\frac{30033990443}{440401920}\pi^2\right)\nu+\left(\frac{3}{35}q_{45}^{\nu^2}-\frac{472730717}{588000}+\frac{6466655}{114688}\pi^2\right)\nu^2+\left(-\frac{7356287}{245760}\pi^2-\frac{2}{5}\bar d_5^{\nu^2}+\frac{965323}{1440}\right)\nu^3$\\
&$+\left(\frac{5535}{512}\pi^2-\frac{57457}{128}\right)\nu^4+\frac{657}{64}\nu^5 $\\ 
$ q_{6E,\rm EG}^0$ &$   \left(-\frac{69733}{350}+\frac{541363}{10240}\pi^2\right)\nu+\left(a_6^{\nu^2}-\frac{5083}{60}+\frac{37537}{5120}\pi^2+\frac{1}{5}\bar d_5^{\nu^2}\right)\nu^2+\left(\frac{1775}{12}-\frac{287}{64}\pi^2\right)\nu^3-\frac{11}{8}\nu^4 $\\ 
$ q_{6E,\rm EG}^1$ &$   \left(\frac{100876235443}{165150720}\pi^2-\frac{195178823647}{47628000}+\frac{45303}{524288}\pi^4\right)\nu+\left(\frac16 \bar d_5^{\nu^2}-\frac{109708309}{196000}+\frac{11}{70}q_{45}^{\nu^2}+\frac{7}{6}a_6^{\nu^2}+\frac{1754467}{28672}\pi^2+\frac16 \bar d_6^{\nu^2}\right)\nu^2$\\
&$+\left(-\frac{963359}{30720}\pi^2+\frac{186533}{360}-\frac{9}{10}\bar d_5^{\nu^2}-\frac52 a_6^{\nu^2}\right)\nu^3+\left(\frac{205}{16}\pi^2-\frac{42773}{96}\right)\nu^4+\frac{73}{16}\nu^5 $\\
$q_{7E,\rm EG}^0$  & $ \left(-\frac{5556443}{524288}\pi^4-\frac{37569023551}{4762800}+\frac{1163042866561}{990904320}\pi^2\right)\nu
+\left(\frac16 \bar d_6^{\nu^2}+\frac{39167441}{44100}+\frac{25}{6}a_6^{\nu^2}-\frac{1}{30}\bar d_5^{\nu^2}+\frac{1}{14}q_{45}^{\nu^2}-\frac{348401}{53760}\pi^2
+a_7^{\nu^2}\right)\nu^2$\\
&$+\left(a_7^{\nu^3}-\frac{49799}{144}+\frac{46379}{12288}\pi^2-\frac12 \bar d_5^{\nu^2}-\frac52 a_6^{\nu^2}\right)\nu^3+\left(\frac{615}{128}\pi^2-\frac{4921}{32}\right)\nu^4+\frac{13}{16}\nu^5 $\\
\end{tabular}
\end{ruledtabular}
\end{table*}


\begin{table*}
\caption{\label{table_Hreal} Coefficients entering the 6PN real EOB Hamiltonian \eq{Heobfin}. 
}
\begin{ruledtabular}
\begin{tabular}{lll}
Coefficient & Powers & Value \\
\hline
$C^{(0)}_{14}(\nu)$ & $p_r^0 p^{14} u^0$ & $\frac{33}{2048}\nu+\frac{33}{2048}\nu^2+\frac{45}{2048}\nu^3+\frac{15}{512}\nu^4+\frac{35}{1024}\nu^5+\frac{63}{2048}\nu^6+\frac{33}{2048}\nu^7$\\
$C^{(0)}_{12}(\nu)$ & $p_r^0 p^{12} u^1$ & $\frac{21}{1024}\nu+\frac{21}{1024}\nu^2+\frac{21}{1024}\nu^3-\frac{35}{512}\nu^5-\frac{189}{1024}\nu^6-\frac{231}{1024}\nu^7$\\
$C^{(0)}_{10}(\nu)$ & $p_r^0 p^{10} u^2$ & $-\frac{7}{512}\nu-\frac{7}{512}\nu^2-\frac{15}{512}\nu^3-\frac{5}{64}\nu^4-\frac{35}{256}\nu^5+\frac{63}{512}\nu^6+\frac{693}{512}\nu^7$\\ 
$C^{(0)}_{8}(\nu)$  & $p_r^0 p^{8} u^3$ & $\frac{5}{256}\nu-\frac{5}{256}\nu^2-\frac{1}{256}\nu^3+\frac{7}{128}\nu^4+\frac{45}{128}\nu^5+\frac{385}{256}\nu^6-\frac{1155}{256}\nu^7$\\
$C^{(0)}_{6}(\nu)$  & $p_r^0 p^{6} u^4$ & $-\frac{5}{128}\nu+\left(\frac{385}{384}-\frac{41}{1024}\pi^2\right)\nu^2
+\left(\frac{373}{384}-\frac{41}{1024}\pi^2\right)\nu^3+\frac{1}{32}\nu^4+\left(-\frac{925}{192}+\frac{205}{1024}\pi^2\right)\nu^5-\frac{595}{128}\nu^6+\frac{1155}{128}\nu^7$\\
$C^{(0)}_{4}(\nu)$  & $p_r^0 p^{4} u^5$ & $\frac{7}{64}\nu+\left(\frac{1141}{480}-\frac{1619}{8192}\pi^2\right)\nu^2+\left(\frac{1153}{240}-\frac{2275}{8192}\pi^2\right)\nu^3+\left(\frac{10105}{8192}\pi^2-\frac{21721}{960}\right)\nu^4+\left(-\frac{123}{128}\pi^2+\frac{347}{16}\right)\nu^5$\\
&&$+\frac{357}{64}\nu^6-\frac{693}{64}\nu^7$ \\
$C^{(0)}_{2}(\nu)$  & $p_r^0 p^{2} u^6$ &$-\frac{21}{32}\nu+\left(\frac{254113}{12288}\pi^2-\frac{8478053}{50400}\right)\nu^2+\left(\frac{7466063}{50400}-\frac{234685}{12288}\pi^2+\frac14 a_6^{\nu^2}\right)\nu^3+\left(\frac{10861}{240}-\frac14 a_6^{\nu^2}-\frac{6169}{2048}\pi^2\right)\nu^4$\\
&&$+\left(-\frac{507}{16}+\frac{369}{256}\pi^2\right)\nu^5-\frac{91}{32}\nu^6+\frac{231}{32}\nu^7$ \\
$C^{(0)}_{0}(\nu)$  & $p_r^0 p^{0} u^7$ & $-\frac{33}{16}\nu+\left(-\frac{2081602903}{1814400}-\frac{2800873}{524288}\pi^4+\frac{756731519}{3538944}\pi^2\right)\nu^2+\left(\frac{1}{2} a_6^{\nu^2}+\frac{133421}{3072}\pi^2+\frac12 a_7^{\nu^2}-\frac{4823719}{12600}\right)\nu^3$\\
&&$+\left(\frac12 a_7^{\nu^3}+\frac12 a_6^{\nu^2}+\frac{6169}{2048}\pi^2-\frac{11251}{240}\right)\nu^4
+\left(\frac{167}{12}-\frac{41}{64}\pi^2\right)\nu^5+\frac{7}{16}\nu^6-\frac{33}{16}\nu^7$\\
\hline
$C^{(2)}_{10}(\nu)$ & $p_r^2 p^{10} u^1$ &$\frac{63}{256}\nu+\frac{63}{256}\nu^2+\frac{21}{64}\nu^3+\frac{105}{256}\nu^4+\frac{105}{256}\nu^5+\frac{63}{256}\nu^6$\\
$C^{(2)}_{8}(\nu)$  & $p_r^2 p^{8} u^2$ &$\frac{35}{128}\nu+\frac{35}{32}\nu^2+\frac{135}{128}\nu^3+\frac{55}{64}\nu^4-\frac{25}{128}\nu^5-\frac{105}{64}\nu^6$\\
$C^{(2)}_{6}(\nu)$  & $p_r^2 p^{6} u^3$ &$-\frac{5}{32}\nu-\frac{185}{32}\nu^2-\frac{81}{16}\nu^3-\frac{239}{32}\nu^4-\frac{269}{32}\nu^5+\frac{135}{32}\nu^6$\\
$C^{(2)}_{4}(\nu)$  & $p_r^2 p^{4} u^4$ &$\frac{3}{16}\nu+\left(\frac{611}{48}-\frac{25729}{8192}\pi^2\right)\nu^2
+\left(-\frac{13921}{8192}\pi^2-\frac{779}{24}\right)\nu^3+\left(-\frac{13921}{8192}\pi^2-\frac{1189}{48}\right)\nu^4+\left(-\frac{69}{16}+\frac{369}{256}\pi^2\right)\nu^5-\frac{45}{8}\nu^6$\\
$C^{(2)}_{2}(\nu)$  & $p_r^2 p^{2} u^5$ &$-\frac{5}{16}\nu+\left(\frac{36359}{2048}\pi^2-\frac{447313}{1400}\right)\nu^2+\left(-\frac{4267103}{8400}-\frac14 \bar d_5^{\nu^2}+\frac{47511}{2048}\pi^2\right)\nu^3+\left(\frac{45409}{2048}\pi^2-\frac14 \bar d_5^{\nu^2}-\frac{16871}{48}\right)\nu^4$\\
&&$+\left(-\frac{41}{16}\pi^2+\frac{767}{48}\right)\nu^5+\frac{75}{16}\nu^6$\\
$C^{(2)}_{0}(\nu)$  & $p_r^2 p^{0} u^6$ &$\frac{7}{8}\nu+\left(-\frac{137379058337}{12700800}+\frac{271118011}{196608}\pi^2+\frac{135909}{524288}\pi^4\right)\nu^2+\left(\frac{8602849}{12600}+\frac12 a_6^{\nu^2}-\frac32 \bar d_5^{\nu^2}-\frac{238105}{6144}\pi^2+\frac12 \bar d_6^{\nu^2}\right)\nu^3$\\
&&$+\left(-\frac{15}{2} a_6^{\nu^2}-\frac{99253}{6144}\pi^2+\frac{18713}{144}\right)\nu^4+\left(\frac{517}{24}-\frac{41}{64}\pi^2\right)\nu^5-\frac{9}{4}\nu^6$\\
\hline 
$C^{(4)}_{6}(\nu)$  & $p_r^4 p^{6} u^2$ &$\frac{35}{32}\nu-\frac{5}{32}\nu^2+\frac{35}{32}\nu^3+\nu^4+\frac{31}{32}\nu^5+\frac{15}{16}\nu^6$\\
$C^{(4)}_{4}(\nu)$  & $p_r^4 p^{4} u^3$ &$\frac{15}{16}\nu+\frac{141}{16}\nu^2-6\nu^3-\frac{63}{16}\nu^4-\frac{57}{8}\nu^5-\frac{15}{4}\nu^6$\\
$C^{(4)}_{2}(\nu)$  & $p_r^4 p^{2} u^4$ &$-\frac{3}{8}\nu+\left(\frac{93031}{6144}\pi^2-\frac{924983}{6300}\right)\nu^2+\left(\frac{23773}{1575}-\frac{467}{1536}\pi^2\right)\nu^3+\left(-\frac{469}{12}-\frac{21793}{2048}\pi^2\right)\nu^4+\left(\frac{615}{128}\pi^2-\frac{903}{8}\right)\nu^5+\frac{15}{4}\nu^6$\\
$C^{(4)}_{0}(\nu)$  & $p_r^4 p^{0} u^5$ &$\frac{1}{4}\nu+\left(-\frac{18410298107}{4233600}+\frac{83902033}{131072}\pi^2\right)\nu^2+\left(\frac{2968391}{6300}+\frac12 q_{45}^{\nu^2}-\frac{201851}{3072}\pi^2\right)\nu^3+\left(-\frac{799151}{432}-\frac{7}{3}\bar d_5^{\nu^2}+\frac{555389}{9216}\pi^2\right)\nu^4$\\
&&$+\left(\frac{205}{16}\pi^2-\frac{4153}{12}\right)\nu^5$\\
\hline
$C^{(6)}_{4}(\nu)$  & $p_r^6 p^{4} u^2$ &$-\frac{27}{80}\nu^2-\frac{27}{20}\nu^3-\frac{9}{40}\nu^4+\frac{9}{80}\nu^5+\frac{9}{8}\nu^6$\\
$C^{(6)}_{2}(\nu)$  & $p_r^6 p^{2} u^3$ &$\frac{5}{4}\nu-\frac{331}{40}\nu^2-\frac{57}{40}\nu^3-\frac{243}{10}\nu^4-\frac{309}{20}\nu^5-\nu^6$\\
$C^{(6)}_{0}(\nu)$  & $p_r^6 p^{0} u^4$ &$\frac{1}{2}\nu+\left(-\frac{109101883}{588000}-\frac{9733841}{655360}\pi^2\right)\nu^2+\left(\frac{156397}{2560}\pi^2-\frac{23052881}{42000}\right)\nu^3+\left(-\frac{29665}{512}\pi^2+\frac{29551}{60}\right)\nu^4$\\
&&$+\left(\frac{287}{16}\pi^2-\frac{32533}{60}\right)\nu^5-\frac{5}{2}\nu^6$\\
\hline
$C^{(8)}_{2}(\nu)$  & $p_r^8 p^{2} u^2$ &$-\frac{3}{14}\nu^2-\frac{6}{7}\nu^3-\frac{3}{2}\nu^4+\frac{9}{14}\nu^5+\frac{3}{2}\nu^6$\\
$C^{(8)}_{0}(\nu)$  & $p_r^8 p^{0} u^3$ &$-\frac{1787}{224}\nu^2-\frac{7311}{560}\nu^3-\frac{837}{140}\nu^4-\frac{921}{14}\nu^5+6\nu^6$\\
\hline
$C^{(10)}_{0}(\nu)$  & $p_r^{10} p^{0} u^2$ &$-\frac{11}{42}\nu^2-\frac{11}{14}\nu^3-\frac{10}{7}\nu^4-\frac{5}{6}\nu^5+3\nu^6$\\
\end{tabular}
\end{ruledtabular}
\end{table*}

\section{Values of the  6PN-accurate $f$- route local scattering angle at PM orders $G^3$, $G^4$, $G^5$ and $G^6$} \label{scatteringangle}

Having determined most of the coefficients parametrizing the f-route local Hamiltonian we can write down the (PN-expanded) values
of  the corresponding successive $n$-PM contributions, $\chi_n$, to the scattering angle.  
The results are more compactly expressed when writing them in terms of the difference between the
energy-rescaled angle \eq{tildechin} and the corresponding test-mass (Schwarzschild value).

Let us first recall that the {\it exact} values of the Schwarzschild scattering angle coefficients are
\begin{eqnarray}
\chi_1^{\rm Schw}(p_\infty)&=& \frac{1}{p_\infty}+2 p_\infty 
\,,\nonumber\\
\chi_2^{\rm Schw}(p_\infty)&=& \pi \left(\frac32 + \frac{15}{8}  p_\infty^2 \right)
\,,\nonumber\\
\chi_3^{\rm Schw}(p_\infty)&=& -\frac{1}{ 3 p_\infty^3}+\frac{4}{p_\infty} +24 p_\infty + \frac{64}{3}  p_\infty^3
\,,\nonumber\\ 
\chi_4^{\rm Schw}(p_\infty)&=& \pi \left(\frac{105}{8}+  \frac{315}{8} p_\infty^2+\frac{3465}{128} p_\infty^4\right)
\,,\nonumber\\
\chi_5^{\rm Schw}(p_\infty)&=&\frac{1}{5 p_\infty^5}-\frac{2}{p_\infty^3}+\frac{ 32 }{p_\infty} +320p_\infty\nonumber\\
&& + 640 p_\infty^3 
+\frac{1792}{5} p_\infty^5
\,,\nonumber\\
\chi_6^{\rm Schw}(p_\infty)&=& \pi\left(
\frac{1155}{8}+\frac{45045}{64} p_\infty^2  + \frac{135135}{128} p_\infty^4\right.\nonumber\\
&&\left.
+\frac{255255}{512}p_\infty^6\right)
\,,\nonumber\\
\chi_7^{\rm Schw}(p_\infty)&=&
-\frac{1}{7 p_\infty^7}+\frac{8}{5 p_\infty^5}  -  \frac{16}{p_\infty^3}  + \frac{320}{p_\infty} +4480 p_\infty\nonumber\\
&& +14336 p_\infty^3 +\frac{86016}{5}p_\infty^5 + \frac{49152}{7}p_\infty^7\,.\nonumber\\
\end{eqnarray}
 We then find that the differences $\widetilde \chi_n^{\rm loc, f} - \chi_n^{\rm Schw}$ (recalling the definition Eq. \eq{tildechin}) read
\begin{widetext}
\begin{eqnarray} \label{chi2-chi7}
\pi^{-1}\left(\widetilde \chi_2^{\rm loc}-\chi_2^{\rm Schw}\right)&=& 0
\,,\nonumber\\
\nu^{-1}\left(\widetilde \chi_3^{\rm loc}-\chi_3^{\rm Schw}\right)&=&  -\frac{1}{3 p_\infty}\eta^2-\frac{47}{12}\eta^4 p_\infty-\frac{313}{24}\eta^6 p_\infty^3-\frac{749}{320} p_\infty^5\eta^8-\frac{7519}{4480}\eta^{10} p_\infty^7+\frac{211469}{161280}\eta^{12} p_\infty^9 
\,,\nonumber\\
\nu^{-1}\pi^{-1}\left(\widetilde \chi_4^{\rm loc, f}-\chi_4^{\rm Schw}\right)&=& 
-\frac{15}{4}\eta^4+\left(-\frac{557}{16}+\frac{123}{256}\pi^2\right)\eta^6 p_\infty^2
+\left(-\frac{4601}{96}+\frac{33601}{16384}\pi^2\right)\eta^8 p_\infty^4\nonumber\\
&&
+\left(-\frac{3978707}{134400}+\frac{93031}{32768}\pi^2\right)\eta^{10}p_\infty^6
+\left(\frac{29201523}{33554432}\pi^2+\frac{5058313}{627200}\right)\eta^{12}p_\infty^8
\,,\nonumber\\
\nu^{-1}\left(\widetilde \chi_5^{\rm loc,f}-\chi_5^{\rm Schw}\right)&=& 
\frac{2}{5p_\infty^3 }\eta^2
+\left(-\frac{121}{10} 
+\frac15\nu\right)\frac{\eta^4}{p_\infty}
+\left(-\frac{19457}{60}+\frac{59}{10}\nu+\frac{41}{8}\pi^2\right)\eta^6 p_\infty\nonumber\\
&&+\left(-\frac{41}{24}\nu\pi^2+\frac{10681}{144}\nu+\frac{5069}{144}\pi^2-\frac{4572503}{4320}\right)\eta^8 p_\infty^3\nonumber\\
&&+\left(\frac{111049}{960}\pi^2+\frac{23407}{5760}\nu\pi^2-\frac{4}{15}\nu \bar d_5^{\nu^2}  -\frac{55558621}{33600}-\frac{573577}{4320}\nu\right)\eta^{10}p_\infty^5\nonumber\\
&&
+\left(-\frac{4}{35}\nu q_{45}^{\nu^2}  -\frac{184881}{4480}\pi^2+\frac{1219303}{20160}\nu\pi^2-\frac{16844006729}{21168000}\nu+\frac{15827493497}{42336000}\right)\eta^{12} p_\infty^7
\,,\nonumber\\
\nu^{-1}\pi^{-1}\left(\widetilde \chi_6^{\rm loc, f}-\chi_6^{\rm Schw}\right)&=& 
\left(-\frac{625}{4}+\frac{105}{16}\nu+\frac{615}{256}\pi^2\right)\eta^6  
+\left(-\frac{1845}{512}\nu\pi^2+\frac{257195}{8192}\pi^2+\frac{10065}{64}\nu-\frac{224113}{192}\right)\eta^8 p_\infty^2\nonumber\\
&&
+\left(-\frac{15}{32}\nu a_6^{\nu^2} -\frac{15}{32}\nu \bar d_5^{\nu^2} -\frac{61855}{32768}\nu\pi^2+\frac{2321185}{16384}\pi^2+\frac{4625}{192}\nu-\frac{20420849}{6720}\right)\eta^{10}p_\infty^4\nonumber\\
&&+
\left(-\frac{35}{64}\nu \bar d_5^{\nu^2} -\frac{5}{64}\nu \bar d_6^{\nu^2} -\frac{4911465305}{25165824}\pi^2-\frac{11437991}{8960}\nu-\frac{35}{64}\nu a_6^{\nu^2} -\frac{15}{64}\nu q_{45}^{\nu^2}\right. \nonumber\\
&& \left. +\frac{2363865}{65536}\nu\pi^2
-\frac{679545}{16777216}\pi^4+\frac{1343882527}{10160640}\right)\eta^{12}p_\infty^6
\,,\nonumber\\
\nu^{-1}\left(\widetilde \chi_7^{\rm loc,f}-\chi_7^{\rm Schw}\right)&=& 
-\frac3{7 p_\infty^5}\eta^2+\left(\frac{227}{28}-\frac37 \nu\right)\frac{\eta^4}{p_\infty^3}
+\left(-\frac{60377}{168}+\frac{339}{14}\nu+\frac{41}{8}\pi^2-\frac17 \nu^2\right)\frac{\eta^6}{p_\infty}\nonumber\\
&&+\left(\frac{33131}{192}\pi^2-\frac{221}{28}\nu^2+\frac{158129}{112}\nu-\frac{152237341}{20160}-\frac{123}{4}\nu\pi^2\right)\eta^8p_\infty\nonumber\\
&&+\left(\frac{378953}{384}\pi^2-\frac{18343}{128}\nu\pi^2+\frac{41}{8}\nu^2\pi^2-\frac{64315}{336}\nu^2-8\nu a_6^{\nu^2} -4\nu \bar d_5^{\nu^2}\right. \nonumber\\
&&\left. +\frac{2208701}{480}\nu-\frac{6769922309}{201600}\right)\eta^{10}p_\infty^3\nonumber\\
&&+
\left(\frac45 \bar d_5^{\nu^2}\nu^2-\frac{284141687}{69120}\pi^2
-\frac{3474679}{3840}\nu\pi^2-\frac{64}{5}\nu \bar d_5^{\nu^2} -\frac{196222844821}{16934400}-\frac{1811850763}{1008000}\nu\right. \nonumber\\
&&+\frac{596213}{720}\nu^2-\frac{128}{5}\nu a_6^{\nu^2} -\frac85 \nu a_7^{\nu^2} -\frac85 \nu^2a_7^{\nu^3} -\frac85 \nu \bar d_6^{\nu^2}\nonumber\\
&&\left. -\frac{12}{5}\nu q_{45}^{\nu^2} -\frac{11471}{384}\nu^2\pi^2+\frac{666241}{40960}\pi^4\right)\eta^{12}p_\infty^5\,.
\end{eqnarray}
\end{widetext}

These results for the  scattering angle provide a lot of new information that offers gauge-invariant checks for future 
independent computations of the dynamics of binary systems. 

In particular, using the fact (explicitly proven in Ref.~\cite{Bini:2017wfr}) that the nonlocal dynamics starts contributing to the
scattering angle only at $O(G^4)$, so that $\chi_3^{\rm tot}= \chi_3^{\rm loc}$,
our result above for $\chi_3^{\rm loc}$ actually describes the total 3PM-level scattering angle. The corresponding
 explicit 6PN-accurate expression of the unrescaled, and unsubtracted 3PM-level scattering angle (which is
 equivalent to the simpler rescaled, subtracted result above) reads
\begin{eqnarray} \label{chi36pn}
 \chi_3
&=&
-\frac{1}{3p_\infty^3}+\frac{4}{p_\infty} +(24-8\nu)p_\infty \nonumber\\
&&+\left(\frac{64}{3}-36\nu+8\nu^2\right)p_\infty^3\nonumber\\
&&+\left(-\frac{91}{5}\nu+34\nu^2-8\nu^3\right)p_\infty^5\nonumber\\
&&+\left(\frac{69}{70}\nu+\frac{51}{5}\nu^2-32\nu^3+8\nu^4\right)p_\infty^7\nonumber\\
&& +\left(\frac{1447}{5040}\nu-\frac{93}{56}\nu^2-\frac{27}{10}\nu^3+30\nu^4-8\nu^5\right)p_\infty^9\nonumber\\
&&+O(p_\infty^{11})\,.
\end{eqnarray}
 This  result is in agreement with the corresponding 6PN-level term in the PN expansion
of the 3PM-level  recent result of ~\cite{Bern:2019nnu,Bern:2019crd}. It has also been recently obtained in 
Refs. \cite{Blumlein:2020znm,Cheung:2020gyp}.

Let us emphasize that our results also provide a complete,  6PN-accurate value for the 4PM-level scattering angle 
$\chi_4= \chi_4^{\rm loc, f} + \chi_4^{\rm nonloc, f}$. We will discuss separately the 6PN-accurate nonlocal contribution
$\chi_4^{\rm nonloc, f}$. Let us, for completeness, exhibit the unrescaled, unsubtracted value of $\chi_4^{\rm loc, f}$.
It reads
\begin{widetext}
\begin{eqnarray} \label{chi4xpl}
\pi^{-1}\chi_4^{\rm  loc, f} &=& \left(\frac{105}{8}-\frac{15}{4}\nu\right) \nonumber\\
&+&\left[\frac{315}{8}+\left(-\frac{109}{2}+\frac{123}{256}\pi^2\right)\nu+\frac{45}{8}\nu^2\right] p_\infty^2 \nonumber\\
&+&\left[\frac{3465}{128}+\left(\frac{33601}{16384}\pi^2-\frac{19597}{192}\right)\nu+\left(\frac{4827}{64}-\frac{369}{512}\pi^2\right)\nu^2-\frac{225}{32}\nu^3\right] p_\infty^4 \nonumber\\
&+&\left[\left(-\frac{1945583}{33600}+\frac{93031}{32768}\pi^2\right)\nu+\left(\frac{1937}{16}-\frac{94899}{32768}\pi^2\right)\nu^2+\left(-\frac{2895}{32}+\frac{1845}{2048}\pi^2\right)\nu^3+\frac{525}{64}\nu^4\right] p_\infty^6 \nonumber\\
&+&\left[\left(\frac{3879719}{313600}+\frac{29201523}{33554432}\pi^2\right)\nu+\left(\frac{4843207}{89600}-\frac{469191}{131072}\pi^2\right)\nu^2
+\left(\frac{444975}{131072}\pi^2-\frac{15875}{128}\right)\nu^3\right. \nonumber\\
&+&\left.\left(\frac{104755}{1024}-\frac{4305}{4096}\pi^2\right)\nu^4
-\frac{4725}{512}\nu^5\right] p_\infty^8+O(p_\infty^{10})\,.
\end{eqnarray}
\end{widetext}
Finally, concerning our results above for the 5PM, 6PM and 7PM local scattering angles,  
 if we transcribe them in terms of the unrescaled coefficients,
$\chi_5^{\rm loc, f}$, $\chi_6^{\rm loc, f}$, $\chi_7^{\rm loc, f}$, they contain (in spite of the presence of undetermined parameters at the $O(\nu^2)$ level) a lot of new information, both for the linear-in-$\nu$ contributions, and for many terms involving higher powers of $\nu$. 

\section{Radial action and its hidden structure}

In Ref. \cite{Bini2020} we pointed out the existence of a hidden simplicity in the mass-ratio-dependence
of the (rescaled) radial action
\beq \label{Ir}
I_r(\g,j) = \frac1{2\pi} \oint p_r dr\,,
\eeq
when it is expressed in terms of the EOB effective energy $\g= \e$ (or equivalently $p_{\infty}$) and of the rescaled
angular momentum $j=J/(GM\mu)$.
We work here with dimensionless scaled variables $I_r = I_r ^{\rm phys}/(GM\mu)$, $p_r = p_r ^{\rm phys}/\mu$, 
$r = r ^{\rm phys}/GM$.

This hidden simplicity consists in noting the remarkably simple $\nu$-dependence of the coefficients
$I_{n}(\g; \nu)$ entering the following way of writing the 6PN-accurate expression for $I_r$:
\begin{eqnarray} \label{Irxp}
I_r^{\rm loc,f}(\g,j) &=& -j+I^S_0(\g) +\frac{I_1^S(\g)}{hj}+\frac{I_3(\g; \nu)}{(hj)^3}\nonumber\\
&&+ \frac{I_5(\g; \nu)}{(hj)^5}+\frac{I_7(\g; \nu)}{(hj)^7}\nonumber\\
&&+\frac{I_9(\g; \nu)}{(hj)^9}+\frac{I_{11}(\g; \nu)}{(hj)^{11}}\,.
\end{eqnarray}
First, the second term $I_0(\g) $  in this expression is independent of $\nu$ and equal to 
the analytic continuation (in $\g$) of $\chi_1$ \cite{Kalin:2019inp}
\beq \label{IS0}
I^S_0(\gamma) = \frac{ 2\gamma^2-1 }{\sqrt{1-\gamma^2}}\,,
\eeq
and, second, and most importantly, after having factored out the same power of $\frac1{h}$ as the power of $\frac1j$,
the numerator $I_{2n+1}(\g; \nu)$ is a {\it polynomial} in $\nu$ of degree $n$:
\beq \label{Cn3}
I_{2n+1}(\g; \nu)=  I^S_{2n+1}(\g) +\sum_{k=1}^{n}  I_{2n+1}^{\nu^k}(\g) \nu^k\,.
\eeq
The latter polynomial structure was not pointed out in previous discussions \cite{Damour:1988mr,Kalin:2019inp} of the radial action. 
Several conditions are needed to reveal it: the use of the effective EOB energy $\e$ as energy variable, and a PN-complete account 
of each coefficient $I_{2n+1}(\e; \nu)$. We note in this respect that Eq. (3.10) of Ref. \cite{Damour:1988mr}  used the  specific binding
energy $ (H-Mc^2)/\mu$  as energy variable, and that Eq. (4.29) of Ref. \cite{Kalin:2019inp} is a 
{\it PN-incomplete} 2PM truncation of $I_r$, which does not satisfy the simple rule \eq{Cn3}.

As pointed out (and proven) in our previous work \cite{Bini2020}, the $\nu^0$ terms (corresponding to the $\nu \to 0$ limit) in Eq. \eq{Cn3} can be exactly computed (for all
values of $n$) because they correspond (like the term $I^S_0(\gamma) $) to the test-mass dynamics,
described by a Schwarzschild metric of mass $M=m_1+m_2$. The {\it exact} values of the
most 6PN-relevant $\nu^0$,  Schwarzschildlike, terms read
\bea
\label{Irxp2}
I_1^S(\gamma) &=& -\frac{3}{4}+\frac{15}{4}\gamma^2, \nonumber\\
I_3^S(\gamma) &=& \frac{35}{64}-\frac{315}{32}\gamma^2+\frac{1155}{64}\gamma^4, \nonumber\\
I_5^S(\gamma) &=&  -\frac{231}{256}+\frac{9009}{256}\gamma^2-\frac{45045}{256}\gamma^4+\frac{51051}{256}\gamma^6 \nonumber\\
I_7^S(\gamma) &=&  \frac{32175}{16384}-\frac{546975}{4096}\gamma^2+\frac{10392525}{8192}\gamma^4\nonumber\\
&&-\frac{14549535}{4096}\gamma^6+\frac{47805615}{16384}\gamma^8\nonumber\\
I_9^S(\gamma) &=& -\frac{323323}{65536}+\frac{33948915}{65536}\gamma^2-\frac{260275015}{32768}\gamma^4\nonumber\\
&&+\frac{1301375075}{32768}\gamma^6-\frac{5019589575}{65536}\gamma^8\nonumber\\
&&+\frac{3234846615}{65536}\gamma^{10}\,.\nonumber\\ 
\eea
Let us only cite the $\g\to1$ value of the last Schwarzschildlike coefficient entering Eq. \eq{Irxp} 
(which suffices at the 6PN accuracy)
\beq
I_{11}^S(\gamma) =\frac{14196819}{256}+ O(\g^2-1)\,.
\eeq

The most useful consequence of the expression \eq{Irxp} for the radial action is that it condenses the {\it irreducible} (post-test-mass)
information about the 6PN local dynamics in a rather small number of coefficients, namely the fifteen  energy-dependent coefficients
$ I_{2n+1}^{\nu^k}(\g)$, with $1\leq k \leq n$ and $1\leq n \leq 5$.  Our 6PN-accurate computation yields
these coefficients in the form of a PN expansion, \ie, an expansion in powers of $p_\infty^2 \equiv \g^2-1$. [Note that the
so-defined quantity $p_\infty^2$ is negative for bound states.]
We found, for example,
\bea
I_3^{\nu^1}(\g) &=&  -\frac{5}{2}\eta^4+\left(\frac{41}{128}\pi^2-\frac{557}{24}\right)p_\infty^2\eta^6\nonumber\\
&+& \left(-\frac{4601}{144}+\frac{33601}{24576}\pi^2\right)p_\infty^4\eta^8\nonumber\\
&+&\left(-\frac{3978707}{201600}+\frac{93031}{49152}\pi^2\right)p_\infty^6\eta^{10} \nonumber\\
&+&
\left(\frac{9733841}{16777216}\pi^2+\frac{5058313}{940800}\right)p_\infty^8\eta^{12}   
\,.
\eea
The other $\nu$-dependent contributions can be read off Table \ref{various_I_n}, which lists the PN expansions
of the full coefficients $I_{2n+1}(\g; \nu)=  I^S_{2n+1}(\g) +\sum_{k=1}^{n}  I_{2n+1}^{\nu^k}(\g) \nu^k$.

We recall that the periastron-advance  parameter is derived from the radial action as follows \cite{Damour:1988mr}:
\beq
K \equiv 1+k \equiv \frac{\Phi}{2\pi}= - \partial_j I_r(\g,j)\,.
\eeq
Inserting the expression \eq{Irxp} in the latter formula yields
\begin{eqnarray} \label{kxp}
k(\g,j)&=&   \frac{I_1^S(\g)}{hj^2}+3\frac{I_3(\g; \nu)}{ h^3j^4}\nonumber\\
&+&5\frac{I_5(\g; \nu)}{h^5j^6}+7\frac{I_7(\g; \nu)}{h^7j^8}\nonumber\\
&+&9\frac{I_9(\g; \nu)}{h^9 j^{10}}+11\frac{I_{11}(\g; \nu)}{h^{11}j^{12}}\,,\nonumber\\
\end{eqnarray}
where the various coefficients $I_n(\g, \nu)$ are listed in Table \ref{various_I_n}.

\begin{table*}[h]
\caption{\label{various_I_n} List of the various coefficients $I_n(\g, \nu)$ (expressed in terms of $p_\infty^2 \equiv \g^2-1<0$)
entering the expression \eqref{Irxp} of the radial action.}
\begin{ruledtabular}
\begin{tabular}{l|l}
\hline
$I_0(p_\infty)$&$\frac1{\sqrt{-p_\infty^2}}(1+2\eta^2p_\infty^2)$\\
$I_1(p_\infty)$&$3\eta^2+\frac{15}{4}\eta^4p_\infty^2$\\
$I_3(p_\infty;\nu)$&$\frac{35}{4}\eta^4+\frac{105}{4}p_\infty^2\eta^6+\frac{1155}{64}p_\infty^4\eta^8$\\
&$+\left[-\frac{5}{2}\eta^4+\left(\frac{41}{128}\pi^2-\frac{557}{24}\right)p_\infty^2\eta^6+\left(-\frac{4601}{144}+\frac{33601}{24576}\pi^2\right)p_\infty^4\eta^8+\left(-\frac{3978707}{201600}+\frac{93031}{49152}\pi^2\right)p_\infty^6\eta^{10}\right.$\\
&$\left.
+\left(\frac{9733841}{16777216}\pi^2+\frac{5058313}{940800}\right)p_\infty^8\eta^{12}\right]\nu$\\
$I_5(p_\infty;\nu)$&$\frac{231}{4}\eta^6+\frac{9009}{32}p_\infty^2\eta^8+\frac{27027}{64}p_\infty^4\eta^{10}+\frac{51051}{256}p_\infty^6\eta^{12}$\\
&$+\left[\left(\frac{123}{128}\pi^2-\frac{125}{2}\right)\eta^6+\left(-\frac{224113}{480}+\frac{51439}{4096}\pi^2\right)p_\infty^2\eta^8+\left(\frac{464237}{8192}\pi^2-\frac{20420849}{16800}\right)p_\infty^4\eta^{10}\right.$\\
&$\left.
+\left(-\frac{135909}{8388608}\pi^4-\frac{982293061}{12582912}\pi^2+\frac{1343882527}{25401600}\right)p_\infty^6\eta^{12}\right]\nu$\\
&$
+\left[\frac{21}{8}\eta^6+\left(-\frac{369}{256}\pi^2+\frac{2013}{32}\right)p_\infty^2\eta^8+\left(\frac{925}{96}-\frac{3}{16}a_6^{\nu^2}-\frac{3}{16}{\bar d}_5^{\nu^2}-\frac{12371}{16384}\pi^2\right)p_\infty^4\eta^{10}\right.$\\
&$\left.
+\left(-\frac{7}{32}{\bar d}_5^{\nu^2}-\frac{7}{32}a_6^{\nu^2}-\frac{3}{32}q_{45}^{\nu^2}-\frac{1}{32}{\bar d}_6^{\nu^2}-\frac{11437991}{22400}+\frac{472773}{32768}\pi^2\right)p_\infty^6\eta^{12}\right]\nu^2$\\
$I_7(p_\infty;\nu)$&$\frac{32175}{64}\eta^8+\frac{109395}{32}p_\infty^2\eta^{10}+\frac{2078505}{256}p_\infty^4\eta^{12}$\\
&$
+\left[\left(\frac{425105}{24576}\pi^2-\frac{248057}{288}\right)\eta^8+\left(\frac{2310485}{16384}\pi^2-\frac{99111883}{13440}\right)p_\infty^2\eta^{10}\right.$\\
&$\left.
+\left(-\frac{109665759605}{75497472}\pi^2-\frac{28658940509}{3386880}+\frac{81987555}{8388608}\pi^4\right)p_\infty^4\eta^{12}\right]\nu$\\
&$
+\left[\left(-\frac{1025}{256}\pi^2+\frac{18925}{96}\right)\eta^8+\left(-\frac{1290275}{24576}\pi^2+\frac{1089349}{576}-\frac{15}{8}a_6^{\nu^2}-\frac{5}{8}{\bar d}_5^{\nu^2}\right)p_\infty^2\eta^{10}\right.$\\
&$\left.
+\left(-\frac{103473815}{196608}\pi^2-\frac{15}{32}q_{45}^{\nu^2}-\frac{405}{32}a_6^{\nu^2}-\frac{15}{16}a_7^{\nu^2}+\frac{832072211}{161280}-\frac{135}{32}{\bar d}_5^{\nu^2}-\frac{15}{32}{\bar d}_6^{\nu^2}+\frac{25215}{65536}\pi^4\right)p_\infty^4\eta^{12}\right]\nu^2$\\
&$
+\left[-\frac{45}{16}\eta^8+\left(-\frac{7595}{64}+\frac{3075}{1024}\pi^2\right)p_\infty^2\eta^{10}+\left(\frac{15}{32}{\bar d}_5^{\nu^2}-\frac{15}{16}a_7^{\nu^3}+\frac{5585}{32}-\frac{477255}{65536}\pi^2+\frac{15}{32}a_6^{\nu^2}\right)p_\infty^4\eta^{12}\right]\nu^3$\\
$I_9(p_\infty;\nu)$&$\frac{323323}{64}\eta^{10}+\frac{11316305}{256}p_\infty^2\eta^{12}$\\
&$
+\left[\left(\frac{121807}{1024}\pi^2-\frac{6817563}{640}\right)\eta^{10}+\left(-\frac{551913398477}{113246208}\pi^2+\frac{387365405}{8388608}\pi^4-\frac{23711330921}{345600}\right)p_\infty^2\eta^{12}\right]\nu$\\
&$
+\left[\left(-\frac{7}{16}{\bar d}_5^{\nu^2}+\frac{572999}{128}-\frac{35}{16}a_6^{\nu^2}-\frac{1755159}{16384}\pi^2\right)\eta^{10}\right.$\\
&$\left.
+\left(-\frac{385}{32}{\bar d}_5^{\nu^2}-\frac{21}{32}q_{45}^{\nu^2}-\frac{35}{8}a_7^{\nu^2}+\frac{176505}{32768}\pi^4-\frac{122488583}{49152}\pi^2-\frac{1925}{32}a_6^{\nu^2}-\frac{35}{32}{\bar d}_6^{\nu^2}+\frac{6532266163}{115200}\right)p_\infty^2\eta^{12}\right]\nu^2$\\
&$
+\left[\left(\frac{10045}{1024}\pi^2-\frac{42665}{96}\right)\eta^{10}+\left(\frac{105}{16}a_6^{\nu^2}+\frac{35}{16}{\bar d}_5^{\nu^2}+\frac{13076035}{98304}\pi^2-\frac{35}{8}a_7^{\nu^3}-\frac{11754113}{2304}\right)p_\infty^2\eta^{12}\right]\nu^3$\\
&$
+\left[\frac{385}{128}\eta^{10}+\left(\frac{291655}{1536}-\frac{10045}{2048}\pi^2\right)p_\infty^2\eta^{12}\right]\nu^4$\\
$I_{11}(p_\infty;\nu)$&$\frac{14196819}{256}\eta^{12}$\\
&$
+\left[\left(-\frac{3236467169}{30240}-\frac{188085303629}{50331648}\pi^2+\frac{350055909}{8388608}\pi^4\right)\eta^{12}\right]\nu$\\
&$
+\left[\left(-\frac{21}{32}{\bar d}_6^{\nu^2}+\frac{529515}{65536}\pi^4+\frac{2062272503}{22400}-\frac{1911}{32}a_6^{\nu^2}-\frac{9}{32}q_{45}^{\nu^2}-\frac{63}{16}a_7^{\nu^2}-\frac{179354853}{65536}\pi^2-\frac{273}{32}{\bar d}_5^{\nu^2}\right)\eta^{12}\right]\nu^2$\\
&$
+\left[\left(-\frac{63}{16}a_7^{\nu^3}+\frac{315}{32}a_6^{\nu^2}+\frac{24980025}{65536}\pi^2-\frac{978061}{64}+\frac{63}{32}{\bar d}_5^{\nu^2}\right)\eta^{12}\right]\nu^3$\\
&$
+\left[\left(-\frac{38745}{2048}\pi^2+\frac{428085}{512}\right)\eta^{12}\right]\nu^4$\\
&$
+\left[-\frac{819}{256}\eta^{12}\right]\nu^5$\\
\end{tabular}
\end{ruledtabular}
\end{table*}

Recently, Ref. \cite{Kalin:2019inp} pointed out that the periastron precession $\Phi(\gamma,j) - 2 \pi= 2\pi k(\g,j)$
could (under some conditions) be identified with  a suitably defined analytic
continuation of $\chi(\gamma,j)+ \chi(\gamma,-j)$. The $\nu$-structure of the formula  \eq{kxp} is then seen
to be a consequence of the rule, Eqs. \eq{Cn1}, \eq{Cn2}, found in Ref. \cite{Damour2019}, about the polynomial $\nu$-structure of the energy-rescaled
scattering angle $h^{n-1}\chi_n(\g,\nu)$. We then tried to replace the imposition of the constraint \eq{Cn2} by the imposition of the
polynomiality constraint \eq{Cn3} directly to the radial action [or, equivalently to the periastron precession $k(\g,j)$, Eq. \eq{kxp}].
However,   imposing the polynomiality constraints \eq{Cn3} or \eq{kxp} is not equivalent,
and, in fact, significantly {\it weaker} than  imposing  the conditions \eq{Cn2}.  Imposing the conditions
\eq{Cn3} or \eq{kxp} at the 6PN level leaves undetermined many more coefficients than imposing \eq{Cn2}.  This non equivalence
essentially follows from the fact that $I_{2n+1}(\g; \nu)$ is proportional to $\chi_{2n+2}(\gamma,\nu)$ and therefore misses
the $\nu$-information contained in the odd scattering-angle coefficients $\chi_{2n+1}(\gamma,\nu)$.

Finally, let us recall the well-known fact that the gauge-invariant relation between energy and angular momentum along
circular orbits can be conveniently obtained by setting $I_r=0$  in Eq. \eq{Irxp}. The resulting equation,
\bea
j &=& I^S_0(\g) +\frac{I_1^S(\g)}{hj}+\frac{I_3(\g; \nu)}{(hj)^3}\nonumber\\
&&+ \frac{I_5(\g; \nu)}{(hj)^5}+\frac{I_7(\g; \nu)}{(hj)^7}\nonumber\\
&&+\frac{I_9(\g; \nu)}{(hj)^9}+\frac{I_{11}(\g; \nu)}{(hj)^{11}}\,,
\eea
can then be easily perturbatively solved to get either $\frac1{j^2}$ as an expansion in powers of $\pinf^2$,
or  $\pinf^2 $ as an expansion in powers of $\frac1{j^2}$, say 
\bea
1-\gamma^2&=&-p_\infty^2=\frac{1}{j^2}+\frac{2}{j^4} +
  (9-2\nu)\frac{1}{j^6} \nonumber\\
&& +\left(54-\frac{154}{3}\nu+\frac{41}{32}\pi^2\nu\right)\frac{1}{j^8}   +\cdots\,,
\eea
or, equivalently,
\bea
\g\equiv \widehat E^{\rm loc,f, circ}_{\rm eff}&=& 1-\frac{1}{2j^2}-\frac{9}{8j^4} +\left(-\frac{81}{16}+\nu \right)\frac{1}{j^6}\nonumber\\
&&+\left(-\frac{3861}{128}+\frac{157}{6}\nu-\frac{41}{64}\pi^2\nu\right)\frac{1}{j^8}\nonumber\\
&& +\cdots\,.
\eea
The local contribution to the circular energy then straightforwardly follows:
\beq
E^{\rm loc,f, circ}(j)=M \sqrt{1+2\nu(\widehat E^{\rm loc,f, circ}_{\rm eff}-1)}\,.
\eeq
Here, we have simply indicated the 3PN-accurate beginning of these expansions. It is easy to use our results to derive the 
6PN-accurate local circular energy. We leave to future work the completion of these results to the full 6PN level, obtained
by adding the 4+5+6PN nonlocal contribution.

\section{Post-Minkowskian view of the determination of the local dynamics.} \label{PMview}

At any given PN accuracy, our new method is able to determine most of the structure of the two-body dynamics
except for a relatively small number of numerical coefficients. When working at the 5PN accuracy, only two numerical coefficients
are left undetermined in the local dynamics, namely $a_6^{\nu^2}$ and $\bar d_5^{\nu^2}$. When working at the 6PN
accuracy, four more numerical coefficients are left undetermined, namely
$a_7^{\nu^2}, a_7^{\nu^3}, \bar d_6^{\nu^2}, \,{\rm and}\;  q_{45}^{\nu^2}$. Let us clarify the basic reason underlying 
this incompleteness, in a way which will allow us to anticipate the number and structure of the higher-order analogs of these
undetermined parameters. This is easily done by working within a PM-expanded scheme, and by using  some of the structural 
results of PM gravity discussed in Ref. \cite{Damour2019}. It was found there that the general structure of the 
PM coefficients $q_n^E(\g,\nu)$ of the EOB $Q$ potential in the E-type energy gauge was
\bea \label{qngnu}
q_n^E(\g,\nu) &&= q_{n,0}(\g) + \frac{q_{n,1}(\g)}{h(\g,\nu)} +\ldots + \frac{q_{n,n-1}(\g)}{h^{n-1}(\g,\nu)} \nonumber\\
&&=\sum_{k=0}^{n-1} \frac{q_{n,k}(\g)}{h^k(\g,\nu)}\,,
\eea
with the constraint
\beq \label{sumq}
\sum_{k=0}^{n-1} q_{n,k}(\g)=0.
\eeq
The important structural information in the expression \eq{qngnu} is the fact that the $\nu$-dependence of $q_n^E(\g,\nu)$ 
is entirely described through the powers of $h(\g,\nu)$ entering the denominators. All the corresponding numerators $q_{n,k}(\g)$
depend only on the EOB effective energy $\g=\e$. The constraint \eq{sumq} expresses the fact that 
\beq
\lim_{\nu \to 0}q_n^E(\g,\nu)=0\,,
\eeq
\ie, the basic feature of the EOB formalism that the $\nu \to 0$ limit of the EOB mass-shell
condition reduces to a geodesic in a Schwarzshild metric of mass $M$. 
Let us also note that the behavior of the
coefficients $q_n^E(\g,\nu)$ in the $\g^2\to 1$ limit,
\begin{eqnarray} 
\label{PNqEn}
q^{E\,\rm loc, f}_{n\rm EG}(\gamma;\nu)&=&q_{n\rm E, EG}^0(\nu)+q_{n\rm E,  EG}^1(\nu)(\gamma^2-1)\nonumber\\
&+&q_{n\rm E, EG}^2(\nu)(\gamma^2-1)^2+\ldots\,.
\end{eqnarray}
is smooth, \ie, the expansion coefficients, and notably the first one, $q_{n\rm E, EG}^0(\nu)$, are all finite (and $O(\nu)$).

As explicitly discussed in the 3PM case, $n=3$, in Ref. \cite{Damour2019}, there are more constraints on the $n$ energy-dependent
coefficients $q_{n,k}(\g)$ which determine some of them in terms of the lower PM orders. Indeed, let us first insert
the decomposition \eq{qngnu} in the expressions \eq{chinvsqn} relating the PM-expansion coefficients $q_n^E(\g,\nu)$ 
of the EOB potential to the PM-expansion coefficients $\chi_n(\g,\nu)$ of the scattering angle. This yields explicit expressions
for the $\chi_n(\g,\nu)$'s in terms of the $q_n^E(\g,\nu)$'s. For instance, at the lowest PM order ($n=2$) we have
\bea
&&\chi_2(\gamma,\nu) = \chi^{\rm Schw}_2(\gamma) -\frac{\pi}{4}q_2(\gamma,\nu)\nonumber\\
&&\qquad =\chi^{\rm Schw}_2(\gamma)  -\frac{\pi}{4} \left(q_{2,0}(\g) + \frac{q_{2,1}(\g)}{h(\g,\nu)} \right),
\eea
so that
\beq
\widetilde \chi_2(\gamma,\nu)\equiv h  \chi_2= h \left( \chi^{\rm Schw}_2(\gamma)  -\frac{\pi}{4} q_{2,0}(\g) \right)  -\frac{\pi}{4} q_{2,1}(\g).
\eeq
Imposing the condition that $\widetilde \chi_2$ is independent of $\nu$ reduces to imposing that the coefficient of $h(\g, \nu)$ 
on the right-hand side vanishes. This yields the constraint
\beq
 \chi^{\rm Schw}_2(\gamma)  -\frac{\pi}{4} q_{2,0}(\g)=0\,,
\eeq
which determines $q_{2,0}(\g)$ in terms of $\chi^{\rm Schw}_2(\gamma) $. The summed constraint \eq{sumq} then determines
$q_{2,1}(\g)= - q_{2,0}(\g)$. One then recovers the result \cite{Damour:2016gwp}
\beq \label{q2chiS}
q_2(\g,\nu)= \frac{4}{\pi}  \chi^{\rm Schw}_2(\gamma) \left(1-\frac1{h} \right)=\frac32 (5 \g^2-1) \left( 1-\frac1{h}\right).
\eeq
In other words, the 2PM dynamics is entirely determined by the test-mass scattering angle.

At the 3PM level the three coefficients entering
\beq \label{q3structure}
q_3(\g,\nu) = q_{3,0}(\g) + \frac{q_{3,1}(\g)}{h(\g,\nu)} + \frac{q_{3,2}(\g)}{h^{2}(\g,\nu)}\,,
\eeq
satisfy two constraints. First, the sum constraint \eq{sumq}, \ie,
\beq \label{sumq3}
 q_{3,0}(\g) + q_{3,1}(\g) + q_{3,2}(\g)=0\,,
\eeq
and then the condition that $\widetilde \chi_3(\g,\nu) \equiv h^2 \chi_3(\g,\nu)$ be linear in $\nu$.
The second Eq. \eq{chinvsqn} allows one to express $\chi_3(\g,\nu)$ in terms of the $ q_{3,k}(\g)$'s. 
It is easily seen that inserting the  expression \eq{q2chiS} of $q_2(\g,\nu)$ in the second Eq. \eq{chinvsqn}
yields  $\widetilde \chi_3(\g,\nu) $ in the form of  a polynomial in $h$, namely,
\bea
\widetilde \chi_3(\g,\nu) &=&h^2  \chi^{\rm Schw}_3(\gamma)  -\frac{2\gamma^2-1}{\sqrt{\gamma^2-1}}h^2 q_2(\gamma,\nu)\nonumber\\
&-&\sqrt{\gamma^2-1}h^2 q_3(\gamma,\nu)\nonumber\\
&=&h^2  \chi^{\rm Schw}_3 -  \frac{2\gamma^2-1}{\sqrt{\gamma^2-1}}\frac32 (5 \g^2-1)  \left( h^2-h\right)\nonumber\\
&-& \sqrt{\gamma^2-1} \left(   h^2 q_{3,0}(\g) + h\, q_{3,1}(\g) + q_{3,2}(\g) \right). \nonumber\\
\eea
As $h^2 = 1 + 2\nu(\g-1)$, the condition to be linear in $\nu$ (at a fixed value of $\g$) is equivalent (for a 
polynomial in $h$ with $\g$-dependent coefficients) to having the structure $c_0+c_2 h^2$. This gives one constraint, namely
the vanishing of the coefficient of $h^1$. This constraint determines the coefficient $q_{3,1}(\g)$ to have the value \cite{Damour2019}
\beq \label{q31}
q_{3,1}(\g) = \frac32 \frac{(2\gamma^2-1) (5\gamma^2-1)}{\gamma^2-1}\,.
\eeq
The two remaining coefficients $q_{3,0}(\g) $,  $q_{3,2}(\g) $ then satisfy the single sum constraint \eq{sumq3}.
The conclusion is that the general solution of the 3PM constraints is a $Q$ potential of the form
\bea \label{q3}
q_3(\g,\nu) &=& q_{3,1}(\g) \left(\frac{1}{h(\g,\nu)}-1\right)\nonumber\\
&&+ q_{3,2}(\g) \left(\frac{1}{h^{2}(\g,\nu)}-1\right)\,,
\eea
where $q_{3,1}(\g)$ is determined from \eq{q31}, and where $q_{3,2}(\g) $ is, at this stage, left undetermined by
the general PM-EOB constraints of Ref. \cite{Damour2019}. On the other hand, let us assume that one has somehow determined
(maybe to some limited PN accuracy) the value of the gauge-invariant 3PM scattering angle, which must
have the structure
\beq \label{chi3structure}
\chi_3(\g,\nu) =   \chi_{3,0}(\g) +  \frac{\chi_{3,2}(\g)}{h^{2}(\g,\nu)}\,.
\eeq

Let us now insert in the second Eq. \eq{chinvsqn} the expressions of $\chi_3(\g,\nu)$
and $q_3(\g,\nu)$ as polynomials in $\frac1h$ (with coefficients depending only on $\g$), \ie, Eqs. 
\eq{q3structure} and \eq{chi3structure}. 
As both sides are polynomials in $\frac1h$, we can identify the coefficients of  $\frac1{h^2}$ on both sides.
Indeed, we are dealing here with expressions depending on $\nu$ only through the energy parameter $h(\g,\nu)$.
Therefore, two functions
of $\g$ and $\nu$, which can written as polynomials in $\frac1h$,
can be equal only if all the $\g$-dependent (but crucially $\nu$-independent) coefficients of the various powers of $\frac1h$
agree with each other. This yields the simple link:
\beq \label{chi32vsq32}
\chi_{3,2}(\g) = - \sqrt{\gamma^2-1} \, q_{3,2}(\g)\,.
\eeq
In addition, using the fact that $  \chi_{3}^{\rm Schw}(\g)=\chi_3(\g,0) +    \chi_{3,2}(\g)$,
we can rewrite Eq. \eq{chi3structure} as
 \beq \label{chi3structure1}
\chi_3(\g,\nu) =\chi_{3}^{\rm Schw}(\g) - 2 \nu  \frac{ (\g-1)  \chi_{3,2}(\g)}{h^{2}(\g,\nu)}\,.
\eeq
This formula shows that the function $\chi_{3,2}(\g)$ parametrizes the deviation of $\chi_3(\g,\nu)$
away from its test-mass limit $\lim_{\nu \to 0} \chi_3(\g,\nu)=\chi_{3}^{\rm Schw}(\g)$.
Let us again  emphasize (following Ref. \cite{Damour2019})  that even if one knows only 
 the linear-in-$\nu$ (1SF) expansion of the 3PM scattering angle, Eq. \eq{chi3structure1}
shows that this suffices to fully determine the function $\chi_{3,2}(\g)$. 
Then having extracted the function $\chi_{3,2}(\g)$ from the 1SF expansion of $\chi_3(\g,\nu)$,
we can compute $q_{3,2}(\g) $ from Eq. \eq{chi32vsq32}, and thereby obtain the full 3PM dynamics
by using Eqs. \eq{q31}, \eq{q3}.

As our method, when applied at any PN approximation, 
determines (in particular) the 1SF expansion of the local dynamics, we see that it will determine the function 
$q_{3,2}(\g) $ with the PN accuracy with which we work. This is why we could determine the 6PN expansion
of $q_{3,2}(\g) $, \ie, of the local 3PM dynamics. 

In view of the several independent 6PN-accurate confirmations
(\cite{Blumlein:2020znm,Cheung:2020gyp}, and the present work) of the value of  $\chi_{3}(\g,\nu) $ derived in 
Refs.~\cite{Bern:2019nnu,Bern:2019crd}, we shall assume in the following that $q_{3,2}(\g) $ is exactly known,
namely (using the notation\footnote{The coefficients $q_{3,0}(\g), q_{3,1}(\g), q_{3,2}(\g)$
are respectively denoted $A(\g), B(\g),C(\g)$ there, with $\overline C(\g)\equiv - (\g-1) C(\g)$.} of \cite{Damour2019})
\beq \label{q32}
q_{3,2}(\g)= - \frac{\overline C^B(\g)}{\g-1},
\eeq
with
\begin{eqnarray} \label{bCB}
{\overline C}^{B}(\gamma)&=&\frac23\gamma(14\gamma^2+25)\nonumber\\
&+&
4\frac{4\gamma^4-12\gamma^2-3}{\sqrt{\gamma^2-1}}{\rm arcsinh}\left(\sqrt{\frac{\gamma-1}{2}}\right)\,.\nonumber\\
\end{eqnarray}
This assumption will allow us to  simplify the discussion of the determination of the higher PM-order coefficients.

Let us indeed indicate how the above  2PM and 3PM results extend at higher PM levels. This will allow us to clarify the 
effectiveness (associated with a partial ineffectiveness) of our method in determining (or leaving undetermined) 
the parameters entering the local dynamics.

The structure of the 4PM-level EOB $Q$ potential is
\beq \label{q4}
q^E_4(\g,\nu) = q_{4,0}(\g) + \frac{q_{4,1}(\g)}{h(\g,\nu)} + \frac{q_{4,2}(\g)}{h^{2}(\g,\nu)}+ \frac{q_{4,3}(\g)}{h^{3}(\g,\nu)}\,,
\eeq
with the usual constraint
\beq \label{sumq4}
q_{4,0}(\g) + q_{4,1}(\g) + q_{4,2}(\g)+ q_{4,3}(\g)=0\,.
\eeq
The third Eq. \eq{chinvsqn} leads to an expression for $\chi_4(\g,\nu)$ of the form
\bea \label{chi4q1}
\chi_4(\g,\nu)&=& \chi_4^{\rm Schw}(\g) - \frac{3 \pi}{8} (\g^2-1) q^E_4(\g,\nu)\nonumber\\
&&+ K\left[q_2(\g,\nu), q_3(\g,\nu)\right]\,,
\eea
where $K[q_2,q_3]$ denotes some {\it known} terms, namely
\bea
K[q_2,q_3]&=&\pi \left[\frac{3}{16}q_2^2(\gamma,\nu)-\frac{9}{16}(5\gamma^2-1)q_2(\gamma,\nu)\right. \nonumber\\
&&\left. -\frac38 (3\gamma^2-1)q_3(\gamma,\nu)  \right]\,.
\eea
Inserting the $\frac1h$ parametrization \eq{q4} of $q^E_4(\g,\nu)$, 
together with the above explicit expressions of $q_2(\g,\nu)$, and $q_3(\g,\nu)$ (as polynomials in $\frac1h$),
 then leads to an expression for $\chi_4(\g,\nu)$ having also the structure of a polynomial in $\frac1h$, say
\beq \label{chi4q2}
\chi_4(\g,\nu)=  \chi_{4,0}^q(\g) + \frac{\chi_{4,1}^q(\g)}{h(\g,\nu)} + \frac{\chi_{4,2}^q(\g)}{h^{2}(\g,\nu)}+ \frac{\chi_{4,3}^q(\g)}{h^{3}(\g,\nu)}\,,
\eeq
where the superscript $q$ means that all the coefficients $ \chi_{4,k}^q(\g)$ are explicit expressions in the $q_{n,k}$'s.

The rule found in \cite{Damour2019} restricts $h^3\chi_4(\g,\nu)$ to be linear in $\nu$. 
This is equivalent to the following  {\it restricted polynomial structure} for $\chi_4(\g,\nu)$:
\beq \label{chi4structure}
\chi_4(\g,\nu)=   \frac{\chi_{4,1}(\g)}{h(\g,\nu)} +  \frac{\chi_{4,3}(\g)}{h^{3}(\g,\nu)}\,,
\eeq
with the constraint
\beq
\chi_{4,1}(\g)+ \chi_{4,3}(\g)=  \chi_4^{\rm Schw}(\g).
\eeq
For the general reason already explained above, the equality (for all values of $\nu$) between  two functions
of $\g$ and $\nu$, which can both be written as polynomials in $\frac1h$ with $\g$-dependent (but crucially $\nu$-independent) coefficients,
implies the equality of the $\g$-dependent coefficients of all  the various powers of $\frac1h$.
We therefore conclude that the coefficients $q_{4,k}$ must satisfy the two equations
\beq \label{eqsq40q42}
\chi_{4,0}^q(\g)=0 \quad; \quad \chi_{4,2}^q(\g)=0 \,.
\eeq
In view of Eq. \eq{chi4q1}, the latter two equations are respectively linear in $q_{4,0}(\g)$ and $q_{4,2}(\g)$, and contain
``source terms'' provided both by  $\chi_4^{\rm Schw}(\g)$ and  by $K\left[q_2(\g,\nu), q_3(\g,\nu)\right]$.
We can then solve  the system of the two equations  \eq{eqsq40q42} for $q_{4,0}(\g)$,  and $q_{4,2}(\g)$.
This yields the (unique) solution
\bea \label{solq4042}
q_{4,2}(\g) &=& a_2(\g) + b_2(\g) q_{3,2}(\g) \,,\nonumber\\
q_{4,0}(\g) &=& -  a_1(\g)-  a_2(\g) -  b_2(\g) q_{3,2}(\g)\,,
\eea
where we denoted
\begin{eqnarray}
a_2(\gamma)&=& \frac{9}{8}\frac{(5\gamma^2-1)^2}{\gamma^2-1}\,, \nonumber\\ 
b_2(\gamma)&=&  -\frac{ 3\gamma^2-1 }{\gamma^2-1}\,, \nonumber\\
a_1(\gamma)&=& -\frac{1875 \gamma^6 -2529 \gamma^4 +905 \gamma^2-59}{16 (\gamma^2-1)^2}  \,.  \nonumber\\ 
\end{eqnarray}
Let us now consider  the sum constraint, Eq. \eq{sumq4}. The latter constraint, together with the solution \eq{solq4042},
yields the following expression for $q_{4,1}(\g)$:
\beq
q_{4,1}(\g) = a_1(\g) - q_{4,3}(\g) \,.
\eeq
In other words, the exact structure of the 4PM $Q$ coefficient is
\bea \label{q4xpl}
q^E_4(\g,\nu)&=& a_1(\g) \left(\frac{1}{h(\g,\nu)}-1\right) \nonumber\\
&+& ( a_2(\g) + b_2(\g) q_{3,2}(\g)) \left(\frac{1}{h^2(\g,\nu)}-1\right) \nonumber\\
&+& q_{4,3}(\g) \left(\frac{1}{h^3(\g,\nu)}-1\right)\,.
\eea
In this expression,  $q_{3,2}(\g)$ can (as far as we know) be replaced by \eq{q32}, so that the only undetermined
function of $\g$ is the last coefficient $q_{4,3}(\g) $. The latter can be determined by the knowledge of $\chi_4(\g,\nu)$.
Indeed, the link \eq{chi4q1} between $q_4$ and $\chi_4$ implies that the coefficient $\chi_{4,3}$ of $\frac1{h^3}$
in the $\frac1h$-polynomial expression \eq{chi4structure} of $\chi_4$ is directly linked to $q_{4,3}(\g)$ via
\beq \label{chi43q32}
 \chi_{4,3}(\g)= - \frac{3 \pi}{8} (\g^2-1) q_{4,3}(\g)\,.
\eeq
In addition, we note that Eq. \eq{chi4structure}  can be rewritten as
\bea \label{chi4vschi43}
\chi_4(\g,\nu) &=&   \frac{\chi_4^{\rm Schw}(\g)}{h(\g,\nu)} + \chi_{4,3}(\g)\left(  \frac{1}{h^{3}(\g,\nu)}-\frac{1}{h(\g,\nu)}\right) \nonumber\\
 &=&   \frac{\chi_4^{\rm Schw}(\g)}{h(\g,\nu)} - 2\nu (\g-1) \frac{ \chi_{4,3}(\g)}{h^3(\g,\nu)}\,.
\eea
The latter expression clearly shows that the knowledge of the $O(G^4)$ scattering angle $\chi_4(\g,\nu)$ 
at the {\it linear} order in $\nu$ (1SF order)
suffices to determine the exact function $\chi_{4,3}(\g)$, and thereby to have the full $\nu$ dependence of the scattering angle,
as defined by the expression \eq{chi4vschi43}.

We have derived above the 4PM scattering angle $\chi_4(\g,\nu)$ with 6PN accuracy. Using the representation \eq{chi4vschi43}
we can transcribe our results  into the
following corresponding 6PN knowledge of  the more primitive function $\chi_{4,3}(\g)$: 
\bea \label{chi436pn}
\chi_{4,3}(\g) &=& \pi\Bigg[\frac{15}{4p_\infty^2}+\frac{391}{8}-\frac{123}{256}\pi^2 \nonumber\\
&&
+\left(\frac{4597}{48}-\frac{35569}{16384}\pi^2\right)p_\infty^2\nonumber\\
&&
+\left(\frac{372943}{5600}-\frac{217695}{65536}\pi^2\right)p_\infty^4\nonumber\\
&&
+\left(-\frac{4976527}{1881600}-\frac{49220339}{33554432}\pi^2\right)p_\infty^6\nonumber\\
&& 
+O(p_\infty^8)\Bigg]\,,
\eea
where $p_\infty^2=\gamma^2-1$.
The latter result can then be transcribed in a corresponding 6PN-accurate knowledge of the function $q_{4,3}(\g)$,
and thereby of the full 4PM $Q$ potential $q^E_4(\g,\nu)$, using Eq. \eq{q4xpl}.
We note in passing that the result \eq{chi436pn} implies for $q_{4,3}(\g)$ a behavior in the small $\pinf$ limit of the form
\bea
q_{4,3}(\g)&=& -\frac{10}{p_\infty^4} +\left(-\frac{391}{3}+\frac{41}{32}\pi^2\right)\frac{1}{p_\infty^2} \nonumber\\
&&+\left(-\frac{4597}{18}+\frac{35569}{6144}\pi^2\right) \nonumber\\
&&+\left(-\frac{372943}{2100}+\frac{72565}{8192}\pi^2\right)p_\infty^2 \nonumber\\
&&+\left(\frac{4976527}{705600}+\frac{49220339}{12582912}\pi^2\right)p_\infty^4
+O(p_\infty^6)\,,\nonumber\\
\eea
so that the corresponding contribution to $q^E_4(\g,\nu)$, Eq. \eq{q4xpl}, reads
\begin{widetext}
\bea
q_{4,3}(\g) \left(\frac{1}{h^3(\g,\nu)}-1\right)&=& 
\frac{15\nu}{p_\infty^2} +\left[\left(\frac{767}{4}-\frac{123}{64}\pi^2\right)\nu-\frac{75}{4}\nu^2\right] \nonumber\\
&+&\left[\left(\frac{4033}{12}-\frac{33601}{4096}\pi^2\right)\nu+\left(-235+\frac{615}{256}\pi^2\right)\nu^2+\frac{175}{8}\nu^3\right]p_\infty^2 \nonumber\\
&+&\left[\left(\frac{6514457}{33600} -\frac{93031}{8192}\pi^2\right)\nu+\left(-\frac{69605}{192} +\frac{158165}{16384}\pi^2\right)\nu^2+\left(\frac{25795}{96}-\frac{1435}{512}\pi^2\right)\nu^3\right.\nonumber\\
&-&\left.\frac{1575}{64}\nu^4\right]p_\infty^4 
\nonumber\\
&+&
\left[\left(-\frac{6859063}{156800}-\frac{29201523}{8388608}\pi^2\right)\nu
+\left(-\frac{1114333}{6720}+\frac{781985}{65536}\pi^2 \right)\nu^2\right. \nonumber\\
&+&\left.
\left(-\frac{1038275}{98304}\pi^2 +\frac{411425}{1152}\right)\nu^3
+\left(-\frac{37905}{128}+\frac{12915}{4096}\pi^2\right)\nu^4
+\frac{3465}{128}\nu^5\right] p_\infty^6
+O(p_\infty^8) \,. \nonumber\\
\eea
\end{widetext}

 This contribution is singular as $\pinf \to 0$. However, it is easily checked that the other contributions to $q^E_4(\g,\nu)$ in Eq. \eq{q4xpl}
 cancell this low-velocity singularity and leave a finite result,
 \beq
 q^E_4(\g,\nu) =  \left(\frac{175}{3}-\frac{41}{32}\pi^2\right)\nu-\frac{7}{2}\nu^2 + O(\pinf^2)\,,
 \eeq
 in agreement with the result listed in Table \ref{EEG_coeffs}.

Let us sketch the extension of these results to the $\geq 5$ PM orders. [See Appendix \ref{A} for more technical details.]
Again the basic trick is to express all dynamical
functions as polynomials in $\frac1h$, with $\g$-dependent coefficients. This trick is efficient because the
PM-EOB results Eqs. \eq{chinvsqn} involve no explicit $\nu$ dependence. In turn, this property follows from the basic fact that the
1PM-accurate EOB dynamics is $\nu$-independent when expressed in terms of the EOB effective energy $\g=\e$ \cite{Damour:2016gwp}.

The structure of the 5PM $Q$ potential reads
\bea \label{q5}
&&q^E_5(\g,\nu)= q_{5,0}(\g) + \frac{q_{5,1}(\g)}{h(\g,\nu)} \nonumber\\
&&\qquad + \frac{q_{5,2}(\g)}{h^{2}(\g,\nu)}+ \frac{q_{5,3}(\g)}{h^{3}(\g,\nu)}+ \frac{q_{5,4}(\g)}{h^{4}(\g,\nu)}\,,
\eea
with the usual constraint
\beq \label{sumq5}
q_{5,0}(\g) + q_{5,1}(\g) + q_{5,2}(\g)+ q_{5,3}(\g) +  q_{5,4}(\g)=0\,.
\eeq
The fourth Eq. \eq{chinvsqn} leads to a corresponding expression for $\chi_5(\g,\nu)$ of the form
\beq \label{chi5q1}
\chi_5(\g,\nu)= \chi_5^{\rm Schw}(\g) -\frac{4 (\gamma^2-1)^{3/2}}{3}q^E_5(\gamma) + K[q_2,q_3,q_4]\,, 
\eeq
where  the ``known" contribution, $K[q_2,q_3,q_4]$, which involves previous PM orders, $q_2$, $q_3$ and $q_4$, will be found in Appendix \ref{A}.

The rule restricting the $\nu$ structure of $\chi_5(\g,\nu)$ \cite{Damour2019}  is equivalent to imposing:
\beq \label{chi5structure}
\chi_5(\g,\nu)=  \chi_{5,0}(\g) +\frac{\chi_{5,2}(\g)}{h^2(\g,\nu)} +  \frac{\chi_{5,4}(\g)}{h^{4}(\g,\nu)}\,,
\eeq
with the constraint
\beq
\chi_{5,0}(\g)+ \chi_{5,2}(\g) + \chi_{5,4}(\g)=  \chi_5^{\rm Schw}(\g).
\eeq
Imposing this structure on the expression following from Eq. \eq{chi5q1} then yields two constraints expressing
the vanishing of the terms $\propto \frac1h$ and $\propto \frac1{h^3}$. This yields two equations of the type
\bea \label{eqsq51q53}
q_{5,1}(\g) &=& {\rm known}\,, \nonumber\\
q_{5,3}(\g) &=& {\rm known}\,,
\eea
whose explicit form will be found in Appendix \ref{A}.

In addition, we have the third equation \eq{sumq5}. The latter equation yields an expression for  $q_{5,0}(\g)$
of the form
\beq
q_{5,0}(\g)= - q_{5,2}(\g)  - q_{5,4}(\g) +  {\rm known}\,.
\eeq
At the end of the day, we have a general expression for $q_5(\g,\nu)$ of the form
\bea \label{q5xpl2}
q^E_5(\g,\nu)&=& {\rm known} + q_{5,2}(\g) \left(\frac{1}{h^2(\g,\nu)}-1\right)\nonumber\\
&&+ q_{5,4}(\g) \left(\frac{1}{h^4(\g,\nu)}-1\right)\,,
\eea
where \lq\lq known" means here 
\beq
 {\rm known} =q_{5,1}(\g)\left(\frac{1}{h}-1\right)
+q_{5,3}(\g)\left(\frac{1}{h^3}-1\right)\,,
\eeq 
with $q_{5,1}(\g)$ and $q_{5,3}(\g)$ given in Eqs. \eqref{q_51_53}.

The expression \eq{q5xpl2}  involves only two  undetermined (at this stage) parameters  $q_{5,2}(\g)$ and $q_{5,4}(\g)$.
As before ({\it mutatis mutandis}), the two remaining parameters $q_{5,2}(\g)$ and $q_{5,4}(\g)$ would be determined 
by the knowledge of the two corresponding coefficients in $\chi_5(\g,\nu)$, Eq. \eq{chi5structure}, namely
$\chi_{5,2}(\g)$ and $\chi_{5,4}(\g)$. Indeed, we have the two equations
\bea
\chi_{5,2}(\g) &=& \frac{45\g^4-34\g^2+7}{3(\g^2-1)^{1/2}} q_{3,2}(\g)\nonumber\\
&&-\frac43  (\g^2-1)^{3/2} q_{5,2}(\g) \nonumber\\
&&-\frac{3(14\g^2-5) (-1+5\g^2)^2}{4(\g^2-1)^{1/2}}
\,, \nonumber\\
\chi_{5,4}(\g)&=&-\frac43 (\g^2-1)^{3/2} q_{5,4}(\g)\,,
\eea
where we recall that the 3PM-level function $q_{3,2}(\g)$ is known.

However, there is now a difference with what happened at lower PM orders.
Indeed, we can rewrite Eq. \eq{chi5structure} in the form
\bea \label{chi5structure2}
\chi_5(\g,\nu)&=&  \chi_5^{\rm Schw}(\g) + \chi_{5,2}(\g) \left( \frac{1}{h^2(\g,\nu)}-1\right)\nonumber\\
&& +  \chi_{5,4}(\g) \left( \frac{1}{h^4(\g,\nu)}-1\right)\,,
\eea
or, equivalently,
\bea 
\label{chi5structure3}
\chi_5(\g,\nu) &=&  \chi_5^{\rm Schw}(\g)\nonumber\\
&-&\frac{4 }{h^4} \left[\nu (\g-1) \left(\chi_{5,4}(\g)+\frac12\chi_{5,2}(\g)\right)\right.\nonumber\\
&+&\left.  \nu^2  (\g-1)^2\left( \chi_{5,4}(\g)+ \chi_{5,2}(\g)\right) \right]\,,\nonumber\\
\eea
In other words, after factoring $1/h^4$ the difference $  \chi_5(\g,\nu) - \chi_5^{\rm Schw}(\g)$ has
a $\nu$ structure of the type $\sim \nu+ \nu^2$. By contrast, we previously had a difference $  \chi_4(\g,\nu) - \frac{\chi_4^{\rm Schw}(\g)}{h(\g,\nu)}$ of the type $\sim \nu$. This change of  $\nu$ dependence (from $\sim \nu$ to $\sim \nu+ \nu^2$) implies that, at the 5PM level,
there appear two independent functions of $\g$ parametrizing the scattering function, after having taken into account the
general structural information about its $\nu$ dependence, while there appeared only one function of $\g$ at the 3PM and 4PM levels.
 As a consequence, our method (which completes a linear-in-$\nu$ self-force information by a general $\nu$-dependence information)
 is able to get complete PN-expanded results at the 3PM and 4PM levels (up to the PN accuracy it uses). However, at the 5PM (and also 6PM)
 levels, it can only determine one combination of the two independent functions of $\g$ appearing at these levels (namely the function parametrizing
 the coefficient of $\nu$ among the total  $\sim \nu+ \nu^2$ dependence just mentioned). More specifically, at the 5PM level, 
 one sees from Eq. \eq{chi5structure3} that, when working at some given PN accuracy, our method will be able to determine,
 within this PN accuracy, the PN expansion of the function $\chi_{5,4}(\g) +\frac12\chi_{5,2}(\g)$, but will leave (partially) undetermined 
 that of the complementary combination $\chi_{5,4}(\g) +\chi_{5,2}(\g)$. Using our results, we find
 
\bea
\chi_{5,4}(\g)+\frac12 \chi_{5,2}(\g)&=& \frac{4}{p_\infty^3} +\left(-\frac{41}{16}\pi^2+\frac{587}{3}\right)\frac{1}{p_\infty} \nonumber\\
&+&\left(-\frac{10507}{576}\pi^2+\frac{480263}{540}\right) p_\infty \nonumber\\
&+&\left(-\frac{715139}{11520}\pi^2+\frac{30034567}{18900}\right) p_\infty^3 \nonumber\\
&+&\left(\frac{1160329}{161280}\pi^2+\frac{411639569}{1176000}\right) p_\infty^5 \nonumber\\
&+&O(p_\infty^7)\,, 
\eea
 which is, indeed, fully determined to our 6PN accuracy, while
\bea
&&\chi_{5,4}(\g)+\chi_{5,2}(\g)=
-\frac{8}{p_\infty^3} +\left(-\frac{406}{9}+\frac{41}{24}\pi^2\right)\frac{1}{p_\infty} \nonumber\\
&&+\left(\frac{112333}{270}+\frac{4}{15}\bar d_5^{\nu^2} -\frac{18487}{5760}\pi^2\right) p_\infty \nonumber\\
&&+\left(\frac{4}{35}q_{45}^{\nu^2} +\frac{2}{15}\bar d_5^{\nu^2}+\frac{1993193869}{1323000}-\frac{5049671}{80640}\pi^2\right) p_\infty^3\nonumber\\
&&+O(p_\infty^5)\,,
\eea
involves the undetermined parameters $\bar d_5^{\nu^2}$ and $q_{45}^{\nu^2}$.

 When translating this knowledge in terms of the EOB $Q$ potential (in E-type energy gauge\footnote{The relations we gave above then allow one to translate the $q_n^E$'s into their H-type correspondants $q_n^H$.}), this means
 that our method is able to determine the function $q_{5,4}(\g) +\frac12 q_{5,2}(\g)$, 
 but  leaves partly undetermined the complementary function $q_{5,4}(\g) +q_{5,2}(\g)$.
 Concerning the other coefficient functions $q_{5,k}(\g)$, with $k=0,1,3$, parametrizing $q_5(\g,\nu)$, the generalization of the
 reasoning explained above for $q_4(\g,\nu)$ shows that they are fully determined in terms of the lower PM information.

 One can check that a similar situation occurs at the 6PM level, where the structure
 of the scattering angle reads
 \bea \label{chi6structure}
\chi_6(\g,\nu) &=&  \frac{\chi_6^{\rm Schw}(\g)}{h(\g,\nu)} + \chi_{6,3}(\g) \left( \frac{1}{h^3(\g,\nu)}-\frac{1}{h(\g,\nu)}\right) \nonumber\\
&& +  \chi_{6,5}(\g) \left( \frac{1}{h^5(\g,\nu)}-\frac{1}{h(\g,\nu)}\right)\,.
\eea
The two independent functions $\chi_{6,3}(\g)$, $\chi_{6,5}(\g)$ parametrize a structure $\sim h^{-5}(\nu + \nu^2)$.
Similarly to Eq. \eq{chi5structure3}, this can be made manifest by introducing the following two combinations (with $\g$-dependent coefficients) of $\chi_{6,3}(\g)$ and $\chi_{6,5}(\g)$, say
\bea
\widehat \chi_{6, \nu}(\g)&\equiv & -4(\g-1)\left( \frac12 \chi_{6,3}(\g)+\chi_{6,5}(\g) \right)\,,\nonumber\\
\widehat \chi_{6, \nu^2}(\g)&\equiv & -4(\g-1)^2\left(\chi_{6,3}(\g)+\chi_{6,5}(\g) \right)\,,
\eea
such that 
  Eq. \eq{chi6structure}  reads 
\beq\label{chi6structure1}
\chi_6(\g,\nu) =  \frac{\chi_6^{\rm Schw}(\g)}{h(\g,\nu)} + \frac{\nu \widehat \chi_{6, \nu}(\g)+ \nu^2 \widehat \chi_{6, \nu^2}(\g) }{h^5(\g,\nu)}\,.
\eeq

Again, our method can only determine one combination (namely $\widehat \chi_{6, \nu}(\g)$) of the two functions
$\chi_{6,3}(\g)$, $\chi_{6,5}(\g)$. When translating this knowledge in terms of the EOB 
$Q$ potential (in energy gauge), this means
 that our method will be able to determine only one combination of the two functions $q_{6,3}(\g)$,and $q_{6,5}(\g)$, via the link of
 Eqs. \eq{chi6q6}.
On the other hand, the  other coefficient functions $q_{6,k}(\g)$, with $k=0,1,2,4$, parametrizing $q_6(\g,\nu)$ are fully determined 
in terms of lower PM information.

At  7PM , one finds that there are {\it three} independent functions of $\g$, namely
$ \chi_{7,2}(\g)$,  $ \chi_{7,4}(\g)$ and  $ \chi_{7,6}(\g)$. They are linked to  their EOB counterparts 
$q_{7,2}(\g)$, $q_{7,4}(\g)$ and $q_{7,6}(\g)$ (and to lower PM functions) via the relations Eqs. \eq{chi7q7}.
 The three functions $ \chi_{7,2}(\g)$,  $ \chi_{7,4}(\g)$ and  $ \chi_{7,6}(\g)$ parametrize a $\nu$ dependence of the type
$\sim (\nu+ \nu^2 + \nu^3)/h^6$. More precisely, there are three combinations of 
$ \chi_{7,2}(\g)$,  $ \chi_{7,4}(\g)$ and  $ \chi_{7,6}(\g)$, say
\bea
\widehat \chi_{7, \nu}(\g)&\equiv &  -2(\g -1) (\chi_{7,2}(\g)+2\chi_{7,4}(\g)+3\chi_{7,6}(\g))\,, \nonumber\\
\widehat \chi_{7, \nu^2}(\g)&\equiv&  -4 (\g -1)^2 (2\chi_{7,2}(\g)+3\chi_{7,4}(\g)+3\chi_{7,6}(\g))\,,\nonumber\\
\widehat \chi_{7, \nu^3}(\g)&\equiv &   -8(\g-1)^3 (\chi_{7,2}(\g)+\chi_{7,4}(\g)+\chi_{7,6}(\g))\,,\nonumber\\
\eea
 such that 
\beq\label{chi7structure1}
\chi_7(\g,\nu) =  \chi_7^{\rm Schw}(\g)+ \frac{\nu \widehat \chi_{7, \nu}(\g)+ \nu^2 \widehat \chi_{7, \nu^2}(\g)+  \nu^3 \widehat \chi_{7, \nu^3}(\g)   }{h^6(\g,\nu)}\,.
\eeq

The situation is similar at the 8PM level, with three independent functions of $\g$
$ \chi_{8,3}(\g)$,  $ \chi_{8,5}(\g)$ and  $ \chi_{8,7}(\g)$, related to their corresponding EOB functions
$ q_{8,3}(\g)$,  $ q_{8,5}(\g)$ and  $ q_{8,7}(\g)$ via   the relations Eqs. \eq{chi8q8}.
 The three functions $ \chi_{8,3}(\g)$,  $ \chi_{8,5}(\g)$ and  $ \chi_{8,7}(\g)$ parametrize a $\sim (\nu+ \nu^2 + \nu^3)/h^7$ 
 structure for the difference $\chi_8(\g,\nu) -  \frac1h \chi_8^{\rm Schw}(\g)$.
And again our method can only determine one combination of these three functions.

\begin{figure}
\includegraphics[scale=0.65]{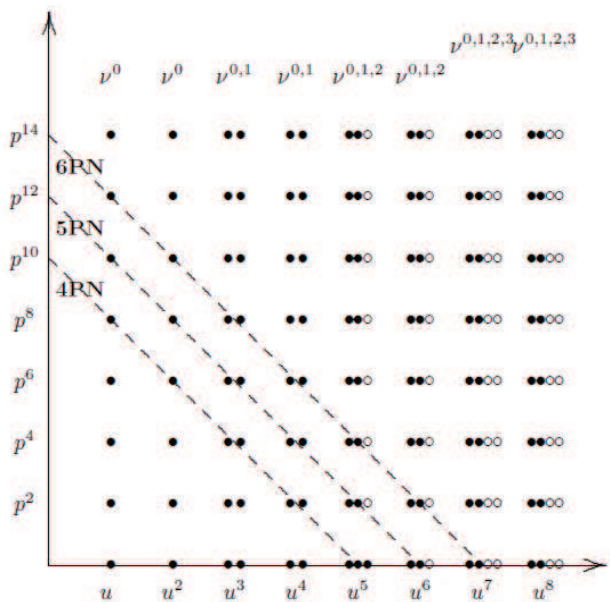}
\caption{Schematic representation of the  irreducible information contained, at each post-Minkowskian level (keyed by a power of $u=GM/r$), 
in the local dynamics. Each vertical column of dots describes the post-Newtonian expansion (keyed by powers of $p^2$) 
of an energy-dependent function parametrizing the scattering angle. The various columns at a given post-Minkowskian level correspond to
increasing powers of the symmetric mass-ratio $\nu$. See text for details.
\label{fig:1}}
\end{figure}

We summarize in a pictorial manner the irreducible information contained, at each PM level, in the local dynamics in Fig. \ref{fig:1}.
The horizontal axis indicates the successive PM orders, while the vertical axis indicates successive PN orders, keyed by powers
of $p^2$ (representing $p_{\infty}^2 \equiv \g^2-1$ when working in the energy-gauge). 
This figure displays the  information contained either in the PM-expansion 
coefficients $ \chi_n(\g,\nu)$ of $\chi$, or in the PM-expansion coefficients $ q^E_n(\g,\nu)$ of  $\widehat Q^E(u,\g;\nu)$.
[We have explained above the (recursive) one-to-one map between these two sequences of coefficients.]
By {\it irreducible} information we mean the building blocks that depend only on $\g$ and that parametrize the 
$\nu$-dependence of the coefficients $ \chi_n(\g,\nu)$ or $ q^E_n(\g,\nu)$. For instance, at the PM level $n=3$ (or $u^3$ in 
$\widehat Q^E(u,\g;\nu)$), the  3PM local  dynamics is fully described by Eq. \eq{chi3structure}, which we write again
for conceptual clarity,
\beq \label{chi3structure2}
\chi_3(\g,\nu)=   \chi_{3,0}(\g) +  \frac{\chi_{3,2}(\g)}{h^{2}(\g,\nu)}\,,
\eeq
\ie, by {\it two} independent functions of $\g$: $ \chi_{3,0}(\g)$ and  $ \chi_{3,2}(\g)$.  One half of this information comes
from the test-mass limit, $\nu \to 0$ (namely $ \chi_{3,0}(\g) +  \chi_{3,2}(\g)= \chi_{3}^{\rm Schw}(\g)$), while the other half is 
encoded in the 1SF (linear in $\nu$) expansion of $\chi_3(\g,\nu)$. This is clear if one rewrites Eq. \eq{chi3structure2} in the form
of Eq. \eq{chi3structure1}, \ie,
\beq \label{chi3structure3}
\chi_3(\g,\nu)=   \chi_{3}^{\rm Schw}(\g) - 2 \nu  \frac{ (\g-1)  \chi_{3,2}(\g)}{h^{2}(\g,\nu)}\,.
\eeq
Here we are talking about the PM expansion. When working within a PN approximation scheme, some of the functions of $\g$
entering as irreducible building blocks are only known in their PN-expanded forms, \ie, only a limited number of terms in their 
expansion in powers of  $p_{\infty}^2 \equiv \g^2-1$ is known. For instance, we derived here, by working at the 6PN
approximation, the first  five  terms of the function $\chi_{3,2}(\g)$,  in the form of the related function
\beq
\overline C(\g)= \frac{\g-1}{p_{\infty}} \chi_{3,2}(\g) = - (\g-1) q_{3,2}(\g)\,,
\eeq
namely
\bea \label{Cb6pn1}
\overline C^{\rm 6PN}(\g) &=& 4 + 18 \pinf^2 + \frac{91}{10} \pinf^4  - \frac{69}{140} \pinf^6  \nonumber\\
&-& \frac{1447}{10080} \pinf^8 + O(\pinf^{10})\,.
\eea
See Eq. \eq{chi436pn} for the analogous result at the 4PM level.


Having in mind this PN-expansion of the $\g$-dependent irreducible PM building blocks $\chi_{n,k}(\g)$,
we represent in Fig. \ref{fig:1} each such building block $\chi_{n,k}(\g)$ by a vertical line of filled circles.
At the 1PM and 2PM levels there is only one irreducible building block, and therefore only one vertical line
of dots. Moreover, these building blocks can be entirely deduced from the test-mass limit, \ie,
they are encoded in the $\nu \to 0$ limit (or Schwarzschild limit) of the scattering angle. At the 3PM 
level, there are two independent irreducible functions of $\g$, represented as two vertical sequences of 
filled circles in the figure. One can think of the left column of dots as being of order $\nu^0$
in the SF expansion, and therefore as being entirely deducible from the Schwarzschild limit. By contrast, 
the right column of dots represents (modulo some $h$-dependent prefactor) a 1SF-level information, 
\ie, it is encoded in the $O(\nu^1)$ term
in the expansion of $\chi_3(\g,\nu)$ in powers of $\nu$. At the 4PM level we have again only two
vertical sequences of dots, say one encoded in the  $\nu \to 0$ limit, and the other one representing
 a fresh 1SF information encoded (modulo some $h$-dependent factor) in the $O(\nu^1)$ term in the $\nu$-expansion of $\chi_4(\g,\nu)$.
[Note in passing that the $\nu$-dependence of the 4PM EOB potential $q_4(\g,\nu)$ deduced from 
$\chi_4(\g,\nu)$ is more involved than the one of $\chi_4(\g,\nu)$. In particular, the $O(\nu^1)$ term
in $q_4(\g,\nu)$ is partly determined by the $O(\nu^1)$ information present at the 3PM level, and
by fresh  $O(\nu^1)$ information contained in $\chi_4(\g,\nu)$.]

At the 5PM and 6PM levels, we have three independent building blocks (see Eqs. \eq{chi5structure3} and \eq{chi6structure1}), 
represented as three vertical columns of dots. Again the left column can be thought of as being $O(\nu^0)$ (and Schwarzschildlike),
the middle column as being $O(\nu^1)$ and 1SF-determined, while the third column is now $O(\nu^2)$,
\ie, encoded at the 2SF level. The only knowledge we currently have of this 5PM third column is its
lowest PN approximation, \ie, the filled circle located at $u^5$ on the horizontal axis. Indeed, this term $\sim \nu^2 u^5 p^0$
was determined by the computation of the 4PN dynamics. In the $p_r$ gauge, it is described by the contribution
$+\left(\frac{41}{32}\pi^2-\frac{221}{6}\right)\nu^2$ to the coefficient $a_5^{\rm loc,f}$ of the $u^5$
term in the EOB radial $A(u,\nu)$ potential.
The 5PN approximation consists of collecting the terms along the second slanted line represented in Fig. \ref{fig:1}.
We see that the slanted 5PN line passes through two of the $O(\nu^2)$ third vertical columns. In the current
implementation of our method, the third (and higher) vertical columns, corresponding to $O(\nu^{\geq 2})$
(2SF and higher) contributions are left undetermined. We highlight this fact by using 
empty circles to represent these columns.
This visually
explains the origin of the two  coefficients left undetermined by our method at 5PN. The empty circle in the $u^5$
column corresponds to $\bar d_5^{\nu^2}$, while the empty circle at the $u^6$ location on the horizontal axis
corresponds to $a_6^{\nu^2}$. 

At the 7PM and 8PM levels, we have four  independent building blocks parametrizing a $\sim \nu^0+\nu^1+\nu^2+\nu^3$
structure (see Eq. \eq{chi7structure1}).
When considering the 6PN, upper slanted line, we now understand clearly why
there were four extra coefficients left undetermined by our method at 6PN. Namely: one in the $u^5$  
$O(\nu^2)$ third vertical column ($ q_{4,5}^{\nu^2} p_r^4  u^5 $); one in the $u^6$ $O(\nu^2)$ third vertical column
($ \bar d_{6}^{\nu^2} p_r^2  u^6 $); and two on the $u^7$ location on the horizontal axis, linked to the
third and fourth columns ($\nu^2 a_{7}^{\nu^2}  u^7$ and  $\nu^3 a_{7}^{\nu^3}  u^7$). 

Looking at Fig. \ref{fig:1}, we can also see what information could give a 7PN-level extension of our method (completed
by a 6.5PN-level purely nonlocal dynamics). It would: (i) provide a 7PN-level test of the 3PM dynamics of 
Refs. \cite{Bern:2019nnu,Bern:2019crd}; (ii) improve the knowledge of the 4PM dynamics at the 7PN level;
(iii) improve the kowledge of the $O(\nu^1)$-encoded local dynamics at the 5PM, 6PM, 7PM and 8PM levels;
but (iv) leave undetermined six numerical coefficients encoding effects of the type 
\beq
\nu^2 (u^5 p^6+ u^6 p^4+ u^7 p^2 +u^8) + \nu^3 (u^7 p^2+ u^8)\,.
\eeq
[In the $p_r$-gauge all the powers of $p$ have to be interpreted as being powers of $p_r$.]
Note that the current lack of determination of coefficients entering  $\nu^2$ and $\nu^3$ effects
is not a conceptual limitation of our method. It is rather a technical limitation of the current development
of SF theory which cannot yet compute any genuine  $O(\nu^2)$ effects. [See, however, \cite{Pound:2019lzj}
for significant progress towards that goal.] The combination of our method with a 2SF-level technology would
allow one to cover, in principle, many more dots in the plane of Fig. \ref{fig:1}.

\section{Concluding remarks}

We have extended the application of a new approach to binary dynamics \cite{Bini2019} to the 6PN level.
Our approach has allowed us to derive an almost complete expression for the 6PN-level
 action, given by the sum of a 4PN+5PN+5.5PN+6PN nonlocal action, Eqs. \eqref{Snonloc00},\eqref{Snonloc_0},
 \eqref{Snonloc}, and of a local one $\int pdq- H_{\rm loc, f}^{4+5+6 \rm PN} dt$. 
 We succeeded in determining the full functional structure of $H_{\rm loc, f}^{\leq {\rm 6PN}}$ (which contains $151$  numerical coefficients), except for four coefficients:   three $\nu^3$-level  coefficients, and one  $\nu^4$-level one (when counting powers
 of $\nu$ in the unrescaled Hamiltonian $H=M c^2 + \ldots$).
One of the crucial tools in our derivation of $H_{\rm loc, f}^{\leq {\rm 6PN}}$ has been the computation of the Detweiler-Barack-Sago
redshift invariant along eccentric orbits in a Schwarzschild spacetime, up to the eight power of the eccentricity and the 9.5-th power
of the inverse semi-latus rectum. This computation alone has been the most time-consuming element of our work, and has extended
the frontier of analytical gravitational self-force theory. 

 We have expressed our final results in five different gauge-invariant ways: 
 (i) in terms of the PM-expanded scattering angle (see Eqs. \eqref{PM_scat_chi}, \eqref{chinvsqn} and discussion in the text
); (ii)  in terms of the PN-expanded radial
 action  (see Eqs. \eqref{Irxp}, \eqref{Irxp2}, with results summarized in Table \ref{various_I_n});
 (iii) in terms of the $p_r$-gauge effective EOB Hamiltonian (see Eqs. \eqref{local_pot_param}-\eqref{local_pot_param6}
 as well as the summary in Table \ref{loc_eob_coeffs}); (iv) in terms of the H-type energy-gauge 
 effective EOB Hamiltonian (see the defining relation in Eq.  \eqref{QEGE} and results listed in Table \ref{HEG_coeffs}); and also, (v) in terms of the irreducible building blocks
 parametrizing a general PM dynamics (see Sec. \ref{PMview}, and notably Eqs. \eq{chi436pn} and \eq{Cb6pn1}). 
 
 Among our new results, let us emphasize: (1) the obtention of the 6PN-accurate $O(G^3)$ scattering angle $\chi_3$
 (see notably Eq. \eq{chi36pn}), in agreement with the PM computation of Refs. \cite{Bern:2019nnu,Bern:2019crd}
 (and with the PN computations of Refs. \cite{Blumlein:2020znm,Cheung:2020gyp}); (2) the obtention (without any
 undetermined parameters) of the 6PN-accurate, 4PM  ($O(G^4)$) {\it local} scattering angle $\chi_4^{\rm  loc, f}$ (see notably
 Eq. \eq{chi4xpl}); (3) the obtention of the linear-in-$\nu$ contributions to the 
  6PN-accurate 5PM, 6PM and 7PM  {\it local} scattering angles $\chi_5^{\rm  loc, f}$, $\chi_6^{\rm  loc, f}$,
  $\chi_7^{\rm  loc, f}$ (see Eqs. \eq{chi2-chi7}); (4) the derivation of the explicit link between the PM-expanded scattering angle
  and the PM-expanded EOB $Q$ potential (in energy gauge) at the 5PM, 6PM and 7PM levels. We leave to future work the derivation
  of the {\it nonlocal} contributions to the scattering angle, and the associated explicit determination of the tuned flexibility
   factor $f(t)$ used here to define the local part of the dynamics.
  
Finally, in Sec. \ref{PMview} we have discussed the synergistic interplay between four approaches to binary dynamics: post-Minkowskian, 
effective-one-body, gravitational self-force, and post-Newtonian (see Fig. \ref{fig:1}). Effective-one-body theory offers  an efficient framework for
 combining gauge-invariant information coming from post-Newtonian, post-Minkowskian, and gravitational self-force results. It has
 also allowed to discover the hidden simplicity of binary dynamics through a deeper understanding of the mass-ratio
 dependence of perturbative results (see, notably Eqs. \eq{Irxp}, \eq{IS0} and \eq{Cn3}, and the discussion of Sec. \ref{PMview}).

\section*{Acknowledgments}
DB thanks the IHES for warm hospitality at various stages during the development of this project.

\appendix

\section{Higher post-Minkowskian links between the scattering angle and the E-type energy-gauge EOB $Q^E$ potential.} 
\label{A}

The terms on the right-hand side of Eq. \eq{q5} read
\bea
\chi_5^{\rm Schw}(\g)&=&\frac1{5(\gamma^2-1)^{5/2}} (1792\gamma^{10}-5760\gamma^8\nonumber\\
&+&6720\gamma^6-3360\gamma^4+630\gamma^2-21) \,,
\eea
and
\begin{eqnarray}
K[q_2,q_3,q_4]&=& q_2^2(\gamma) \frac{(2\gamma^2-1)}{(\gamma^2-1)^{1/2}}\nonumber\\
&&+\left[2 q_3(\gamma) (\gamma^2-1)^{1/2}\right. \nonumber\\
&&\left.
-\frac23 \frac{(64\gamma^6-120\gamma^4+60\gamma^2-5)}{(\gamma^2-1)^{3/2}}\right] q_2(\gamma)\nonumber\\
&&-2 q_3(\gamma) \frac{(8\gamma^4-8\gamma^2+1)}{(\gamma^2-1)^{1/2}}\nonumber\\
&&-\frac43 q_4(\gamma) \frac{(4\gamma^4-5\gamma^2+1)}{(\gamma^2-1)^{1/2}}\,.
\end{eqnarray} 
The explicit form of Eqs. \eq{eqsq51q53} (where  the 4PM-level term $q_{4,3}(\g)$ is considered as being known) is
\begin{eqnarray}
\label{q_51_53}
q_{5,1}(\g) &=& \frac{9(5\g^2-1)}{4(\g^2-1)}  q_{3,2}(\g) 
+\frac{(4\g^2-1)}{(\g^2-1)} q_{4,3}(\g)\nonumber\\
&&+\frac{1}{16 (\g^2-1)^3} (11160\g^8-20193\g^6\nonumber\\
&&+137-2323\g^2+11603\g^4) \,,\nonumber\\
q_{5,3}(\g) &=& -\frac{9(5\g^2-1)}{4(\g^2-1)} q_{3,2}(\g)
-\frac{(4\g^2-1)}{(\g^2-1)} q_{4,3} (\g) \,.\nonumber\\
\end{eqnarray}

At 6PM, the explicit links between the irreducible blocks of the scattering angle and the corresponding
building blocks of the $Q^E$ potential read
\begin{eqnarray} \label{chi6q6}
\chi_{6,3}(\g) &=&  -\frac{15}{128}\pi  (-21+174\g^2-345\g^4)q_{3,2}(\g)\nonumber\\
&&-\frac{15}{128}\pi  (-10+48\g^2-70\g^4)q_{4,3}(\g) \nonumber\\
&&-\frac{15}{32}\pi(\g^2-1)^2q_{6,3}(\g)\nonumber\\
&&-\frac{15}{128}\pi (-9+135\g^2-675\g^4+1125\g^6) \,,\nonumber\\
\chi_{6,5}(\g) &=&  -\frac{15}{32}\pi (\g^2-1)^2  q_{6,5}(\g)\,.
\end{eqnarray}

At 7PM, we have the analogous links:
\begin{eqnarray} \label{chi7q7}
\chi_{7,2}(\g) &=&  \frac{4}{5} q_{3,2}^2(\g)(34\g^2-9) (\g^2-1)^{1/2}\nonumber\\&&
+\frac{1}{20(\g^2-1)^{3/2}} (+2685 \g^8-7692 \g^6\nonumber\\
&&+4626 \g^4-692 \g^2-27
)q_{3,2}(\g) \nonumber\\&&
-\frac{6}{5}q_{4,3}(\g) (34 \g^2-9) (-1+5 \g^2) (\g^2-1)^{1/2}\nonumber\\&&
+\frac{2}{5} q_{5,2} (\g) (99 \g^4-62 \g^2+13) (\g^2-1)^{1/2}\nonumber\\&&
-\frac{8}{5} q_{7,2}(\g) (\g^2-1)^{5/2} \nonumber\\&&
-\frac{3}{40(\g^2-1)^{3/2} } (-1+5 \g^2) (60330 \g^8-114477 \g^6\nonumber\\
&&+65651 \g^4-12773 \g^2+669)  
\,,\nonumber\\
\chi_{7,4}(\g) &=&   -\frac{2}{5(\g^2-1)^{1/2}} q_{3,2}^2(\g)(34 \g^2-9) \nonumber\\&&
+\frac{45}{4}q_{3,2}(\g) (25  \g^4-10 \g^2+1) (\g^2-1)^{1/2}\nonumber\\&&
+\frac{6}{5}q_{4,3}(\g) (170 \g^4-79 \g^2+9) (\g^2-1)^{1/2}\nonumber\\&&
+\frac{2}{5}q_{5,4}(\g) (99 \g^4-62 \g^2+13) (\g^2-1)^{1/2}\nonumber\\&&
-\frac{8}{5} q_{7,4}(\g) (\g^2-1)^{5/2} 
\,,\nonumber\\
\chi_{7,6}(\g) &=&   -\frac{8}{5} q_{7,6}(\g) (\g^2-1)^{5/2}  \,.
\end{eqnarray}

The analogous 8PM links read:
\begin{widetext}
\begin{eqnarray} \label{chi8q8}
\chi_{8,3}(\g) &=&   \frac{945}{128}\pi (\g^2-1) (-1+5 \g^2) q_{3,2}^2(\g) \nonumber\\
&+&\pi
\left[\frac{35}{64}   (\g^2-1) (47 \g^2-11) q_{4,3}(\g)
+\frac{35}{2048 (\g^2-1)}(-18032 \g^2+92698 \g^4  +889+40485 \g^8-139080 \g^6)  \right] q_{3,2}(\g)\nonumber\\
&-&
\frac{35}{1024 \pi (\g^2-1)} (65+2792 \g^6-1590 \g^4+301 \g^8-32 \g^2) q_{4,3}(\g) \nonumber\\
&+&
\frac{105}{256}\pi  (\g^2-1) (-1+5 \g^2) (47 \g^2-11) q_{5,2}(\g)\nonumber\\
&+&
\frac{35}{64} \pi  (33 \g^4-19 \g^2+4) (\g^2-1) q_{6,3}(\g)
-\frac{35}{64}\pi  (\g^2-1)^3 q_{8,3}(\g)\nonumber\\
&-&
\pi\frac{945}{8192 (\g^2-1)}(-1+5 \g^2)^2 (185 \g^6-3359 \g^4+1627 \g^2-85)\,, \nonumber\\
\chi_{8,5}(\g) &=&   -\pi \frac{945}{256} (\g^2-1) (-1+5 \g^2) q_{3,2}(\g)^2
-\pi \frac{35}{128} (47 \g^2-11) (\g^2-1)q_{4,3}(\g)q_{3,2}(\g)\nonumber\\
&+&
\pi \frac{2835}{512} (-1+5 \g^2)^2 (\g^2-1)q_{4,3}(\g)
+\pi \frac{105}{256} (47 \g^2-11) (-1+5 \g^2) (\g^2-1)q_{5,4}(\g) \nonumber\\
&+&
\pi \frac{35}{64} (33 \g^4-19 \g^2+4) (\g^2-1)q_{6,5}(\g)  
-\pi \frac{35}{64}(\g^2-1)^3  q_{8,5}(\g)\,, \nonumber\\
\chi_{8,7}(\g) &=&   -\pi \frac{35}{64}(\g^2-1)^3 q_{8,7}(\g)   \,.
\end{eqnarray}
\end{widetext}


\begin{thebibliography}{99}



\bibitem{Bini2019}
  D.~Bini, T.~Damour and A.~Geralico,
  ``Novel approach to binary dynamics: application to the fifth post-Newtonian level,''
  Phys.\ Rev.\ Lett.\  {\bf 123}, no. 23, 231104 (2019)
  [arXiv:1909.02375 [gr-qc]].
   
\bibitem{Bini2020}
  D.~Bini, T.~Damour and A.~Geralico,
  ``Binary dynamics at the fifth and fifth-and-a-half post-Newtonian orders,''
  arXiv:2003.11891 [gr-qc].
  
	
\bibitem{Wheeler:1949hn} 
  J.~A.~Wheeler and R.~P.~Feynman,
  ``Classical electrodynamics in terms of direct interparticle action,''
  Rev.\ Mod.\ Phys.\  {\bf 21}, 425 (1949).
  
\bibitem{Infeld:1957zz} 
  L.~Infeld,
  ``Equations of Motion in General Relativity Theory and the Action Principle,''
  Rev.\ Mod.\ Phys.\  {\bf 29}, 398 (1957).

\bibitem{Damour:1995kt} 
  T.~Damour and G.~Esposito-Farese,
  ``Testing gravity to second postNewtonian order: A Field theory approach,''
  Phys.\ Rev.\ D {\bf 53}, 5541 (1996)
  [gr-qc/9506063].

\bibitem{Blanchet:1987wq} 
  L.~Blanchet and T.~Damour,
  ``Tail Transported Temporal Correlations in the Dynamics of a Gravitating System,''
  Phys.\ Rev.\ D {\bf 37}, 1410 (1988).
 
\bibitem{Blanchet:1989ki} 
  L.~Blanchet and T.~Damour,
  ``Post-newtonian Generation of Gravitational Waves,''
  Ann.\ Inst.\ H.\ Poincare Phys.\ Theor.\  {\bf 50}, 377 (1989). 
  
\bibitem{Damour:1990ji}
T.~Damour and B.~R.~Iyer,
``PostNewtonian generation of gravitational waves. 2. The Spin moments,''
Ann.\ Inst.\ H.\ Poincare Phys.\ Theor.\  \textbf{54}, 115-164 (1991)
	
\bibitem{Blanchet:1998in} 
  L.~Blanchet,
  ``On the multipole expansion of the gravitational field,''
  Class.\ Quant.\ Grav.\  {\bf 15}, 1971 (1998)
  [gr-qc/9801101].
	
\bibitem{Poujade:2001ie} 
  O.~Poujade and L.~Blanchet,
  ``Post-Newtonian approximation for isolated systems calculated by matched asymptotic expansions,''
  Phys.\ Rev.\ D {\bf 65}, 124020 (2002)
  [gr-qc/0112057].

\bibitem{Blanchet:1985sp} 
  L.~Blanchet and T.~Damour,
  ``Radiative gravitational fields in general relativity I. general structure of the field outside the source,''
  Phil.\ Trans.\ Roy.\ Soc.\ Lond.\ A {\bf 320}, 379 (1986).
  
\bibitem{Foffa:2011np} 
 S.~Foffa and R.~Sturani,
 ``Tail terms in gravitational radiation reaction via effective field theory,''
 Phys.\ Rev.\ D {\bf 87}, no. 4, 044056 (2013)
 [arXiv:1111.5488 [gr-qc]].
	
\bibitem{Galley:2015kus} 
  C.~R.~Galley, A.~K.~Leibovich, R.~A.~Porto and A.~Ross,
   ``Tail effect in gravitational radiation reaction: Time nonlocality and renormalization group evolution,''
  Phys.\ Rev.\ D {\bf 93}, 124010 (2016)
  [arXiv:1511.07379 [gr-qc]].

	
\bibitem{Goldberger:2004jt} 
  W.~D.~Goldberger and I.~Z.~Rothstein,
  ``An Effective field theory of gravity for extended objects,''
  Phys.\ Rev.\ D {\bf 73}, 104029 (2006)
  [hep-th/0409156].
  
\bibitem{Goldberger:2009qd} 
  W.~D.~Goldberger and A.~Ross,
  ``Gravitational radiative corrections from effective field theory,''
  Phys.\ Rev.\ D {\bf 81}, 124015 (2010)
  [arXiv:0912.4254 [gr-qc]].
  
\bibitem{Ross:2012fc} 
  A.~Ross,
  ``Multipole expansion at the level of the action,''
  Phys.\ Rev.\ D {\bf 85}, 125033 (2012)
  [arXiv:1202.4750 [gr-qc]].
  
	
\bibitem{Damour:2014jta} 
  T.~Damour, P.~Jaranowski and G.~Sch\"afer,
  ``Nonlocal-in-time action for the fourth post-Newtonian conservative dynamics of two-body systems,''
  Phys.\ Rev.\ D {\bf 89}, no. 6, 064058 (2014)
  [arXiv:1401.4548 [gr-qc]].

\bibitem{Bernard:2015njp} 
  L.~Bernard, L.~Blanchet, A.~Boh\'e, G.~Faye and S.~Marsat,
  ``Fokker action of nonspinning compact binaries at the fourth post-Newtonian approximation,''
  Phys.\ Rev.\ D {\bf 93}, no. 8, 084037 (2016)
  [arXiv:1512.02876 [gr-qc]].
	
\bibitem{Marchand:2017pir} 
  T.~Marchand, L.~Bernard, L.~Blanchet and G.~Faye,
  ``Ambiguity-Free Completion of the Equations of Motion of Compact Binary Systems at the Fourth Post-Newtonian Order,''
  Phys.\ Rev.\ D {\bf 97}, no. 4, 044023 (2018)
  [arXiv:1707.09289 [gr-qc]].
  
\bibitem{Foffa:2019rdf} 
  S.~Foffa and R.~Sturani,
  ``Conservative dynamics of binary systems to fourth Post-Newtonian order in the EFT approach I: Regularized Lagrangian,''
  Phys.\ Rev.\ D {\bf 100}, no. 2, 024047 (2019)
  [arXiv:1903.05113 [gr-qc]].
  
\bibitem{Foffa:2019yfl} 
  S.~Foffa, R.~A.~Porto, I.~Rothstein and R.~Sturani,
  ``Conservative dynamics of binary systems to fourth Post-Newtonian order in the EFT approach II: Renormalized Lagrangian,''
  Phys.\ Rev.\ D {\bf 100}, no. 2, 024048 (2019)
  [arXiv:1903.05118 [gr-qc]].
	
\bibitem{Damour2010}
  T.~Damour, 2010 (unpublished); cited in L.~Barack, T.~Damour, and N.~Sago, ``Precession effect of the gravitational self-force in a Schwarzschild spacetime and the effective one-body formalism,'' Phys.\ Rev.\ D {\bf 82}, 084036 (2010), which quoted and used some combinations of the (4PN and 5PN) logarithmic contributions to the EOB potentials $A(u)$ and ${\bar D}(u)$.
  
\bibitem{Damour:2015isa} 
  T.~Damour, P.~Jaranowski and G.~Sch\"afer,
   ``Fourth post-Newtonian effective one-body dynamics,''
  Phys.\ Rev.\ D {\bf 91}, no. 8, 084024 (2015)
  [arXiv:1502.07245 [gr-qc]].
	
\bibitem{Damour:1990pi} 
  T.~Damour, M.~Soffel and C.~m.~Xu,
  ``General relativistic celestial mechanics. 1. Method and definition of reference systems,''
  Phys.\ Rev.\ D {\bf 43}, 3273 (1991).
  
\bibitem{Damour:1991yw} 
  T.~Damour, M.~Soffel and C.~m.~Xu,
  ``General relativistic celestial mechanics. 2. Translational equations of motion,''
  Phys.\ Rev.\ D {\bf 45}, 1017 (1992).
  
\bibitem{Foffa:2019eeb} 
  S.~Foffa and R.~Sturani,
  ``Hereditary Terms at Next-To-Leading Order in Two-Body Gravitational Dynamics,''
  arXiv:1907.02869 [gr-qc].

\bibitem{Blanchet:2019rjs} 
  L.~Blanchet, S.~Foffa, F.~Larrouturou and R.~Sturani,
  ``Logarithmic tail contributions to the energy function of circular compact binaries,''
  arXiv:1912.12359 [gr-qc].
  
\bibitem{Levi:2018nxp}
M.~Levi,
``Effective Field Theories of Post-Newtonian Gravity: A comprehensive review,''
[arXiv:1807.01699 [hep-th]].
  
	
\bibitem{Damour:2009sm} 
  T.~Damour,
  ``Gravitational Self Force in a Schwarzschild Background and the Effective One Body Formalism,''
  Phys.\ Rev.\ D {\bf 81}, 024017 (2010)
  [arXiv:0910.5533 [gr-qc]].  
  
\bibitem{Blanchet:2010zd} 
  L.~Blanchet, S.~L.~Detweiler, A.~Le Tiec and B.~F.~Whiting,
   ``High-Order Post-Newtonian Fit of the Gravitational Self-Force for Circular Orbits in the Schwarzschild Geometry,''
  Phys.\ Rev.\ D {\bf 81}, 084033 (2010)
  [arXiv:1002.0726 [gr-qc]].

\bibitem{Blanchet:2013haa} 
  L.~Blanchet,
  ``Gravitational Radiation from Post-Newtonian Sources and Inspiralling Compact Binaries,''
  Living Rev.\ Rel.\  {\bf 17}, 2 (2014)
  [arXiv:1310.1528 [gr-qc]]. 
	
  
\bibitem{Damour:1994pk} 
  T.~Damour and B.~R.~Iyer,
  ``Generation of gravitational waves: The PostNewtonian spin octupole moment,''
  Class.\ Quant.\ Grav.\  {\bf 11}, 1353 (1994)
  Erratum: [Class.\ Quant.\ Grav.\  {\bf 12}, 287 (1995)].
	
\bibitem{Blanchet:1995fr} 
  L.~Blanchet,
  ``Second postNewtonian generation of gravitational radiation,''
  Phys.\ Rev.\ D {\bf 51}, 2559 (1995)
  [gr-qc/9501030].
  
\bibitem{Blanchet:1995fg} 
  L.~Blanchet, T.~Damour and B.~R.~Iyer,
  ``Gravitational waves from inspiralling compact binaries: Energy loss and wave form to second postNewtonian order,''
  Phys.\ Rev.\ D {\bf 51}, 5360 (1995)
  Erratum: [Phys.\ Rev.\ D {\bf 54}, 1860 (1996)]
  [gr-qc/9501029].
	
\bibitem{DD1981a}
 T.~Damour and N.~Deruelle, 
``Lagrangien g\'en\'eralis\'e du syst\`eme de deux masses ponctuelles,
\`a l'appro\-ximation post-post-newtonienne de la relativit\'e
g\'en\'erale,''
 C.R.\ Acad.\ Sc.\ Paris, S\'erie II, {\bf 293}, pp 537-540 (1981).  
 
\bibitem{D1982} 
T.~Damour;
``Probl\`eme des deux corps et freinage de rayonnement en relativit\'e
g\'en\'erale,''
 C.R.\ Acad.\ Sc.\ Paris, S\'erie II, {\bf 294}, pp 1355-1357 (1982).
 
\bibitem{Damour:1990jh}
T.~Damour and G.~Schaefer,
``Redefinition of position variables and the reduction of higher order Lagrangians,''
J.\ Math.\ Phys.\  \textbf{32}, 127-134 (1991)

\bibitem{Schaefer:1986rd}
G.~Schaefer,
``The Gravitational Quadrupole Radiation Reaction Force and the Canonical Formalism of Adm,''
Annals Phys.\  \textbf{161}, 81-100 (1985)
 
\bibitem{Arun:2007rg} 
  K.~G.~Arun, L.~Blanchet, B.~R.~Iyer and M.~S.~S.~Qusailah,
   ``Tail effects in the 3PN gravitational wave energy flux of compact binaries in quasi-elliptical orbits,''
  Phys.\ Rev.\ D {\bf 77}, 064034 (2008)
  [arXiv:0711.0250 [gr-qc]].

\bibitem{Damour:1988mr} 
  T.~Damour and G.~Sch\"afer,
   ``Higher Order Relativistic Periastron Advances and Binary Pulsars,''
  Nuovo Cim.\ B {\bf 101}, 127 (1988).
   
	\bibitem{SW1993} 
 G.~Sch\"afer and N.~Wex, 
 ``Second post-Newtonian motion of compact binaries,'' 
 Phys.\ Lett.\ A, {\bf 174}, 196 (1993)
  Erratum: [Phys.\ Lett.\ A, {\bf 177}, 461 (1993).]

\bibitem{Memmesheimer:2004cv} 
  R.~M.~Memmesheimer, A.~Gopakumar and G.~Schaefer,
  ``Third post-Newtonian accurate generalized quasi-Keplerian parametrization for compact binaries in eccentric orbits,''
  Phys.\ Rev.\ D {\bf 70}, 104011 (2004)
  [gr-qc/0407049].

\bibitem{DD1981}
T.~Damour and N.~Deruelle,
``Radiation Reaction and Angular Momentum Loss in Small Angle Gravitational Scattering,''
Phys.\ Lett.\ A {\bf 87}, 81 (1981)

\bibitem{Damour:2000we}
T.~Damour, P.~Jaranowski, and G.~Sch\"afer,
``On the determination of the last stable orbit for circular general relativistic binaries at the third post-Newtonian approximation,''
Phys.\ Rev.\ D {\bf 62}, 084011 (2000)
[arXiv:gr-qc/0005034].

\bibitem{LeTiec:2011ab} 
  A.~Le Tiec, L.~Blanchet and B.~F.~Whiting,
  ``The First Law of Binary Black Hole Mechanics in General Relativity and Post-Newtonian Theory,''
  Phys.\ Rev.\ D {\bf 85}, 064039 (2012)
  [arXiv:1111.5378 [gr-qc]].

\bibitem{Barausse:2011dq} 
  E.~Barausse, A.~Buonanno and A.~Le Tiec,
  ``The complete non-spinning effective-one-body metric at linear order in the mass ratio,''
  Phys.\ Rev.\ D {\bf 85}, 064010 (2012)
  [arXiv:1111.5610 [gr-qc]].

\bibitem{Tiec:2015cxa}
  A.~Le Tiec,
   ``First Law of Mechanics for Compact Binaries on Eccentric Orbits,''
  Phys.\ Rev.\ D {\bf 92}, no. 8, 084021 (2015)
  [arXiv:1506.05648 [gr-qc]].
 
\bibitem{Detweiler:2008ft} 
  S.~L.~Detweiler,
   ``A Consequence of the gravitational self-force for circular orbits of the Schwarzschild geometry,''
  Phys.\ Rev.\ D {\bf 77}, 124026 (2008)
  [arXiv:0804.3529 [gr-qc]].

\bibitem{Barack:2011ed} 
  L.~Barack and N.~Sago,
   ``Beyond the geodesic approximation: conservative effects of the gravitational self-force in eccentric orbits around a Schwarzschild black hole,''
  Phys.\ Rev.\ D {\bf 83}, 084023 (2011)
  [arXiv:1101.3331 [gr-qc]].
  
\bibitem{Bini:2013zaa} 
  D.~Bini and T.~Damour,
  ``Analytical determination of the two-body gravitational interaction potential at the fourth post-Newtonian approximation,''
  Phys.\ Rev.\ D {\bf 87}, no. 12, 121501 (2013)
  [arXiv:1305.4884 [gr-qc]].

\bibitem{Bini:2013rfa}
D.~Bini and T.~Damour,
``High-order post-Newtonian contributions to the two-body gravitational interaction potential from analytical gravitational self-force calculations,''
Phys.\ Rev.\ D \textbf{89}, no.6, 064063 (2014)
[arXiv:1312.2503 [gr-qc]].

	\bibitem{Bini:2015bla} 
  D.~Bini and T.~Damour,
   ``Detweiler's gauge-invariant redshift variable: Analytic determination of the nine and nine-and-a-half post-Newtonian self-force contributions,''
  Phys.\ Rev.\ D {\bf 91}, 064050 (2015)
  [arXiv:1502.02450 [gr-qc]].
	
	\bibitem{Kavanagh:2015lva} 
  C.~Kavanagh, A.~C.~Ottewill and B.~Wardell,
   ``Analytical high-order post-Newtonian expansions for extreme mass ratio binaries,''
  Phys.\ Rev.\ D {\bf 92}, no. 8, 084025 (2015)
  [arXiv:1503.02334 [gr-qc]].
	
\bibitem{Bini:2015bfb}
  D.~Bini, T.~Damour and A.~Geralico,
  ``Confirming and improving post-Newtonian and effective-one-body results from self-force computations along eccentric orbits around a Schwarzschild black hole,''
  Phys.\ Rev.\ D {\bf 93}, 064023 (2016)
  [arXiv:1511.04533 [gr-qc]].

\bibitem{Bini:2016qtx}
  D.~Bini, T.~Damour and A.~Geralico,
  ``New gravitational self-force analytical results for eccentric orbits around a Schwarzschild black hole,''
  Phys.\ Rev.\ D {\bf 93}, 104017 (2016)
 arXiv:1601.02988 [gr-qc].

\bibitem{Hopper:2015icj}
  S.~Hopper, C.~Kavanagh and A.~C.~Ottewill,
  ``Analytic self-force calculations in the post-Newtonian regime: eccentric orbits on a Schwarzschild background,''
  Phys.\ Rev.\ D {\bf 93}, 044010 (2016)
  [arXiv:1512.01556 [gr-qc]].
	
\bibitem{Damour:2017zjx} 
  T.~Damour,
   ``High-energy gravitational scattering and the general relativistic two-body problem,''
  Phys.\ Rev.\ D {\bf 97}, no. 4, 044038 (2018)
  [arXiv:1710.10599 [gr-qc]].

\bibitem{Damour2019} 
  T.~Damour,
  ``Classical and Quantum Scattering in Post-Minkowskian Gravity,''
  arXiv:1912.02139 [gr-qc].
  
\bibitem{Antonelli:2020aeb}
A.~Antonelli, C.~Kavanagh, M.~Khalil, J.~Steinhoff and J.~Vines,
``Gravitational spin-orbit coupling through third-subleading post-Newtonian order: from first-order self-force to arbitrary mass ratios,''
[arXiv:2003.11391 [gr-qc]].  

\bibitem{Cheung:2018wkq} 
  C.~Cheung, I.~Z.~Rothstein and M.~P.~Solon,
  ``From Scattering Amplitudes to Classical Potentials in the Post-Minkowskian Expansion,''
  Phys.\ Rev.\ Lett.\  {\bf 121}, no. 25, 251101 (2018)
  [arXiv:1808.02489 [hep-th]].

\bibitem{Bern:2019nnu} 
  Z.~Bern, C.~Cheung, R.~Roiban, C.~H.~Shen, M.~P.~Solon and M.~Zeng,
  ``Scattering Amplitudes and the Conservative Hamiltonian for Binary Systems at Third Post-Minkowskian Order,''
  Phys.\ Rev.\ Lett.\  {\bf 122}, no. 20, 201603 (2019)
  [arXiv:1901.04424 [hep-th]].

   
\bibitem{Bern:2019crd}
Z.~Bern, C.~Cheung, R.~Roiban, C.~Shen, M.~P.~Solon and M.~Zeng,
``Black Hole Binary Dynamics from the Double Copy and Effective Theory,''
JHEP \textbf{10}, 206 (2019)
[arXiv:1908.01493 [hep-th]].  
	
\bibitem{Blumlein:2020znm} 
  J.~Bl\"umlein, A.~Maier, P.~Marquard and G.~Sch\"afer,
  ``Testing binary dynamics in gravity at the sixth post-Newtonian level,''
  arXiv:2003.07145 [gr-qc].

\bibitem{Cheung:2020gyp} 
  C.~Cheung and M.~P.~Solon,
  ``Classical Gravitational Scattering at ${\cal O}(G^3)$ from Feynman Diagrams,''
  arXiv:2003.08351 [hep-th].

\bibitem{Antonelli:2019ytb} 
  A.~Antonelli, A.~Buonanno, J.~Steinhoff, M.~van de Meent and J.~Vines,
  ``Energetics of two-body Hamiltonians in post-Minkowskian gravity,''
  Phys.\ Rev.\ D {\bf 99}, no. 10, 104004 (2019)
  [arXiv:1901.07102 [gr-qc]].

\bibitem{Damour:2016gwp} 
  T.~Damour,
  ``Gravitational scattering, post-Minkowskian approximation and Effective One-Body theory,''
  Phys.\ Rev.\ D {\bf 94}, no. 10, 104015 (2016)
  [arXiv:1609.00354 [gr-qc]]. 
	
	
\bibitem{Bini:2017wfr} 
  D.~Bini and T.~Damour,
  ``Gravitational scattering of two black holes at the fourth post-Newtonian approximation,''
  Phys.\ Rev.\ D {\bf 96}, no. 6, 064021 (2017)
  [arXiv:1706.06877 [gr-qc]].
  
		
\bibitem{Kalin:2019inp}
G.~K\"alin and R.~A.~Porto,
``From boundary data to bound states. Part II. Scattering angle to dynamical invariants (with twist),''
JHEP \textbf{02}, 120 (2020)
[arXiv:1911.09130 [hep-th]]. 

	
\bibitem{Pound:2019lzj}
A.~Pound, B.~Wardell, N.~Warburton and J.~Miller,
``Second-Order Self-Force Calculation of Gravitational Binding Energy in Compact Binaries,''
Phys.\ Rev.\ Lett.\  \textbf{124}, no.2, 021101 (2020)
[arXiv:1908.07419 [gr-qc]].

\end{thebibliography}
\end{document}